\documentclass[final,3p,times]{elsarticle}
\usepackage{graphicx} 
\usepackage[utf8]{inputenc}
\usepackage{caption}
\usepackage{subcaption}
\usepackage{siunitx}
\usepackage{float} 
\usepackage{setspace}
\usepackage{amsmath}
\usepackage{cite}
\usepackage{cleveref}
\usepackage[labelfont=bf,labelsep=colon]{caption}
\usepackage{placeins}

\def\tsc#1{\csdef{#1}{\textsc{\lowercase{#1}}\xspace}}
\tsc{WGM}
\tsc{QE}
\tsc{EP}
\tsc{PMS}
\tsc{BEC}
\tsc{DE}

\begin{document}
\title{Integrating Gaussian Random Functions with Genetic Algorithms for the Optimization of Functionally Graded Lattice Structures}

\author{Piyush Agrawal\textsuperscript{a}}
\author{Manish Agrawal\textsuperscript{a}\textsuperscript{,*}}
\ead{manish.agrawal@iitrpr.ac.in}
\cortext[cor1]{Corresponding author.}

\address[1]{Department of Mechanical Engineering, Indian Institute of Technology Ropar, Rupnagar-140001, Punjab, India}

\begin{abstract}
  The properties of lattice-based structures can be enhanced by varying their geometric parameters in a graded manner, and the gradation can be tailored to extremize a particular objective. In this manuscript, we propose a non-gradient-based optimization framework to find the tailor-made graded profiles for lattice-based structures. The key challenge addressed in the work is to ensure the graded nature/smoothness of the underlying structure in a non-gradient-based optimization scheme. As we demonstrate in the manuscript, the conventional implementation of the genetic algorithm provides structures with abrupt changes, leading to issues such as stress concentration. In this work, we propose a Gaussian random function (GRF)/Gaussian process regression (GPR) integrated genetic algorithm to obtain an optimal graded lattice profile for a given objective. The integration of the GRF/GPR along with a projection operator ensures the smoothness of the designs at each stage of the optimization. We present several numerical examples to demonstrate that the proposed framework provides smoother designs that are less susceptible to stress concentration, while ensuring satisfaction of the underlying objective.   
\end{abstract}

\begin{keyword}
    Functionally graded lattice structure, Gaussian random field, Gaussian process regression, Genetic algorithm.
\end{keyword}
 
\maketitle

\section{Introduction}

In recent years, advancements in additive manufacturing (AM) processes have enabled the fabrication of highly complex geometries, including lattice structures. Modern AM processes such as Fused Deposition Modeling \citep{kumar2020supportless}, Directed Energy Deposition \citep{saboori2017overview}, Binder Jetting \citep{xu2023mechanical}, and Powder Bed Fusion \citep{yang2025additive}, have played a crucial role in translating complex designs into practical components, which are nearly impossible to manufacture using conventional manufacturing techniques. In addition, processes like Vat Photopolymerization \citep{golubovic2023comprehensive} and Material Jetting (MJ) have further expanded AM capabilities, offering excellent dimensional accuracy, high surface finish, and the ability to produce highly precise complex geometries. Lattice structures, one of the applications of these advanced manufacturing processes, represent a class of advanced heterogeneous materials composed of periodically or aperiodically repeated unit cells interconnected to form a continuous network \citep{helou2018design}. They are the designs inspired by biological structures, for example, the honeycomb of bees and bones, which provide high energy absorption capacity. Also, the lattice structures of the auxetic type are capable of providing a negative Poisson's ratio in the structures. In general, lattice structures offer a high stiffness-to-weight ratio, making them highly attractive for biomedical implants, aerospace structures, and mechanical systems applications.

To enhance the performance of lattice-based structures, the geometric parameters of the unit cells can be varied across the structure. In particular, the geometric parameters of the unit cells, such as strut thickness, orientation, or spacing, vary gradually across the structure. The gradual variation of the geometric parameters ensures the absence of stress concentration within the structure, while tailor-made gradation for a specific objective enhances the overall functionality of the underlying structure.

In the literature, a wide range of optimization strategies has been developed for functionally graded lattice (FGL) structures, which are broadly classified into gradient-based and non-gradient-based approaches. Gradient-based methods include the solid isotropic material with penalization (SIMP) method \citep{bendsoe1999material,johnsen2013structural,trudel2022penalization}, the method of moving asymptotes (MMA) \citep{cheng2019functionally}, and level set methods (LSM) \citep{liu2022kriging, sivapuram2016simultaneous}. On the other hand, non-gradient-based methods such as particle swarm optimization (PSO) \citep{tikani2024bandgap} and genetic algorithms (GA) \citep{javadi2012design, pokkalla2021isogeometric} have been extensively applied. GA-based techniques have demonstrated strong potential for optimization under broad objectives and diverse constraint conditions. For example, Kappe et al. \citep{kappe2024multi} employed non-dominated sorting genetic algorithm (NSGA-II) to minimize peak crushing force while maximizing energy absorption capacity in body centered cubic (BCC) and re-entrant unit cells under dynamic loading. Similarly, Nian et al. \citep{nian2021energy} applied NSGA-II to BCC-type 3D unit cells with circular geometries, aiming to maximize specific energy absorption while reducing peak impact force. Using GA, Ozdemir et al. \citep{ozdemir2023novel} performed multi-morphology optimization of triply periodic minimal surface (TPMS) structures, while Mahbod and Asgari \citep{mahbod2020multiobjective} optimized double-dodecahedron unit cells to maximize elastic modulus in both the X- and Y-directions. Han and Lu \citep{han2018evolutionary} tailor the deformation pattern of a structure composed of re-entrant unit cells using a GA framework.

A key challenge in non-gradient-based optimization frameworks is generating a smooth profile for lattice parameter variation. The traditional method for profile generation includes simple schemes such as the power law, the B-spline method, and the exponential law. But the design representation of these schemes is very limited, hence they may produce sub-optimal results. On the other end of the spectrum, the profiles can be generated in a totally random manner i.e. each of the design variables is independently generated between a min-max range.  Although this method offers diversity, it produces non-smooth gradation patterns. This non-sooth gradation in turn can produce  higher stress concentrations, leading to a reduction in the strength of the FGL structures. This limitation motivates the need to provide a profile generation algorithm that guarantees smooth profile generation while offering diversity in the design space.

In this manuscript, we propose a profile generation algorithm based on Gaussian Random Field (GRF) and Gaussian Process Regression (GPR). The proposed profile generation scheme, along with providing a large diversity in the design space, ensures a smooth transition of geometric parameters between adjacent unit cells. A length-scale hyperparameter governs the degree of smoothness of the unit-cell geometry. The mathematical foundations of GRF given by Kolmogorov on stochastic processes \citep{Kolmogorov1940}, later extended by Matérn \citep{Matern1960} and Yaglom \citep{Yaglom1962}, who established links between GRF, covariance functions, and spectral representations. In parallel, the machine learning community employed GPR \citep{williams2006gaussian} for regression tasks.  In engineering applications, GPR has been used as a surrogate model to predict stress intensity factors \citep{loghin2019augmenting} and to replace finite element simulations in the reliability analysis of complex domains \citep{su2017gaussian}. Beyond surrogate modeling, recently GPR-based design algorithms have also been employed to determine the material gradation within the complex functionally graded material domains \citep{konda2025fgm}. However, to the best of the authors' knowledge, this is the first attempt to employ the GRF/GPR-based profile generation scheme towards optimization of functionally graded lattice structures.

The GRF/GPR-based profile generation algorithm for lattice structures has been further integrated with a GA to optimize FGL structures under diverse objective conditions. The genetic algorithm involves crossover and mutation operations; applying these operations on smooth profiles might result in non-smooth profiles, even if the parents' profiles are completely smooth in nature. We used an additional projection operator, projecting non-smooth profiles into smooth profiles. To demonstrate the effectiveness of the proposed approach, two widely studied lattice topologies, centered-rectangular and re-entrant unit cells under two-dimensional consideration, are taken as case studies. The centered-rectangular type unit cell is known for its ability to absorb high energy and provide good mechanical properties. While a re-entrant unit cell exhibits auxetic (negative Poisson’s ratio) behavior. The performance of GRF-based designs is compared against conventional implementation in terms of stress distribution, strength, and optimization outcomes.

In summary, the key contributions of this work can be summarized as follows:
\begin{itemize}
\item Development of a GRF-based profile generation algorithm for functionally graded lattice structures.
\item Integration of GRF with a genetic algorithm for geometric parameter optimization.
\item Comparative study of centered-rectangular and re-entrant unit cells under GRF and conventional implementation.
\item Demonstration that GRF-based designs achieve smoother geometric transitions, reduced stress concentration, and enhanced structural strength.
\end{itemize}

The remainder of this manuscript is as follows. Section \ref{model} provides the brief description of the functionally graded lattice structure problem. The detailed explanation of the GRF/GPR-based FGL structures design algorithm is given in the section \ref{PGA}. Section \ref{FEA} presents the finite element formulation used for the FGL simulation. Section \ref{GA} gives the genetic algorithm-based framework modified for the FGL optimization in line with the GRF/GPR-based profile generation scheme. Section \ref{Numerical example} demonstrates the efficacy of the proposed GRF/GPR-based framework with a GA-based optimization algorithm through various numerical examples of the FGL structures composed of the centered-rectangular and re-entrant unit cells. Finally, the conclusion of the manuscript is given in the section \ref{Conclusion}

\section{Model}
\label{model}
\begin{figure}
    \centering
    \begin{subfigure}{0.27\textwidth}
        \includegraphics[width=\linewidth,trim=0.0 0.00 0.0 0.0,clip]{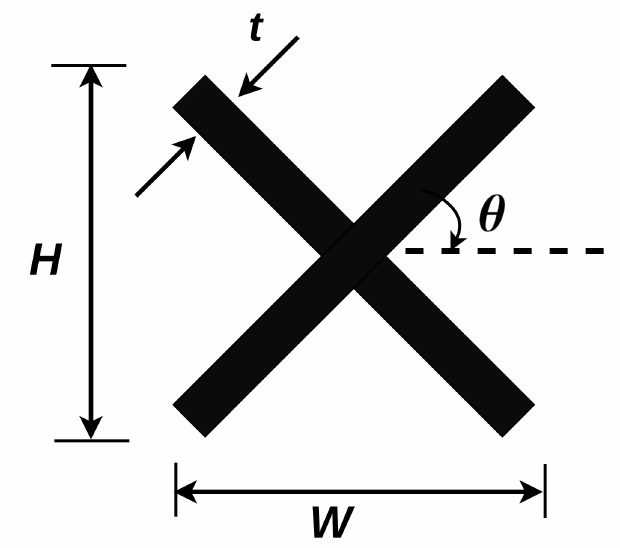}
        \caption{}
        \label{fig:sub1}
    \end{subfigure}
    \hspace{2cm}
    \begin{subfigure}{0.30\textwidth}
        \includegraphics[width=\linewidth,trim=0.0 0.00 0.0 0.0,clip]{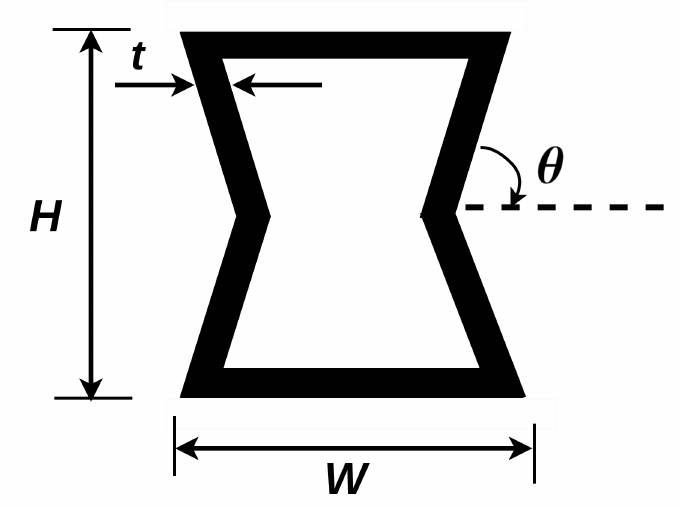}
        \caption{}
        \label{fig:sub2}
    \end{subfigure}
    
    \caption{The geometric parameter characterization of the (a) centered rectangular unit cell and (b) re-entrant unit cell.}
    \label{unit_cells1}
\end{figure}

In the manuscript, we consider the problem of designing functionally graded structures made of a single type of lattice cells. Although the type of lattice cells remains the same throughout the structure, we do not assume any restrictions on the type of unit cells; hence, the unit cell can be of any type, such as centered rectangular, re-entrant, etc.   A unit cell, depending upon the underlying geometry, can be uniquely characterized by various geometric parameters. For example, the centered-rectangular and the re-entrant unit cell can be characterized by the geometric parameters as shown in the Fig~\ref{unit_cells1}. Here, we intend to consider a subset of these geometric parameters as design variables and determine their optimal values to maximize a particular objective function. The main focus here is to ensure, in an optimal design the geometric parameters of the unit cells across the structure vary in a smooth and graded fashion. 

\begin{figure}[ht!]
    \centering
    \begin{subfigure}{0.40\textwidth}
        \includegraphics[width=\linewidth,trim=0.0 0.00 0.0 0.0,clip]{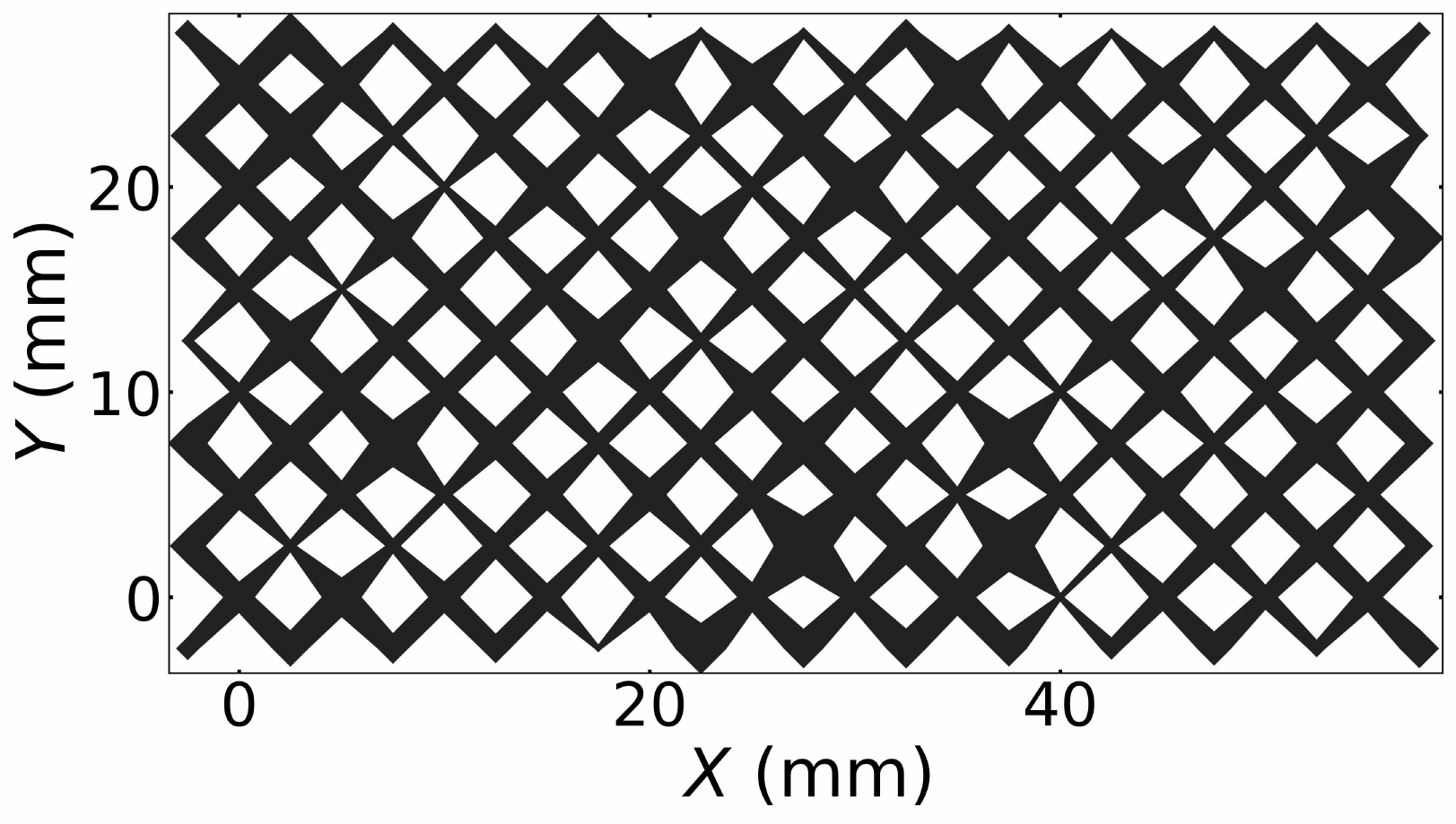}
        \caption{}
        \label{}
    \end{subfigure}
    \hspace{1.2cm}
    \begin{subfigure}{0.40\textwidth}
        \includegraphics[width=\linewidth,trim=0.0 0.00 0.0 0.0,clip]{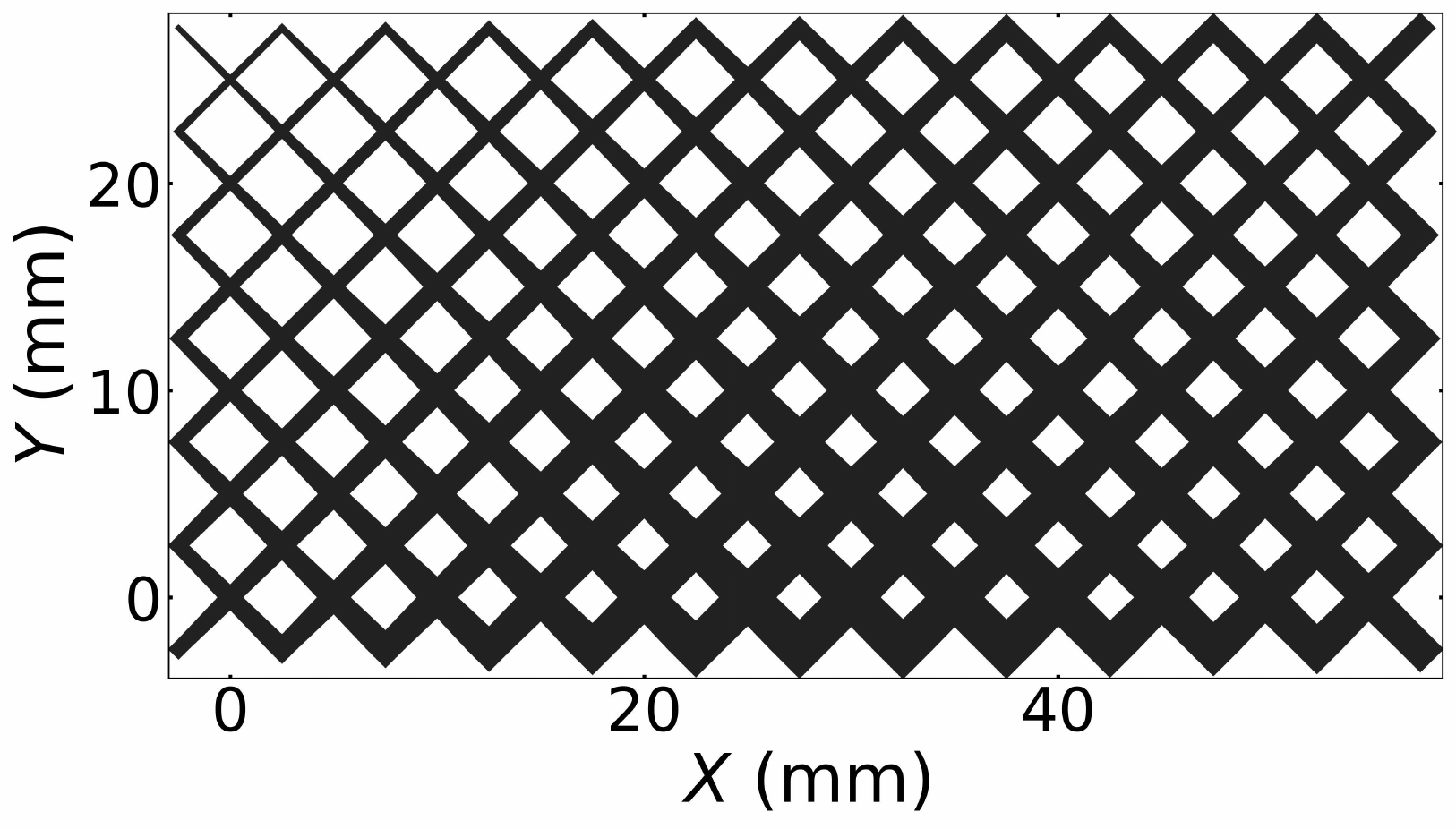}
        \caption{}
        \label{fig:sub2}
    \end{subfigure}    
    \caption{Sample FGL structures composed of centered rectangular unit cells generated by (a) considering design variables are uncorrelated in nature and (b) GRF-based profile generation algorithm (GRF parameters: $l$ = 30 mm and $\sigma$ = 0.60 mm).}
    \label{grf_prof_bcc}
\end{figure}

\begin{figure}
    \centering
    \begin{subfigure}{0.35\textwidth}
        \includegraphics[width=\linewidth,trim=0.0 0.00 0.0 0.0,clip]{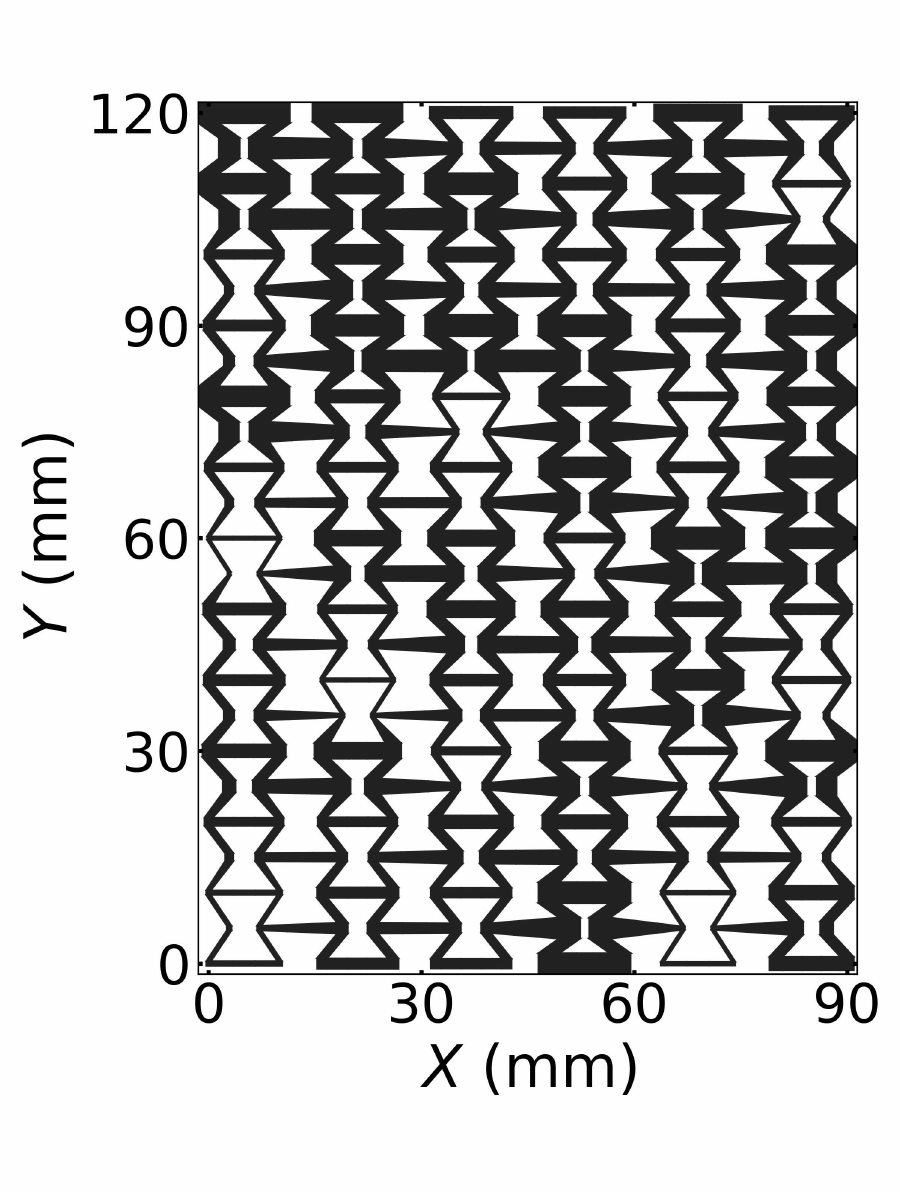}
        \caption{}
        \label{}
    \end{subfigure}
    \hspace{1.0cm}
    \begin{subfigure}{0.35\textwidth}
        \includegraphics[width=\linewidth,trim=0.0 0.00 0.0 0.0,clip]{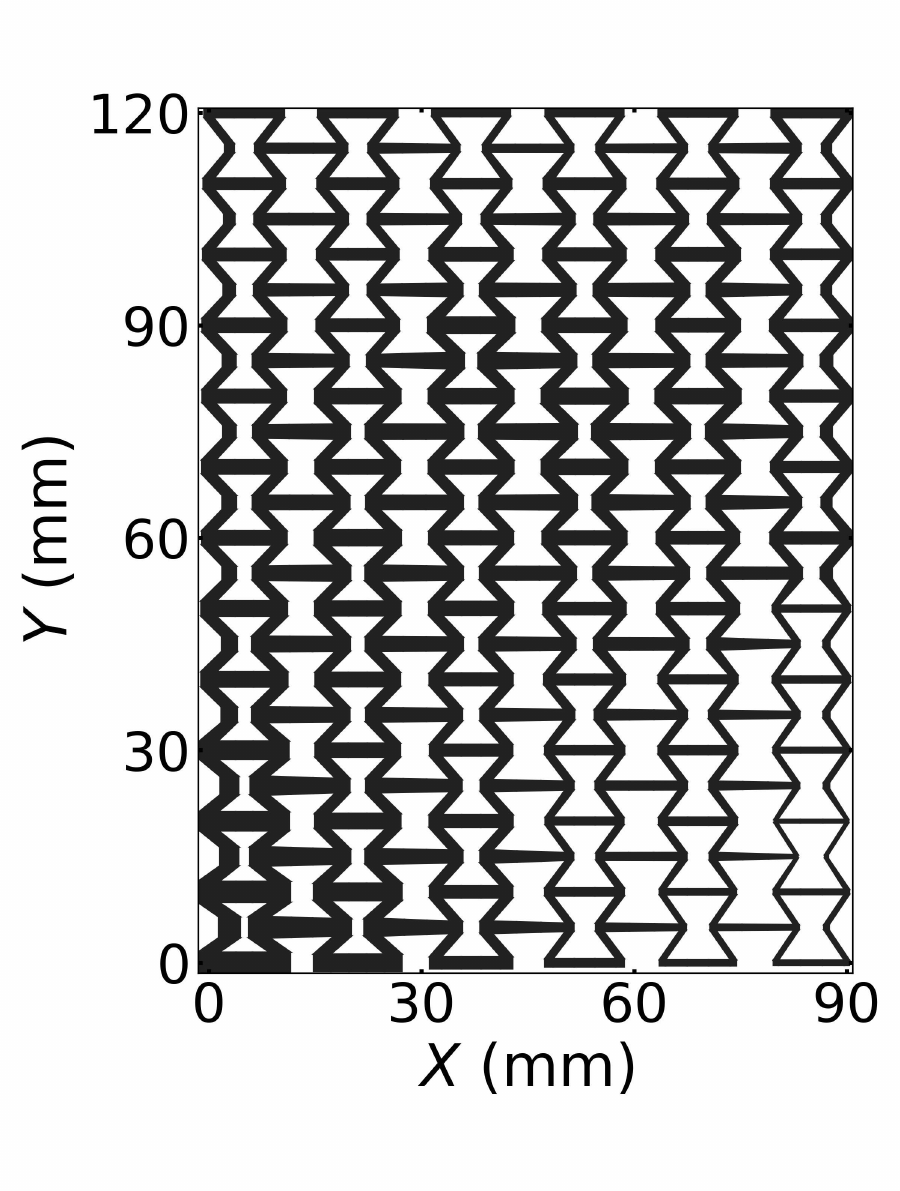}
        \caption{}
        \label{fig:sub2}
    \end{subfigure}    
    \caption{Sample FGL structures composed of re-entrant unit cells generated by (a) considering design variables are uncorrelated in nature and (b) GRF-based profile generation algorithm (GRF parameters: $l$ = 40 mm and $\sigma$ = 0.60 mm).}
    \label{grf_prof_reentrant}
\end{figure}

In order to carry out the optimization of lattice structures, first, we need to formulate the design space. Towards this objective, we consider that the structure is made of \(n\) number of unit cells and each of the cells is uniquely characterized by geometric design variables \((g_1,g_2,...)\). One way to formulate the design space is to consider for each lattice the geometric variables  \(g_i\in[g_{min},g_{max}], i=1,2,...n\) are independently generated from each other. This design space, obtained although diverse and generic in nature, will consist of numerous designs having a non-smooth transition of the geometric variables. The non-smoothness of the geometric variables might lead to issues such as high stress concentration. In the manuscript, we strive to formulate the design space, which is generic but does not contain designs with abrupt changes in geometric parameters. To illustrate the difference between non-smooth and smooth design spaces, an example for both of them has been shown in Figs~\ref{grf_prof_bcc} and~\ref{grf_prof_reentrant}. 

In summary, we strive to formulate the design space with the following properties: (1) large diversity in the design space, (2) the geometric parameters of the unit cells are smoothly varied across the structure, resulting in a continuous graded distribution that defines the final graded architecture, and (3) the incorporation of a length scale parameter ($l$) enables precise control over the smoothness of geometric transitions.

\section{Functionally graded lattice design space based on GRF/GPR}
\label{PGA}
This section demonstrates the strategy for generating the design space for functionally graded lattice structures. In order to satisfy the design space requirements listed in the above section,  we use the Gaussian random field/Gaussian process regression to generate the design space. The main advantages of the GRF/GPR-based approach are that it allows the precise control over the smoothness of geometric transitions by a length scale parameter ($l$).  Additionally, any geometric parameter constraint on the boundaries of the FGL structures can be incorporated using GPR.

\begin{figure}[h!]
    \centering
    \includegraphics[width=0.40\textwidth]{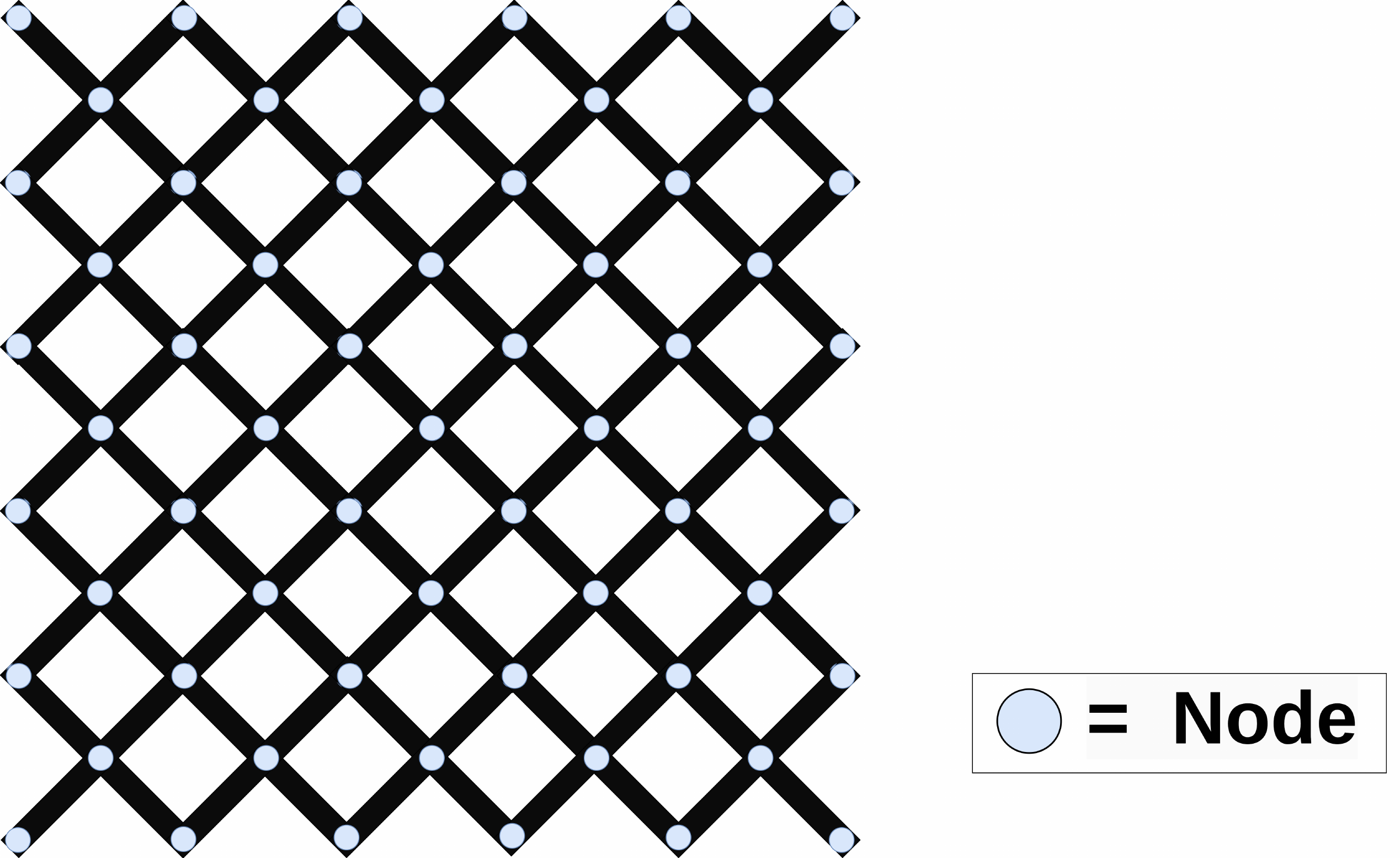} %
    \caption{FGL domain is discretized into the nodes, and each node is assigned a geometric parameter value.}
    \label{GRF}
\end{figure}

To implement the proposed GRF-based design scheme in the lattice structures, we discretize our domain into \(n\) number of nodes, as shown in Fig. \ref{GRF}, where each node is assigned geometric parameter values. In the discretized domain, nodes might represent the center of lattice cells or the location of the lattice members. The nodes are selected in a manner that geometric parameter (such as thickness, orientation, or other defining features) values at these nodes are sufficient to uniquely characterize the lattice structure. The value of the geometric parameters at other points is interpolated from the nodal values.

The spatial coordinates matrix ($\boldsymbol{X}$) serves as input to the GRF, while the geometric parameter at each node can be randomly generated using the multivariate Gaussian distribution, given by Eq. \eqref{GRF_eq}.

\begin{equation}
\begin{aligned}
    P(X) \sim \mathcal{N}(\mu(\boldsymbol{X}),\boldsymbol{K}),
     \label{GRF_eq}
\end{aligned}
\end{equation}

\noindent where,
$P(X)$ represents the distribution function of the geometric parameter in the unit cell, $\mathcal{N}$ is a multivariate Gaussian distribution, while $\mu(X)$ is the mean of the geometric distribution and $\boldsymbol{K}$ is the covariance matrix. In the GRF framework, the covariance matrix is defined using a kernel function to capture the spatial correlation structure. Literature provides several kernel functions, such as linear, polynomial, radial basis function (RBF), and sigmoidal functions. Our work uses the RBF kernel function given by the following equation: 
\begin{equation}
\begin{aligned}
    K_{ij} = k({X}_i, {X}_j) = \sigma^2 \exp\left( -\frac{\|{X}_i - {X}_j\|^2}{2l^2} \right).
    \label{rbf_kernel}
\end{aligned}
\end{equation}
The RBF kernel function is governed by two hyperparameters: length scale ($l$) and standard deviation ($\sigma$). The length scale parameter determines how quickly the correlation between two nodes decreases with distance. As the value of the length scale decreases, the change in the gradation pattern has more fluctuations, while the increase in length scale value results in a smoother gradation. The standard deviation determines the magnitude of the generated values or the span of the obtained values. The choice of the $l$ and $\sigma$ is crucial to develop the geometric parameters of the unit cells using GRF.

\begin{figure}
    \centering
    \begin{subfigure}{0.28\textwidth}
        \includegraphics[width=\linewidth,trim=0.0 0.00 0.0 0.0,clip]{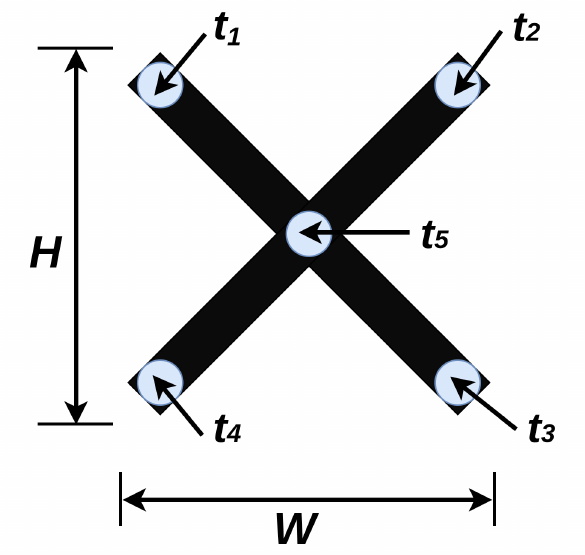}
        \caption{}
        \label{bcc2}
    \end{subfigure}
    \hspace{1.75cm}
    \begin{subfigure}{0.28\textwidth}
        \includegraphics[width=\linewidth,trim=0.0 0.00 0.0 0.0,clip]{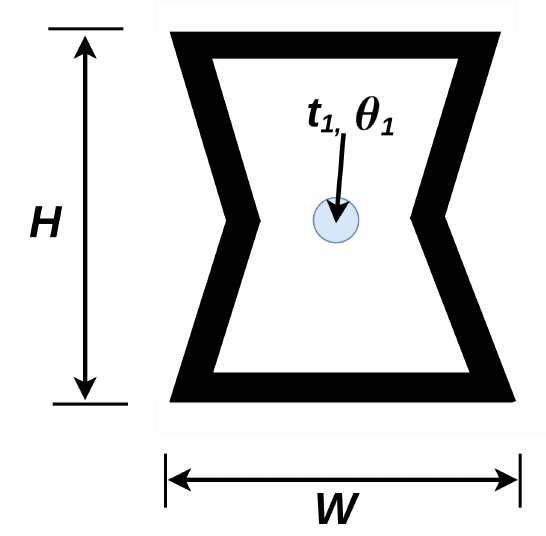}
        \caption{}
        \label{reentrant2}
    \end{subfigure}
    
    \caption{(a) Strut thickness in the centered rectangular unit cell is obtained by the corner nodes and center node, and (b) Strut thickness and angle of re-entrant unit cell are obtained by the center node of the unit cell.}
    \label{unit_cell2}
\end{figure}

\begin{figure}[ht!]
    \centering
    \begin{subfigure}{0.30\textwidth}
        \includegraphics[width=\linewidth,trim=0.0 0.00 0.0 0.0,clip]{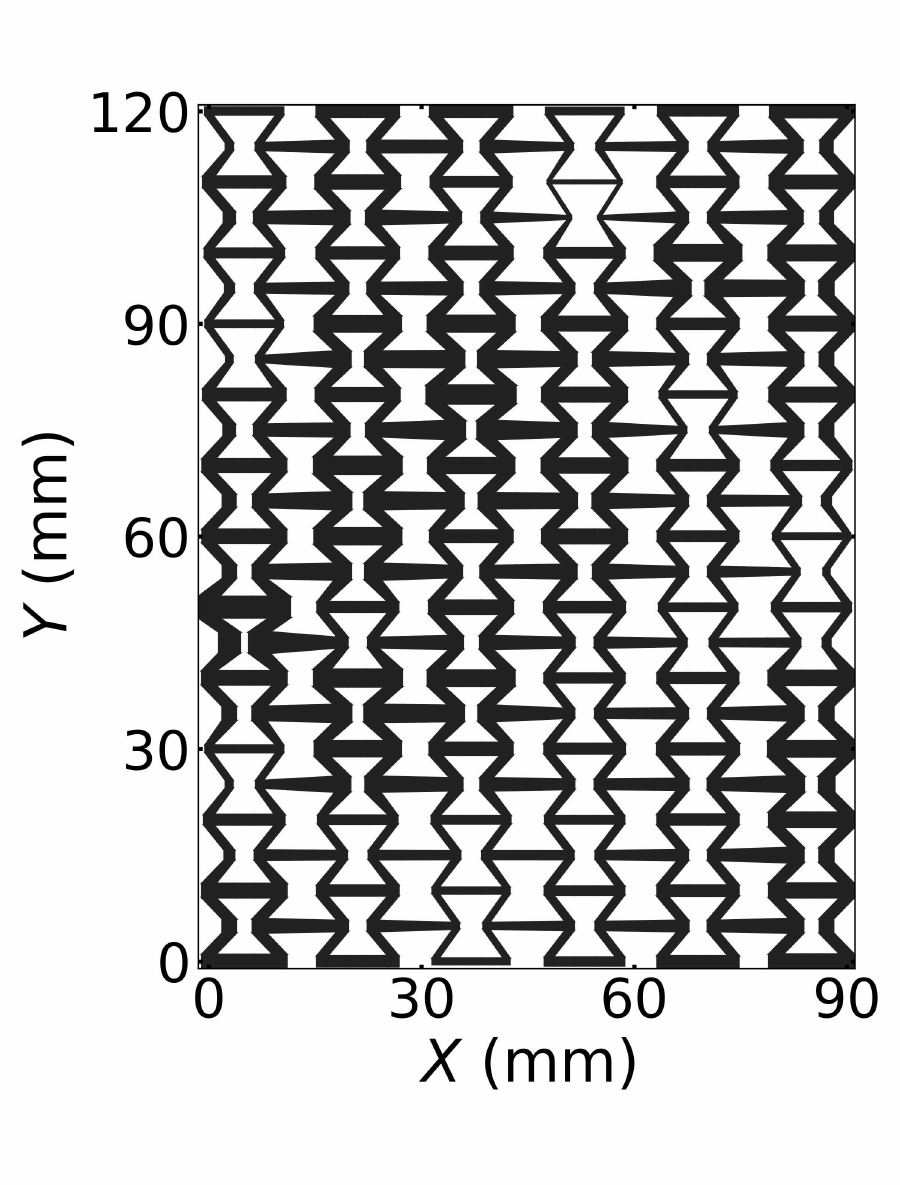}
        \caption{}
        \label{}
    \end{subfigure}
    \hspace{0.3cm}
    \begin{subfigure}{0.30\textwidth}
        \includegraphics[width=\linewidth,trim=0.0 0.00 0.0 0.0,clip]{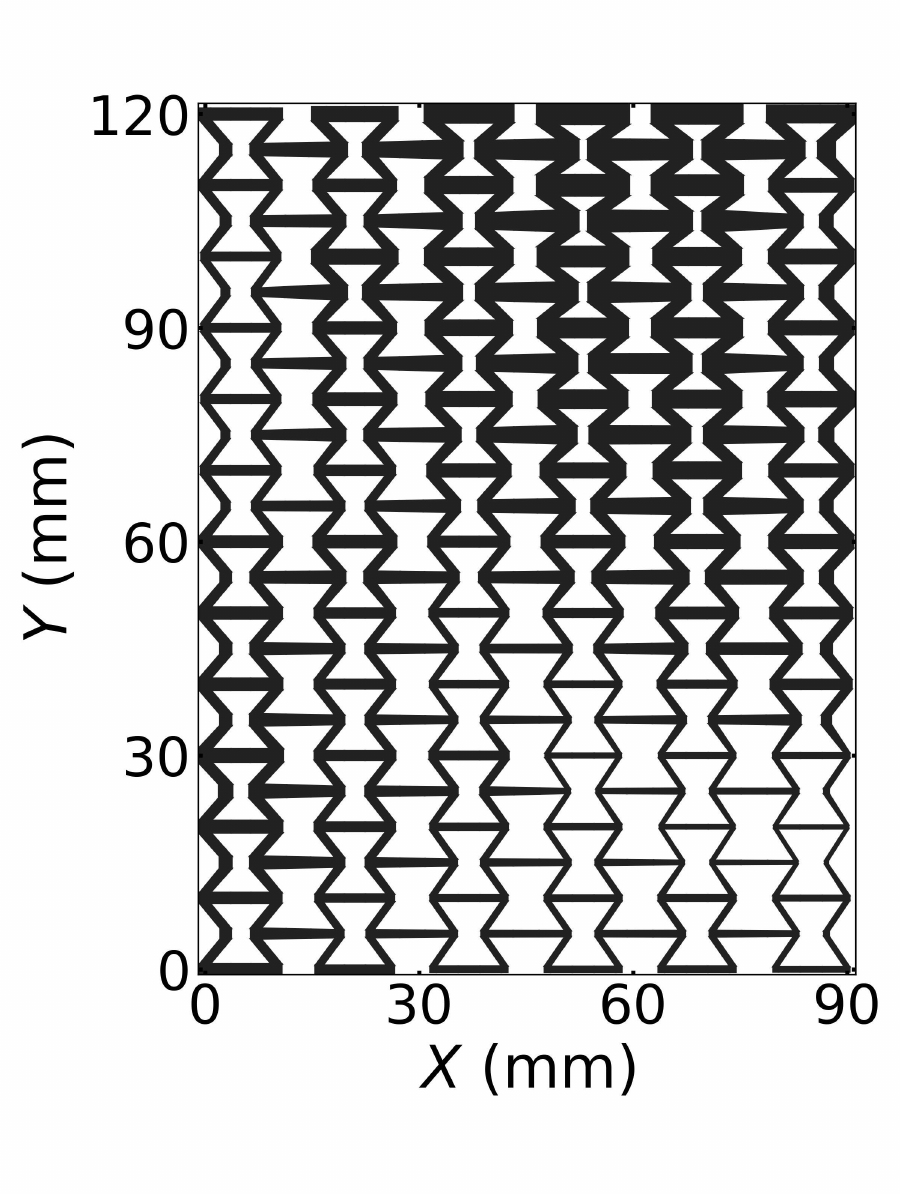}
        \caption{}
        \label{}
    \end{subfigure}
    \hspace{0.3cm}
    \begin{subfigure}{0.30\textwidth}
        \includegraphics[width=\linewidth,,trim=0.0 0.00 0.0 0.0,clip]{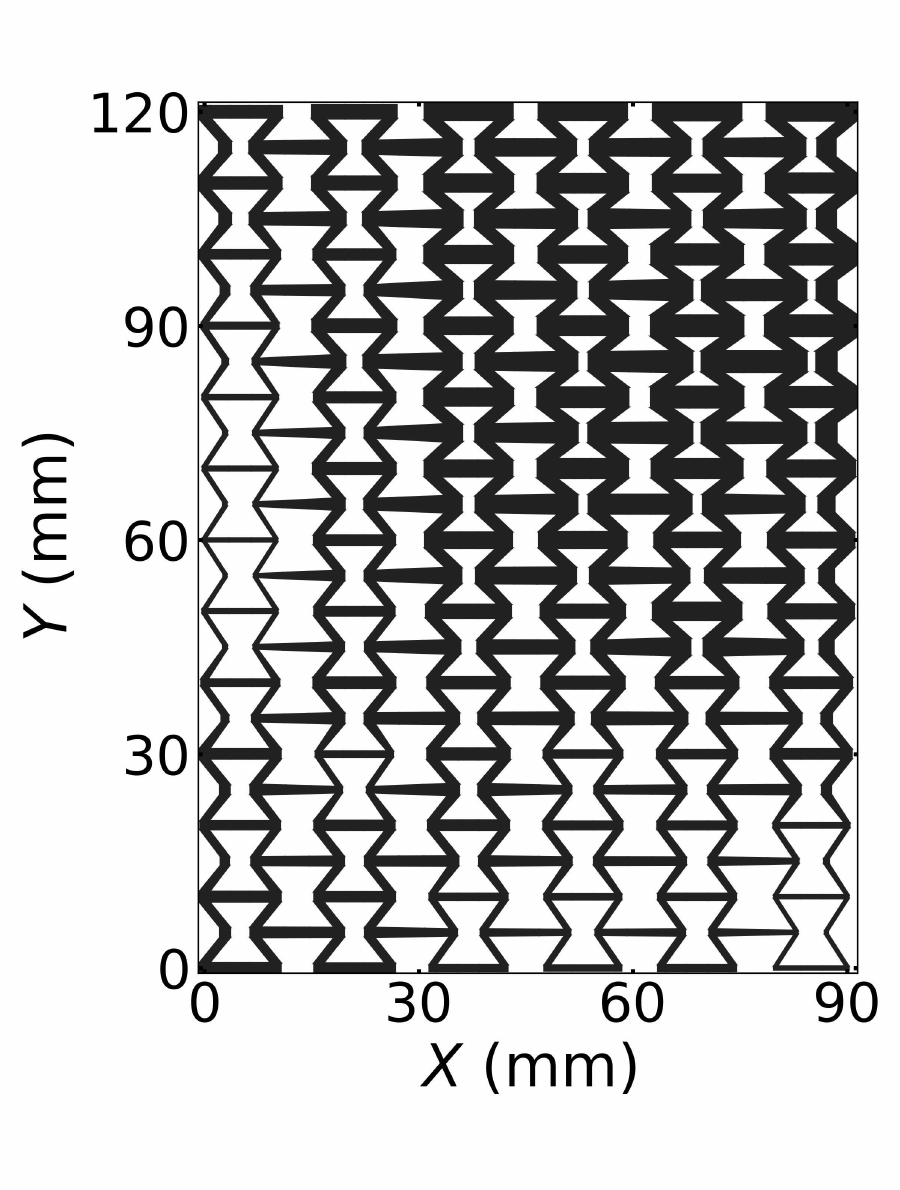}
        \caption{}
        \label{}
    \end{subfigure}
    
    \caption{Sample FGL re-entrant structures generated using GRF-based profile generation algorithm with the length scale of (a) 10 mm, (b) 30 mm, and (c) 50 mm.}
    \label{reentrant_lengthscale}
\end{figure}

In this work, two different unit cell types, 2D functionally graded lattice structures, are generated through a GRF-based profile generation algorithm: the centered rectangular unit cell and the re-entrant unit cell. For the centered rectangular unit cell, the strut thickness value is obtained by the GRF at the nodes located at the corners of each strut and the geometric center of the unit cell, as shown in Fig~\ref{bcc2}. The thickness distribution within each unit cell is then obtained through linear interpolation between these nodal points. In case of the re-entrant unit cells, it is more convenient to define the center of each unit cell as the node location, as shown in Fig~\ref{reentrant2}. The thickness of the strut connecting two adjacent unit cells is determined through linear interpolation of the thickness design variables. Furthermore, in the vertically connected unit cell, with a shared strut, the value of thickness is chosen to be identical to that of the lower unit cell. We illustrate a few sample FGL profiles generated using GRF in Fig.~\ref{reentrant_lengthscale}, consisting of re-entrant unit cells. We have  generated the FGL structure with different values of the length scale to obtain various levels of smoothness in the FGL structure.

\subsection{Gaussian process regression-based design of functionally graded lattice structures subjected to the geometric constraints}

In functionally graded lattice structures, boundary regions are often subject to geometric constraints imposed by design requirements, manufacturing limitations, or the need for compatibility with adjoining components. To design the FGL structure subjected to these boundary constraints, we are using the Gaussian process regression-based profile generation algorithm.

\begin{figure}[ht!]
    \centering
    \begin{subfigure}{0.30\textwidth}
        \includegraphics[width=\linewidth,trim=0.0 0.00 0.0 0.0,clip]{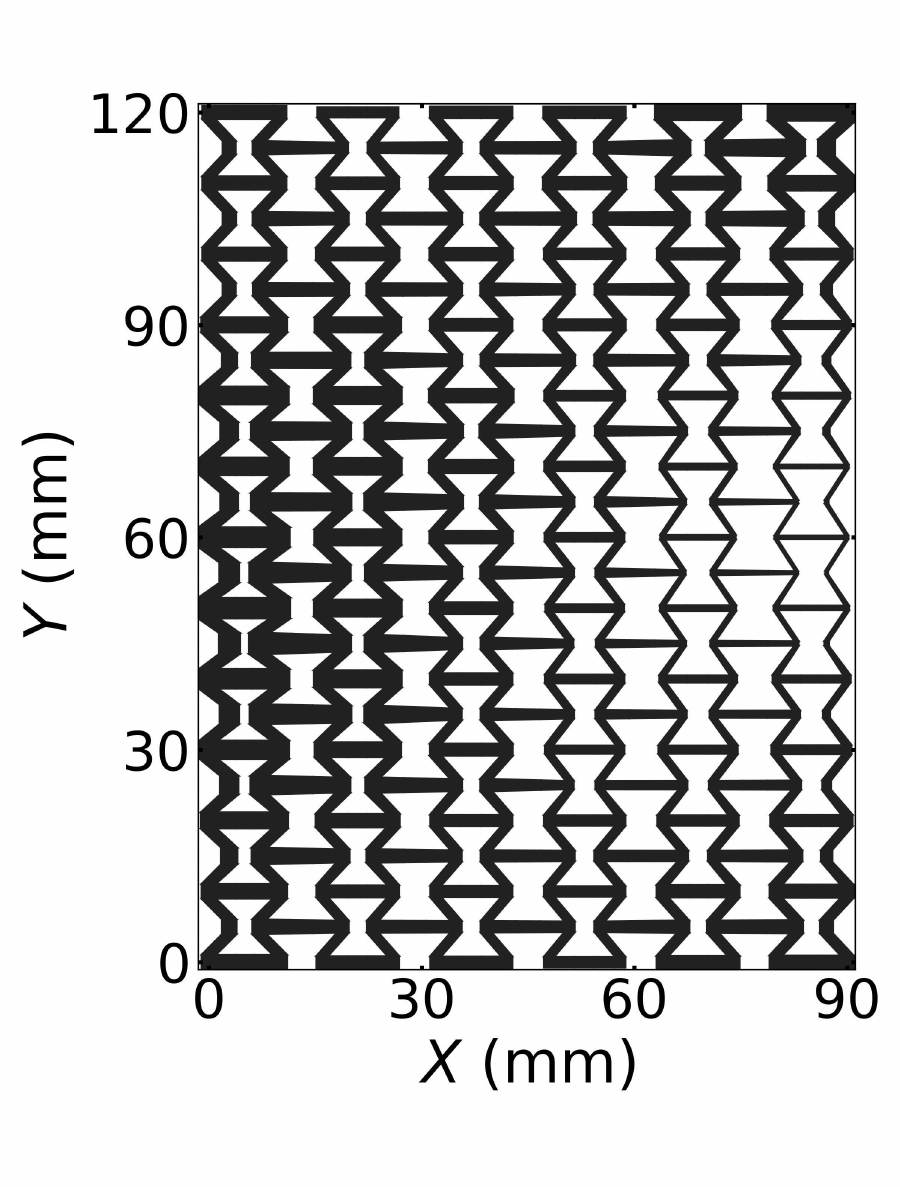}
        \caption{}
        \label{}
    \end{subfigure}
    \hspace{0.3cm}
    \begin{subfigure}{0.30\textwidth}
        \includegraphics[width=\linewidth,trim=0.0 0.00 0.0 0.0,clip]{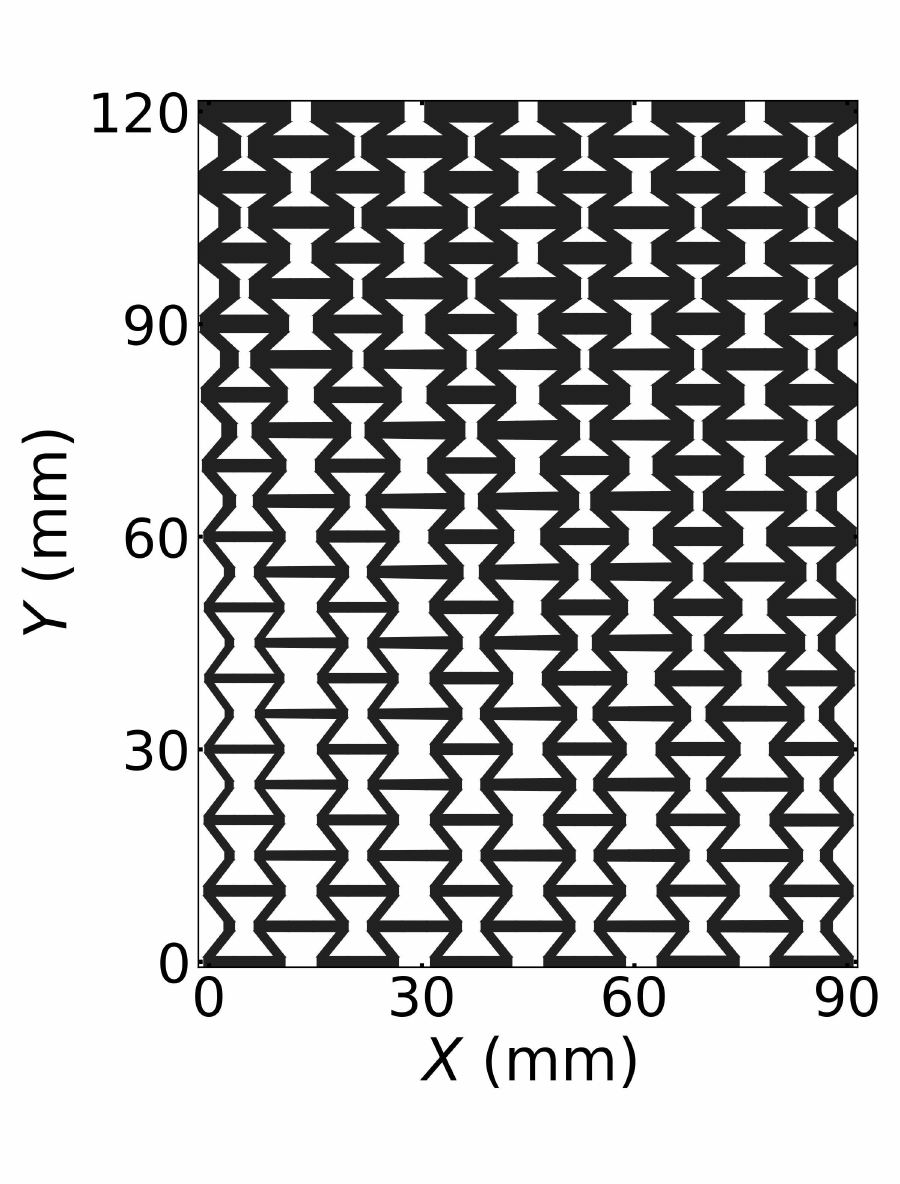}
        \caption{}
        \label{}
    \end{subfigure}
    \hspace{0.3cm}
    \begin{subfigure}{0.30\textwidth}
        \includegraphics[width=\linewidth,,trim=0.0 0.00 0.0 0.0,clip]{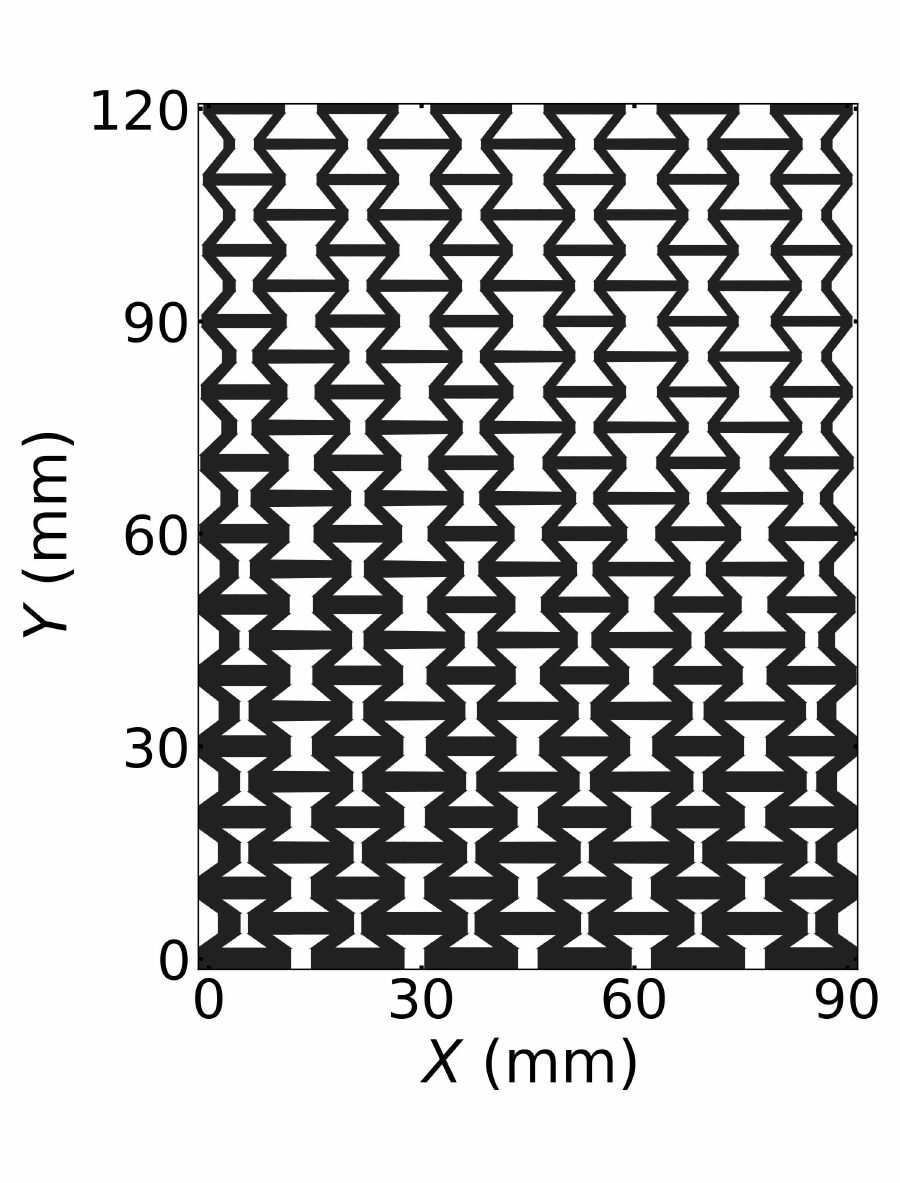}
        \caption{}
        \label{}
    \end{subfigure}
    
    \caption{Sample FGL re-entrant structures generated using GRF-based profile generation algorithm and (b–c) GPR-based profile generation algorithm under maximum-thickness constraints applied to top and bottom unit cells, respectively.}
    \label{gpr_reentrant}
\end{figure}

GPR is the extension of the GRF, in which we use the Bayesian framework to update the prior values based on the information obtained using the boundary nodes. The constraint boundary nodes coordinate ($\boldsymbol{X_{b}}$) and the corresponding geometric parameter values ($\boldsymbol{P_{b}}$) are extracted from the discretized domain. In GPR, the prior distribution is conditioned on the observed input boundary values ($\boldsymbol{P_{b}}$), such that the predictive distribution is expressed as $\boldsymbol{P| P_{b}, X,X_{b} }$.  Where $\boldsymbol{P}$ is the vector containing the geometric parameter values of all the nodes, $\boldsymbol{X}$ denotes the corresponding node coordinate values. The boundary values to obtain the posterior distribution provide the most probable function values at unobserved nodes. The final posterior distribution obtained by GPR is given by the following expression:

 \begin{equation}
\begin{aligned}
    \boldsymbol{P \mid P_{b}}, \boldsymbol{X}, \boldsymbol{X_{b}} &\sim \mathcal{N} \left( \boldsymbol{\mu^{'}}(\boldsymbol{X}),\boldsymbol{K^{'}} \right),
\end{aligned}
\label{eq:f_star_given_f}
\end{equation}
where,
\begin{equation}
\begin{aligned}
    \boldsymbol{\mu^{'}} &= \mu(\boldsymbol{X})+\boldsymbol{K}_*^T (\boldsymbol{K}+ \sigma^{2}_{n}\boldsymbol{I})^{-1}( \boldsymbol{T_{b} - \mu(X_{b})}), \\
    \boldsymbol{K^{'}} &= \boldsymbol{K}_{**} - \boldsymbol{K}_*^T (\boldsymbol{K_{b}}+ \sigma^{2}_{n}\boldsymbol{I})^{-1} \boldsymbol{K}_*,
\end{aligned}
\end{equation}

where $\boldsymbol{K}$ is the covariance matrix between the coordinates of the boundary nodes, $\boldsymbol{K_{*}}$ is the matrix between the coordinates of boundary nodes and all the nodes. $\boldsymbol{K_{**}}$ is the matrix between all the node coordinates. $\boldsymbol{\mu^{'}}$ denotes the mean of the posterior distribution with variance $\boldsymbol{K^{'}}$. The term $\sigma^{2}_{n}$ represents the assumed noise variance, and $\boldsymbol{I}$ is the identity matrix. Also, note that the value of $\sigma^{2}_{n}$ is chosen very small (in our cases $\sigma^{2}_{n}$ = $10^{-12}$) to satisfy the boundary volume fraction constraint. Figs. \ref{gpr_reentrant} demonstrates that some FGL structures consist of the re-entrant unit cells obtained by the GPR algorithm, subjected to geometry constraints on the boundary nodes.

\section{Finite element analysis}
\label{FEA}
This section outlines the finite element scheme employed to analyze functionally graded lattice structures. In general, since the lattice structure can undergo large deformation, the governing equations are written in the reference configuration. Further, since the conventional continuum-based elements are known to be prone to locking issues with structures having high aspect ratios, they might not provide a computationally efficient way of simulating the lattice structures. In literature, the stress-based hybrid elements are shown in literature to alleviate this issue of locking with any additional kinematic assumption and provide a robust and  computationally efficient strategy to simulate such structures. Thus, in this work, we deploy the stress-based hybrid finite elements to carry out the analysis. Here, we describe the hybrid formulation in brief; for more details, readers are requested to refer to Jog and Kelkar \citep{jog2006non}.

\subsection{Governing equation}
The governing equation is given in the reference configuration $\boldsymbol{\Omega}$, with boundary $\boldsymbol{\Gamma}$ is composed of the regions $\boldsymbol{\Gamma_{u}}$ and $\boldsymbol{\Gamma_{t}}$ are given below:

\begin{alignat}{2}
\nabla \cdot ( \boldsymbol{F}\boldsymbol{S}) + \rho_0 \boldsymbol{b}^0 &= 0 \quad && \text{on } \Omega, \label{fe1} \\[6pt]
\boldsymbol{E} &= \hat{\boldsymbol{E}}(\boldsymbol{S}) \quad && \text{on } \Omega, \label{fe2} \\[6pt]
\boldsymbol{t}^0 &= \bar{\boldsymbol{t}}^0 \quad && \text{on } \Gamma_t, \label{fe3}\\[6pt]
\boldsymbol{u} &= \boldsymbol{u_0} \quad && \text{on } \Gamma_u, \label{fe4}
\end{alignat}

where, $\boldsymbol{F}$ is the deformation gradient, given in material coordinates ($\boldsymbol{X}$), $\boldsymbol{S}$ is the second-Piola Kirchhoff stress, $\rho_{0}$ is the density of the material in the reference configuration, $\boldsymbol{b}^0$ is the applied body force in the reference configuration, traction $\boldsymbol{t}^{0}$ is given by $\boldsymbol{FSn}^{0}$, $\boldsymbol{E(U)}$ is the Green strain tensor, given in terms of displacement ($\boldsymbol{u}$) by the following equation:

\begin{equation}
    \boldsymbol{\bar{E}(u)} = \frac{1}{2}[(\nabla\boldsymbol{u})+ (\nabla\boldsymbol{u})^{T} + (\nabla\boldsymbol{u})^{T}(\nabla\boldsymbol{u}) ].
    \label{strain_eq}
\end{equation}

\subsection{Weak formulation of Hybrid elements}
In this subsection, we provide the preliminary details regarding the hybrid elements. For more details, readers are referred to \citep{pian1984rational,jog2006non}. Hybrid finite element formulation is derived from the two-field variational principle, in which the momentum conservation equation in reference configuration (Eqs.~\eqref{fe1},\eqref{fe2},\eqref{fe3}, and \eqref{fe4}), and the relationship between the displacement and strain (Eqs.~\eqref{strain_eq}) are enforced in the weak sense. If:

\[
\begin{aligned}
V_u &:= \{\boldsymbol{u}_\delta : \ \boldsymbol{u}_\delta = \boldsymbol{0}\ \text{on }\Gamma_u\},\\
V_S &:= \{\boldsymbol{S}_\delta : \ \boldsymbol{S}_\delta^{\,t} = \boldsymbol{S}_\delta\ \text{on }\Omega\},
\end{aligned}
\]
where $V_{u}$ denotes the space of variations of the displacement variations and $V_{s}$  denotes the space of second Piola-Kirchhoff stress variations. Applying integration by parts to Eqs.~\eqref{fe1} and ~\eqref{fe2}, the weak form of the two-field variational principle can be expressed as:

\[
\int_{\Omega} \boldsymbol{S} : \boldsymbol{E}_\delta \, d\Omega 
= \int_{\Omega} \rho_0 \boldsymbol{u}_\delta \cdot \boldsymbol{b}^0 \, d\Omega 
+ \int_{\Gamma_t} \boldsymbol{u}_\delta \cdot \boldsymbol{t}^0 \, d\Gamma 
\quad \forall \boldsymbol{u}_\delta \in V_u,
\]

\[
\int_{\Omega} \boldsymbol{S}_\delta : \left[ \boldsymbol{E}(\boldsymbol{u}) - \hat{\boldsymbol{E}}(\boldsymbol{S}) \right] d\Omega 
= 0 \quad \forall \boldsymbol{S}_\delta \in V_S,
\]
where, variation of $\boldsymbol{E(u)}$ is given below:
\[
\boldsymbol{E}_\delta(\boldsymbol{u}, \boldsymbol{u}_\delta) 
= \tfrac{1}{2} \Big[ (\nabla \boldsymbol{u}_\delta) + (\nabla \boldsymbol{u}_\delta)^{t} 
+ (\nabla \boldsymbol{u})^{t} (\nabla \boldsymbol{u}_\delta) 
+ (\nabla \boldsymbol{u}_\delta)^{t} (\nabla \boldsymbol{u}) \Big].
\]
In the case of the hybrid elements, the stresses are interpolated independently of the displacement field. The stress interpolation is discontinuous across elements, and is given by:
\begin{equation}
    \boldsymbol{S}=\boldsymbol{P}\boldsymbol{\beta}
\end{equation}

Here, $P$ are the stress interpolation functions and \(\boldsymbol{\beta}\) are the stress parameters. Since the stress interpolation functions are discontinuous across elements, the stress degrees of freedom (\(\boldsymbol{\beta}\)) can be eliminated at the element level itself. This ensures that the overall size of the global stiffness matrix remains unchanged, compared to the conventional displacement-based formulation. The choice of stress interpolation functions significantly influences the stability and efficiency of hybrid elements. In this manuscript, we have taken the stress interpolation functions presented in~\citep{bombarde2022hellinger}. These interpolation functions have been shown in the literature to provide robust simulation strategies for a wide range of large-deformation problems~\citep{agrawal2019hybrid}.

\section{Genetic algorithm}
\label{GA}
The genetic algorithm belongs to the category of non-gradient-based evolutionary optimization algorithms. It operates on the principle of natural selection and genetics. This work uses the GA for the single objective optimization of FGL structures subjected to different constraints. For the optimization of FGL structures, the design variables are the geometric parameters of the unit cells (thickness/orientation of the struts). The GRF/GPR-based profile generation algorithm is utilized as the initial design generation algorithm within the GA framework. Further, the evolution of the population is done with the help of operator selection, crossover, and mutation, and the conservation of the best solution is carried out using the elite selection. Also, an additional operator named as projection is included to ensure the smoothness of the FGL structures. The detailed explanation of the GA framework (Fig. \ref{GA_flowchart}) is given further in this section:

\begin{figure} 
    \centering
    \includegraphics[width=0.6\linewidth]{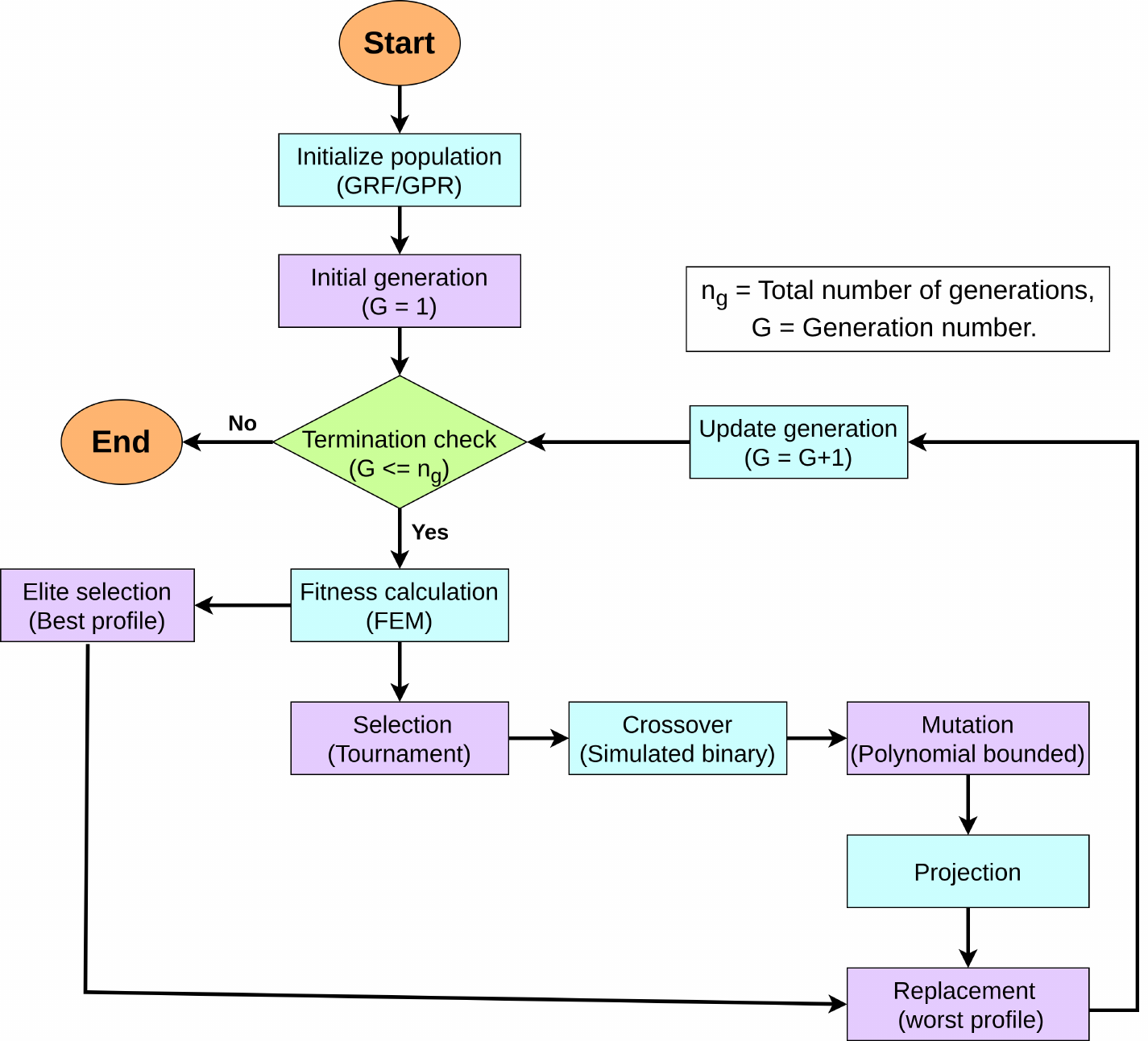} 
    \caption{Flowchart of genetic algorithm framework used for the functionally graded lattice structure optimization.} 
    \label{GA_flowchart}
\end{figure}

\subsection{Fitness calculation and Selection}
In this GA framework, the performance of each candidate profile is evaluated using finite element analysis as mentioned in the section \ref{FEA}. The resulting performance measures are subsequently integrated with the adopted constraint handling scheme \citep{deb2000efficient}, through which the final fitness score of each profile is obtained. In this constraint handling approach, each profile is either feasible or unfeasible, depending on the fulfillment of the constraint criteria. The fitness score of the feasible profile is equal to the objective function value, while the unfeasible solution is determined by adding the maximum value of the objective function within the feasible region, and adding the constraint violation value. A constraint optimization problem is generally given by the following expression:

\begin{equation}
\begin{aligned}
\textbf{Minimize:}\; & \quad f(\vec{\boldsymbol{x}}),\\ 
\textbf{Subject to:}\; & \quad g_{k}(\vec{\boldsymbol{x}})  \ge g_{k}^{*}, \quad k =1,2,..,K,\\
&  \quad x_{i}^{l} \le x_{i} \le x_{i}^{u}, \quad i=1,2,..,n,
\end{aligned}
\end{equation}
where, $f(\vec{\boldsymbol{x}})$  is the objective function, $k$ is greater than-equal-to  type inequality type constraints. $\vec{\boldsymbol{x}}$ is the vector with $n$ number of variables. Further, the fitness score calculation of the feasible or unfeasible solution based on the constrained criteria is given by the following expression:

\begin{equation}
\begin{aligned}
F(\vec{\boldsymbol{x}}) =
\begin{cases}
    f(\vec{\boldsymbol{x}}) & \text{if }\quad g_{k}(\vec{\boldsymbol{x}})  \ge g_{k}^{*}, \quad k =1,2,..,K, \\
    f_{\max} + \sum\limits_{k=1}^{K} |\langle g_k(\vec{\boldsymbol{x}}) \rangle| &   \text{otherwise},
\end{cases}
\end{aligned}
\end{equation}
\noindent where, $f_{\max}$ is worst feasible solution. Furthermore, this study adopts tournament selection as the selection operator. A subset of FGL profiles is randomly sampled from the initial population, and the profiles with the highest fitness score are moved to the next stage of the optimization process.

\subsection{Crossover operator}
\begin{figure} 
    \centering
    \includegraphics[width=0.8\linewidth]{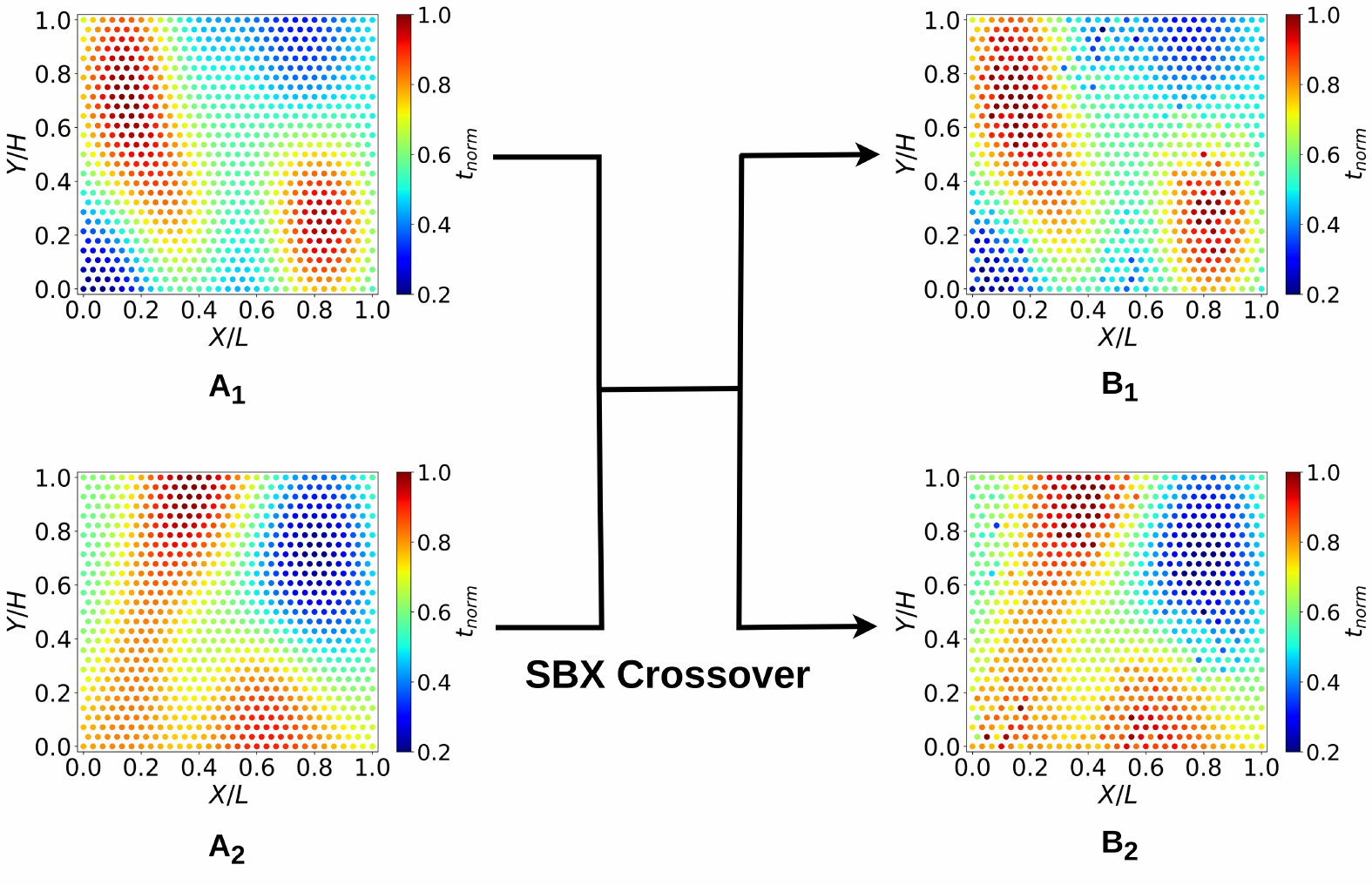} 
    \caption{Sample of the normalized Strut thickness distribution over the FGL structure before ($A_{1}$ and $A_{2}$) and after ($B_{1}$ and $B_{2}$) the crossover operation.} 
    \label{cross}
\end{figure}

This work uses simulated binary (SBX) crossover to explore new areas of the solution space in the FGL design. It combines the information of two parents ($A_{1}$ and $A_{2}$) and produce new offsprings ($B_{1}$ and $B_{2}$). The mathematical expression demonstrating the crossover operation is given below:

\begin{equation}
\begin{aligned}
\boldsymbol{B}_1 &= 0.5\left[(1+\boldsymbol{\beta})\odot \boldsymbol{A}_1 + (1-\boldsymbol{\beta})\odot \boldsymbol{A}_2\right],\\
\boldsymbol{B}_2 &= 0.5\left[(1-\boldsymbol{\beta})\odot \boldsymbol{A}_1 + (1+\boldsymbol{\beta})\odot \boldsymbol{A}_2\right],
\end{aligned}
\label{cross_eq}
\end{equation}
where, $\odot$ represents the node-wise  product, and $\boldsymbol{\beta}$ is the vector of spread factors $\beta_i$ given by the following expressions:

\[
\beta_i =
\begin{cases}
(2r_i)^{\frac{1}{\eta_{c} + 1}}, & \text{if } r_i \leq 0.5, \\
\left(\dfrac{1}{2(1 - r_i)}\right)^{\frac{1}{\eta_{c} + 1}}, & \text{if } r_i > 0.5.
\end{cases}
\]
Here, $\eta_{c}$ denotes the crossover strength parameter that governs the shape of the spread, and $\boldsymbol{r}$ is a random vector with each component $r_i \in [0, 1]$. Further, the Fig. \ref{cross} demonstrates a sample of the the distribution in the normalized strut thickness ($t_{norm}$) before and after the crossover operation.

\subsection{Mutation operator}
The mutation operation is accomplished using the polynomial-bounded mutation operator. It provides a controlled small perturbation in each design variable ($x_{i}$). It provides a mutated value of the design variable ($x_i'$) by adding a perturbation term derived from a polynomial distribution governed by a distribution index $\eta_{m}$. Where each variable $x_i$ is selected for mutation with a probability $p_m$.

For a selected design variable $x_i$, bounded within $[x_{\text{low}}, x_{\text{up}}]$, we define
\begin{equation}
\delta_1 = \frac{x_i - x_{\text{low}}}{x_{\text{up}} - x_{\text{low}}}, 
\quad
\delta_2 = \frac{x_{\text{up}} - x_i}{x_{\text{up}} - x_{\text{low}}},
\quad
m = \frac{1}{\eta_{m} + 1}.
\end{equation}

A random number $r \sim U(0,1)$ determines the mutation step as follows:
\begin{equation}
\Delta_q =
\begin{cases}
\left[ 2r + (1 - 2r)(1 - \delta_1)^{(\eta_{m}+1)} \right]^m - 1, & r < 0.5, \\[8pt]
1 - \left[ 2(1-r) + 2(r-0.5)(1 - \delta_2)^{(\eta_{m}+1)} \right]^m, & r \geq 0.5.
\end{cases}
\end{equation}

The updated variable is then expressed as
\begin{equation}
x_i' = x_i + \Delta_q \cdot (x_{\text{up}} - x_{\text{low}}),
\end{equation}
followed by bounding within the feasible range:
\begin{equation}
x_i' = \min\!\big(\max(x_i', x_{\text{low}}), \, x_{\text{up}}\big).
\end{equation}

\begin{figure} 
    \centering
    \includegraphics[width=0.8\linewidth]{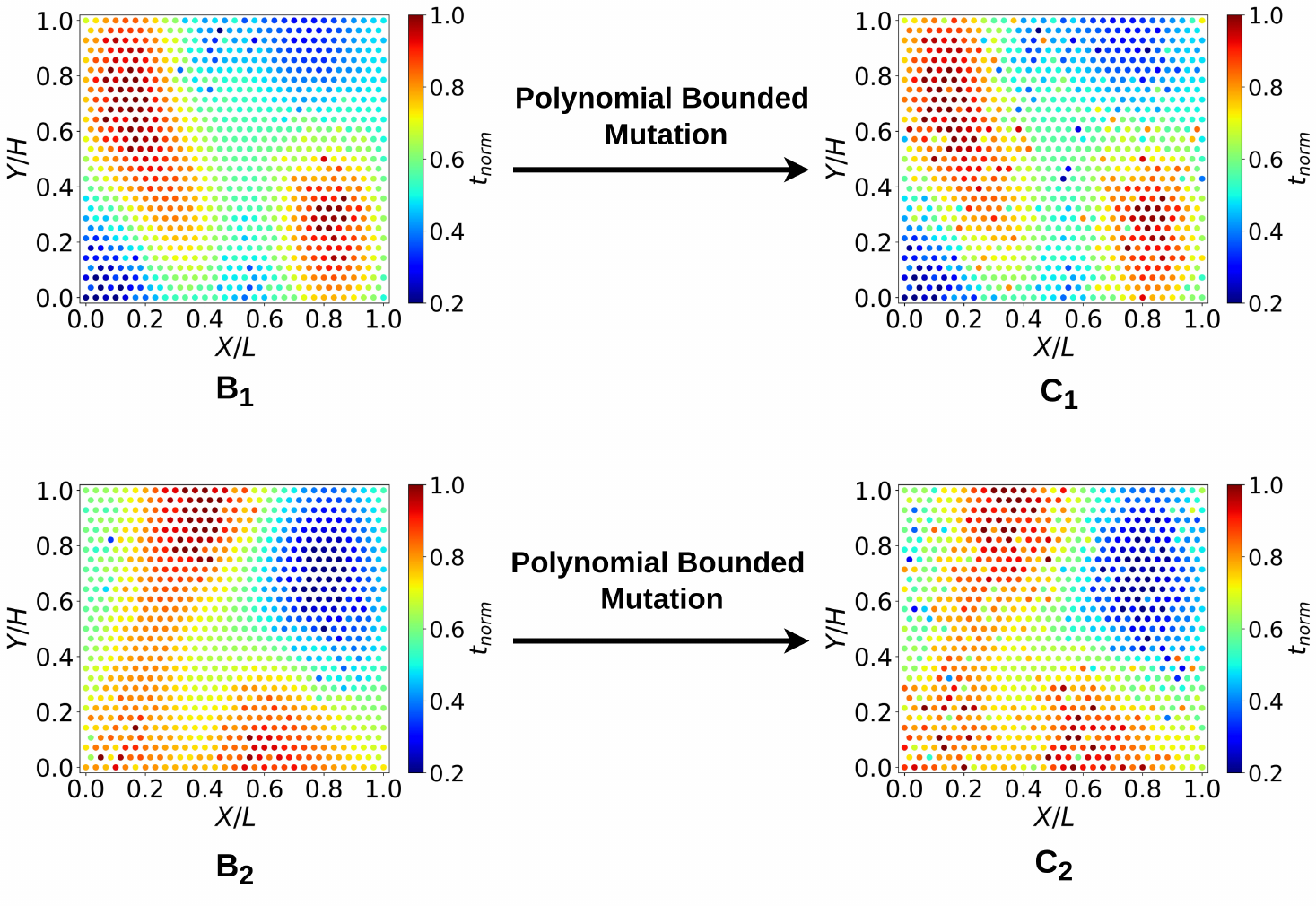} 
    \caption{Sample of the normalized strut thickness distribution over the FGL structure before ($B_{1}$ and $B_{2}$) and after ($C_{1}$ and $C_{2}$) the mutation operation.}
    \label{mutation}
\end{figure}

This operator ensures that the mutated solution always lies within the specified bounds while enabling small and large variations depending on the value of $\eta_{m}$. A higher distribution index $\eta_{m}$ biases the mutation toward more minor changes, while a lower $\eta_{m}$ promotes exploration through larger perturbations. Fig. \ref{mutation} demonstrates a sample of the the distribution in the thickness before and after the mutation in the FGL structures. 

\subsection{Projection operator}
\begin{figure} 
    \centering
    \includegraphics[width=0.8\linewidth]{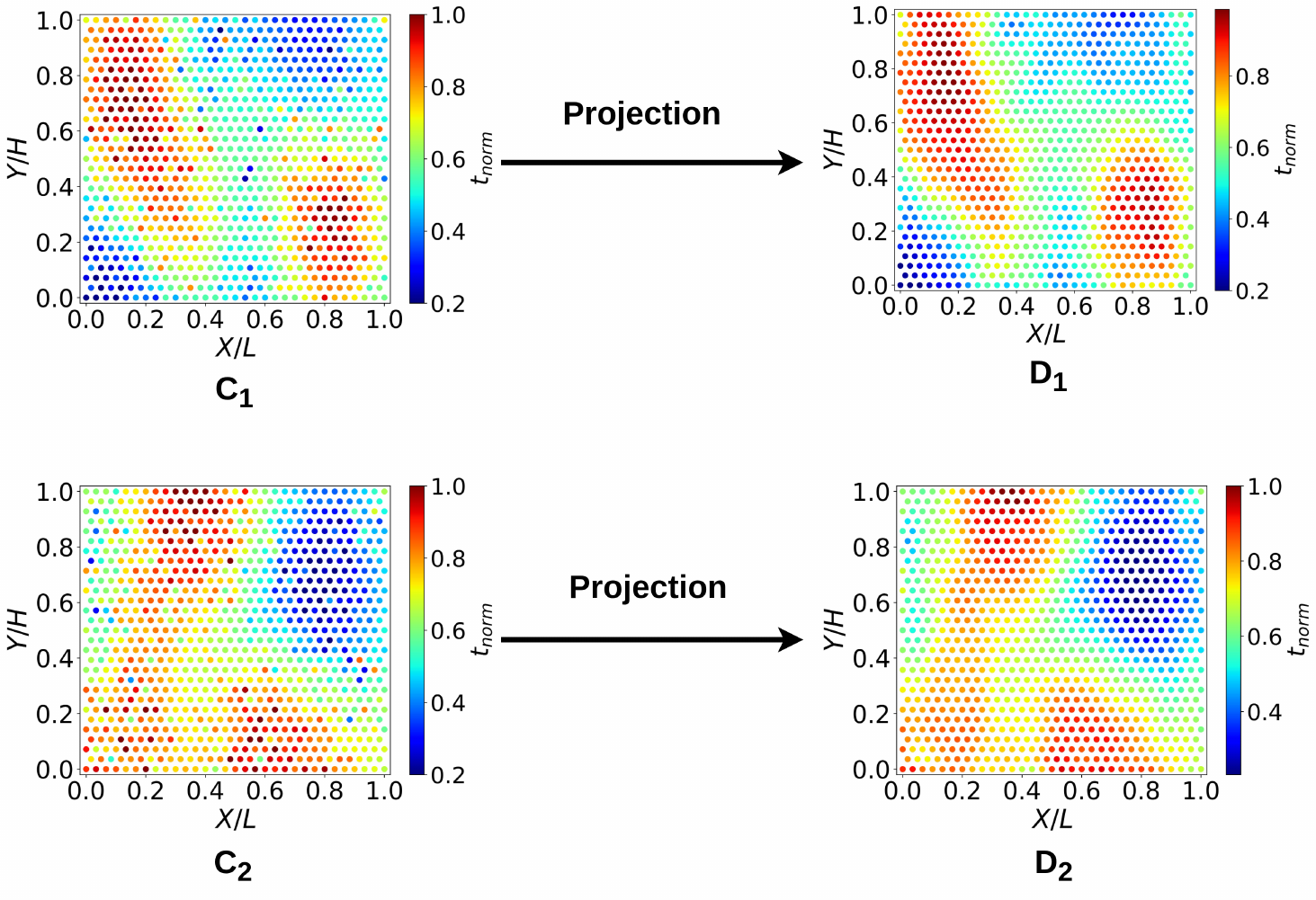} 
    \caption{Sample of the normalized strut thickness distribution over the FGL structure before ($C_{1}$ and $C_{2}$) and after ($D_{1}$ and $D_{2}$) the projection operation.}
    \label{projection}
\end{figure}

The FGL structure obtained after the evolution steps of GA (crossover and mutation), finds abrupt gradation or a non-smooth pattern of geometric parameters distribution over the domain, as shown in  Fig. \ref{mutation} ($C1$ and $C2$). Note the resultant designs after the crossover and mutation operators are non-smooth even if both the underlying parent designs are smooth in nature. To overcome this issue and to ensure the smoothness of the generating designs at each generation, we employ a projection operator in the GA framework. This operator projects the non-smooth FGL profile into a smoother space. The projection operation is given by the following expression:

\begin{equation}
 \boldsymbol{C^{'}} = \boldsymbol{K}(\boldsymbol{K} + \sigma^{2}_{l}\boldsymbol{I})^{-1}\boldsymbol{C},
\end{equation}

where $\boldsymbol{C^{'}}$ is the vector of the projected geometric parameters, $\boldsymbol{K}$ is the covariance matrix obtained by the radial kernel function, given by Eq. \eqref{rbf_kernel}. This covariance matrix is identical to the one used to generate the design space. We demonstrate the effect of the projection operator in  Fig. \ref{projection}, showing a non-smooth distribution of the geometric parameter before projection and a subsequent smooth distribution obtained after the projection. In the above expression, \(\sigma_{l}\) is a small value (order of $10^{-6}-10^{-12} $) and control the smoothness of the underlying projection. Specifically, \(\lambda_1\geq\lambda_2\geq\lambda_3,...\) and \(\boldsymbol{e_1},\boldsymbol{e_2},..\) represents the eigenvalues and eigenvectors of the matrix \(\boldsymbol{K}\). Further, it can be shown that the smoothness of the eigenvectors decreases as the eigenvalue decreases.  The projected space as shown in  Fig~\ref{projection}, consists of \(\boldsymbol{e_1},\boldsymbol{e_2}...,\boldsymbol{e_p}\),where \(\lambda_{p+j}\leq\leq \sigma^2_{l}\). Thus, the projected space only consists of the first \(p\) number of eigenvectors, which are smooth in nature compared to \(\boldsymbol{e_{p+1}},\boldsymbol{e_{p+2}}...,\boldsymbol{e_n}\).

\section{Numerical examples}
\label{Numerical example}
In this section, we demonstrate the utility of the proposed framework in designing graded lattice structures through numerous numerical examples. We present optimization of various two-dimensional structures consisting of centered rectangular and re-entrant auxetic unit cells under diverse loading conditions. Furthermore, the optimal profiles obtained through the GRF-based optimization framework are compared with those generated using standard genetic algorithm combined with a conventional implementation. 

In a conventional implementation, each design variable is generated independently of the others. For each of the design variable, underlying distribution is taken gaussian in nature. In a way, the conventional implementation is same as GRF with a length scale equal to zero. The normalization is carried out to make sure that even in the conventional implementation, the design values remain between the maximum and minimum range. Further, the standard genetic algorithm without any projection operator is used. 

The comparison is carried out by evaluating both the objective function values and the resulting stress levels developed within the optimal FGL structure. These exercises provide a clear assessment of the relative performance of the two optimization approaches.

\subsection{Re-entrant unit cell-based lattice structure}

\begin{figure}[h]
    \centering
    \includegraphics[width=0.3\textwidth]{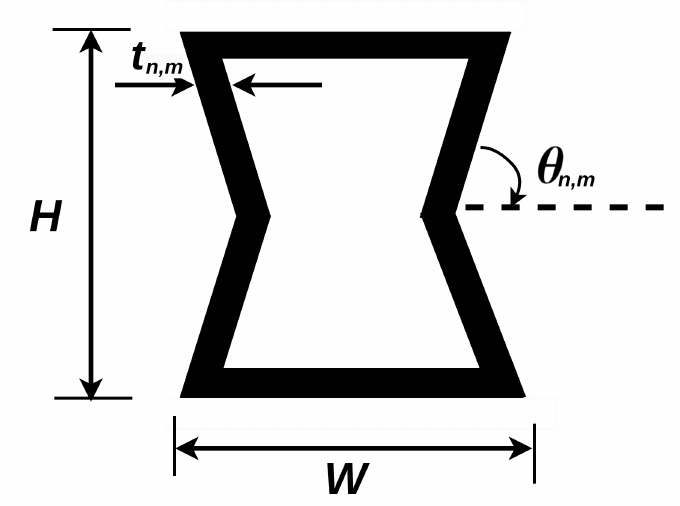}
    \caption{Schematic of a single re-entrant unit cell located at position ($n,m$) in the lattice structure.}
    \label{reentrant_unit_cell}
\end{figure}

In this section, the structures composed of $N \times M$ re-entrant auxetic unit cells is considered. The schematic of the single re-entrant unit cell is demonstrated in Fig.~\ref{reentrant_unit_cell}. Based on this unit-cell geometry, we investigate two different optimization problems: in the first, the strut thickness ($t$) of the unit cells is treated as a design variable; in the second, both strut thickness ($t$) and the re-entrant angle ($\theta$) are treated as design variables. The index set of unit cells is defined as:
\begin{equation}
\mathcal{C}
=
\left\{
(p,q)\;\middle|\;
p = 1,\ldots,N,\;
q = 1,\ldots,M
\right\}.
\end{equation}

Where, $N$ is the number of unit cells along the X-axis and $M$ is the number of unit cells along Y-axis. Each unit cell is characterized by a strut thickness $t_{p,q}$ and an re-entrant angle $\theta_{p,q}$.

\subsubsection{Case 1: Deflection Maximization with strut thicknesses as design variables.}
\label{re_case1}

\begin{figure}[ht!]
    \centering
    \includegraphics[width=0.4\textwidth]{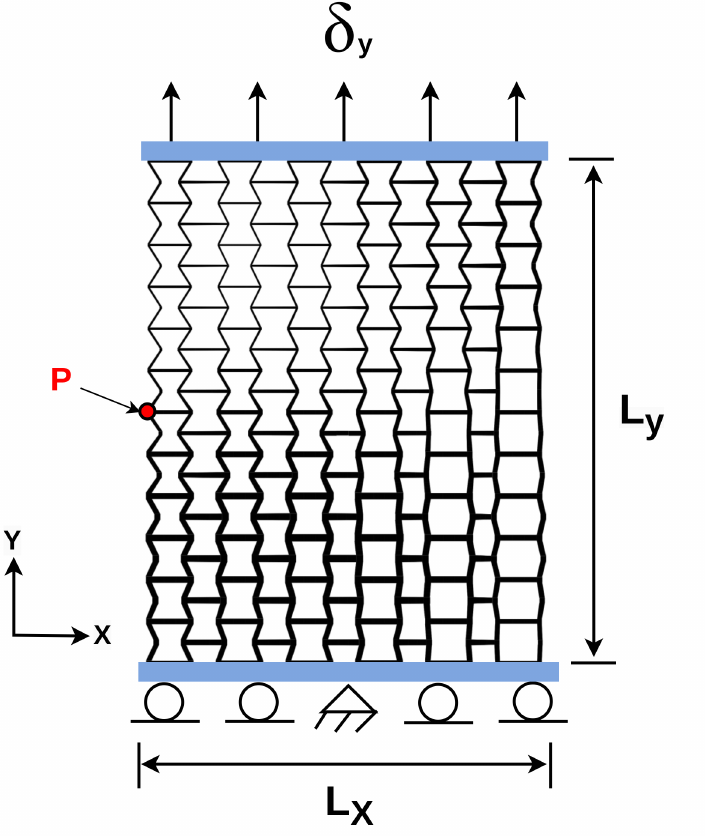}
    \caption{Schematic of the re-entrant unit cell structure subjected to the uniform displacement at the top surface.}
    \label{reentrant_case1_diagram}
\end{figure}

In this problem, we consider a rectangular structure composed of the re-entrant unit cells, as shown in Fig.~\ref{reentrant_case1_diagram}. Each unit cell has the dimensions $H$ = 6 mm and $W$ = 6 mm. For this example, we have taken 6 × 12 unit cells in the underlying structure. The bottom edge of the structure is subjected to the roller-supported boundary condition, with one point fixed at the bottom edge of the structure. A uniform displacement ($\delta_y$) of 3.00 mm is applied at the top surface along the positive Y-axis. We have defined the objective to maximize the deflection ($\delta_{x}$) along the -ve X-axis at point "P", which lies midway along the left edge, as shown in Fig~\ref{reentrant_case1_diagram}. The unit-cell angle is defined as:\(\theta_{p,q} = \theta_0, (p,q) \in \mathcal{C}\), where, $\theta_0$ is taken $60^\circ$ and the strut thickness is the only design variable:\(\boldsymbol{\alpha}^{(I)} = \Big\{ t_{p,q} \;\big|\; (p,q) \in \mathcal{C} \Big\}\). Thus, the optimization problem can be stated as follows:

\begin{equation}
\begin{aligned}
&\textbf{maximize:} && \quad -\delta_{x}^{P}(\boldsymbol{\alpha}^{I}), 
 \\
&\textbf{subject to:} && \quad t_{\min} \le t_{p,q} \le t_{\max}, \quad (p,q) \in \mathcal{C},\\
&&& \quad \frac{1}{|\mathcal{C}|}
\sum_{(p,q)\in\mathcal{C}} t_{p,q}
\ge \bar{t}_{\max}.
\end{aligned}
\end{equation} 

For the present problem, the minimum allowable strut thickness is set to $t_{\min}$ = 0.2 mm, and the maximum allowable thickness is $t_{\max}$ = 2.0 mm.  Apart from this, the minimum average strut thickness is chosen as ${\bar{t}_{\max}}$ = 1.25 mm. The material considered has an elastic modulus of 400 MPa and a Poisson ratio of 0.4.  To predict the displacement field for all designs, 7,164 four-node quadrilateral elements have been used. In the genetic algorithm, in order to generate the initial population for the optimization, the GRF parameters considered are: length scales of 30 mm and 40 mm, and a standard deviation of 0.60 mm. Once the random designs are generated, for each design, the values are normalized to ensure the thickness values remain in the specified minimum and maximum thickness range. The GA parameters are listed in Table~\ref{GA_Para1}, while the termination criteria for the same are 100 generations.

\begin{table}[hb]
    \centering
        \caption{Parameters of the genetic algorithm.}
    \begin{tabular}{c c c c}
        \hline
        \textbf{Parameter} & \textbf{Value} & \textbf{Parameter} & \textbf{Value} \\
        \hline
        Population size & 200 & Tournament size & 4 \\
        Crossover strength parameter, \(\eta_c\) & \(3 \left[1+\frac{1}{2}\left(1-e^{\frac{-g}{100}}\right)\right]\) & Number of Generations & 100 \\
        Mutation strength parameter, \(\eta_m\) & \(10 \left[1+\frac{1}{2}\left(1-e^{\frac{-g}{100}}\right)\right]\) & Mutation probability & 0.4 \\
       
          \hline
    \end{tabular}
    \label{GA_Para1}
\end{table}

\begin{figure}[ht!]
    \centering
    \begin{subfigure}[b]{0.27\textwidth}
        \centering
        \includegraphics[width=\textwidth]{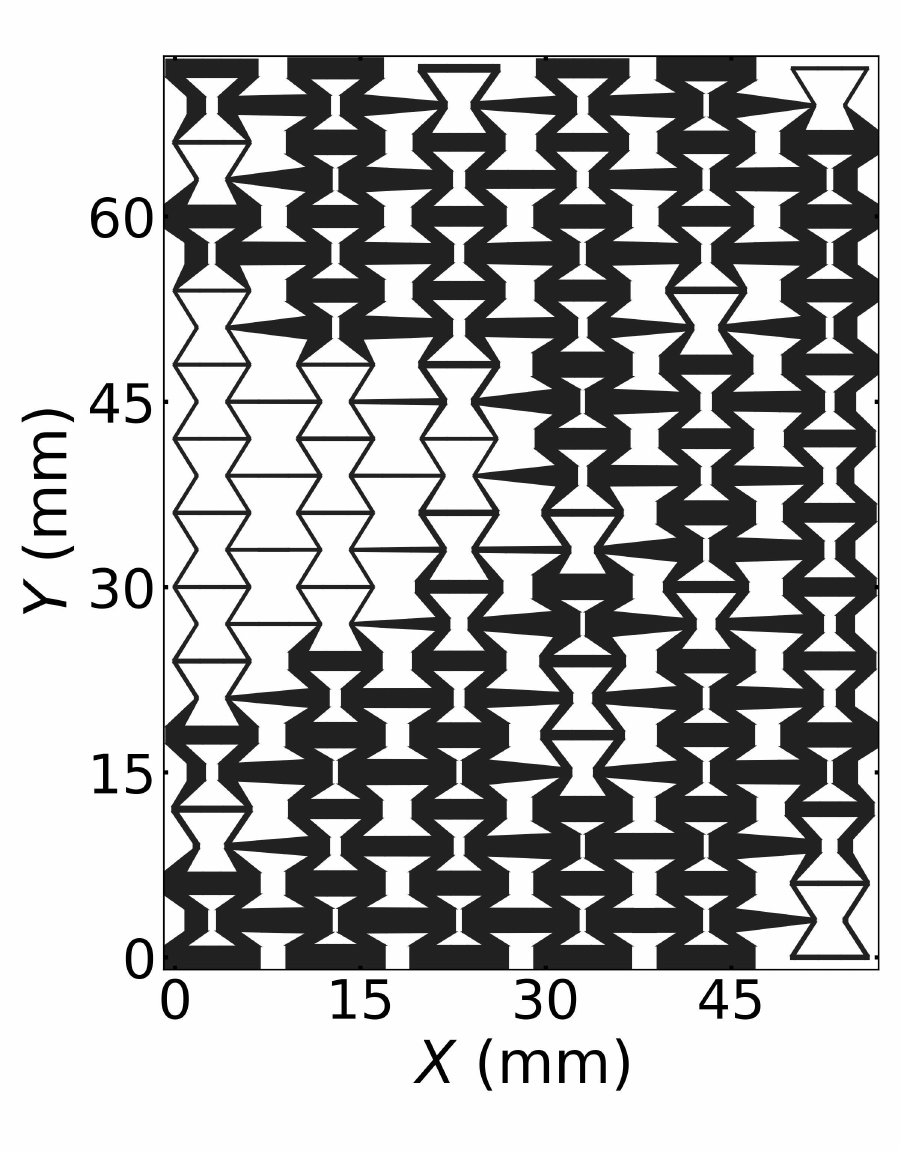}
        \caption{}
        \label{random_re_case1}
    \end{subfigure}
    \hspace{0.7cm}
    \begin{subfigure}[b]{0.27\textwidth}
        \centering
        \includegraphics[width=\textwidth]{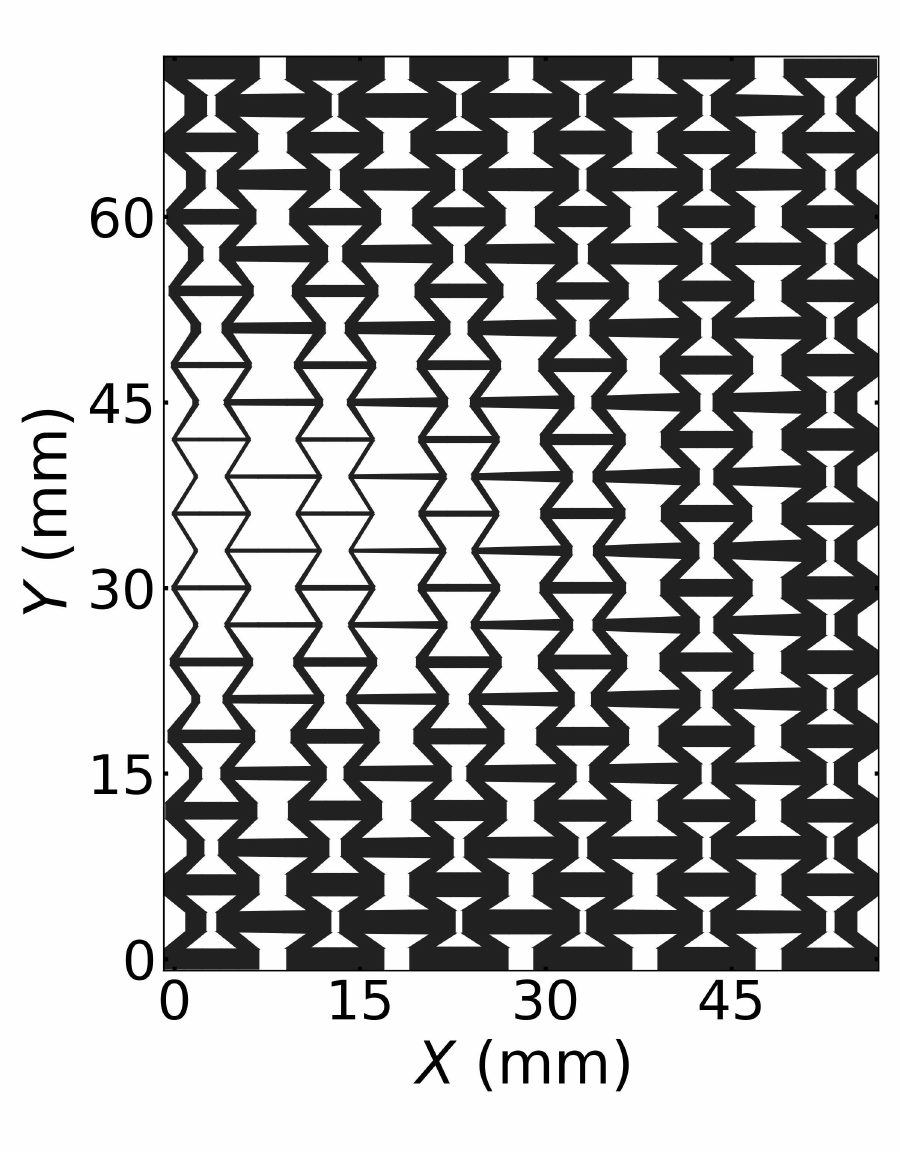}
        \caption{}
        \label{l20_re_case1}
    \end{subfigure}
    \hspace{0.7cm}
    \begin{subfigure}[b]{0.27\textwidth}
        \centering
        \includegraphics[width=\textwidth]{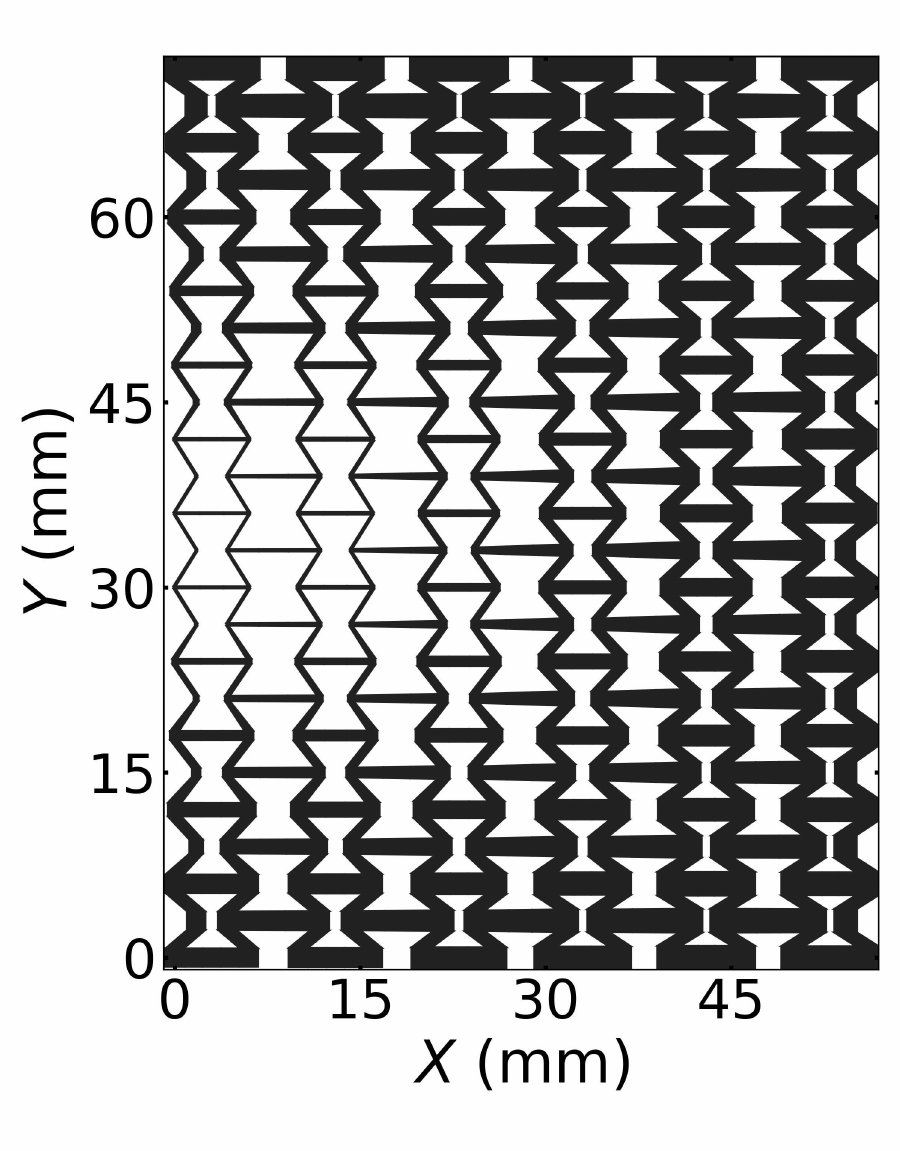}
        \caption{}
        \label{l30_re_case1}
    \end{subfigure}
    
    \caption{Optimized profiles of the lattice structure composed of the re-entrant unit cells generated by (a) conventional implementation, (b) GRF with length scale of 30 mm, and (c) GRF with length scale of 40 mm.}
    \label{re_case1}
\end{figure}

\begin{figure}[ht!]
    \centering
    \begin{subfigure}[b]{0.27\textwidth}
        \centering
        \includegraphics[width=\textwidth]{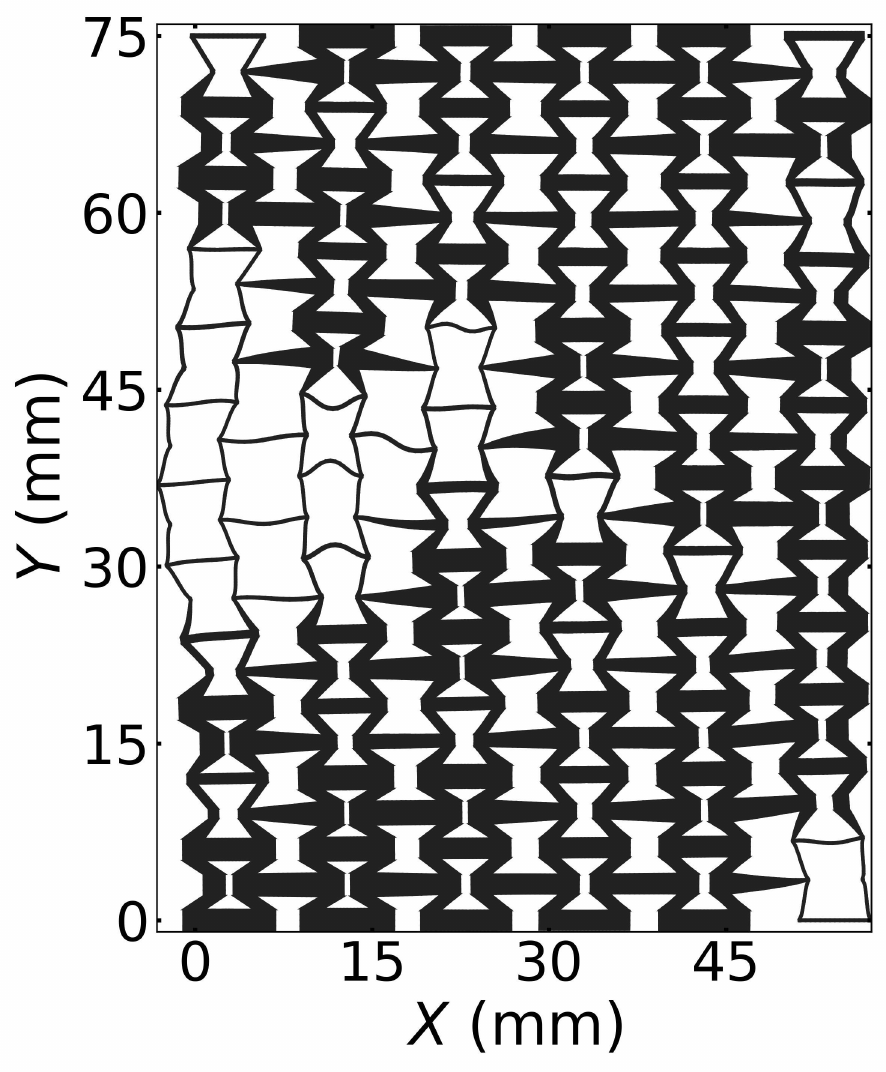}
        \caption{}
        \label{defrandom_re_case1}
    \end{subfigure}
    \hspace{0.7cm}
    \begin{subfigure}[b]{0.27\textwidth}
        \centering
        \includegraphics[width=\textwidth]{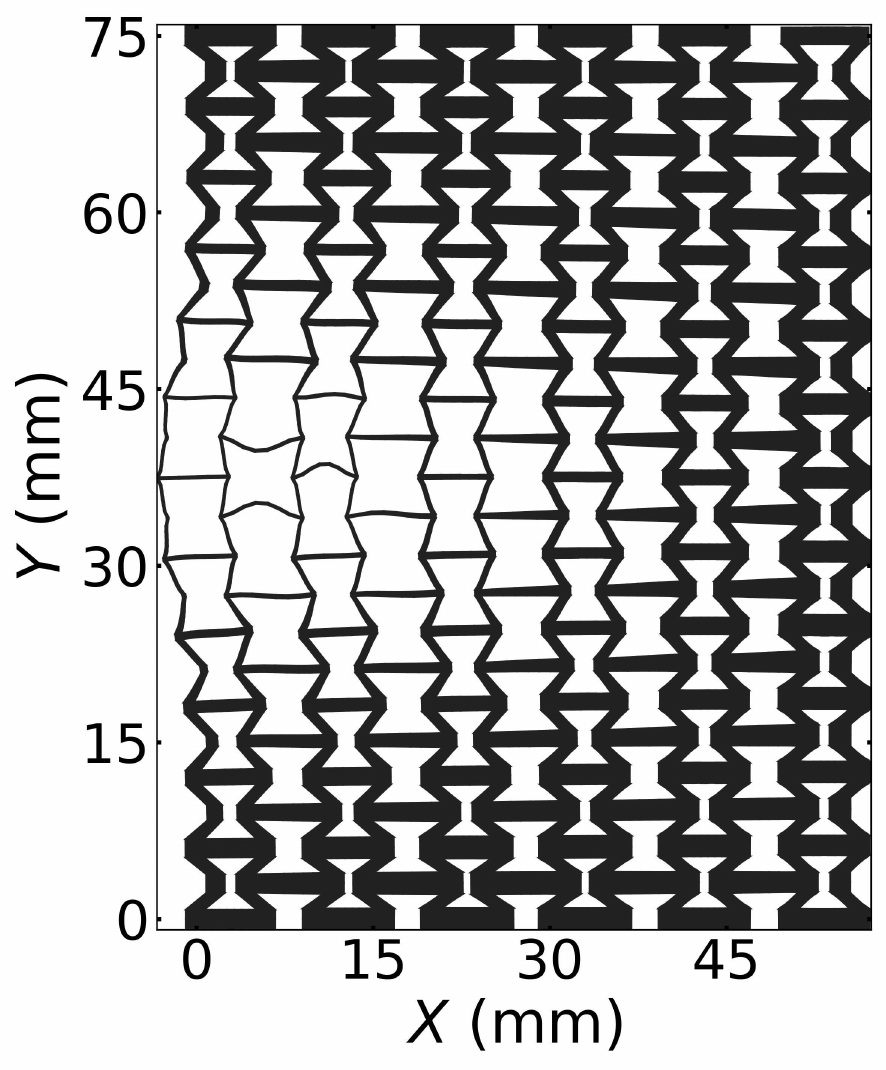}
        \caption{}
        \label{defl20_re_case1}
    \end{subfigure}
    \hspace{0.7cm}
    \begin{subfigure}[b]{0.27\textwidth}
        \centering
        \includegraphics[width=\textwidth]{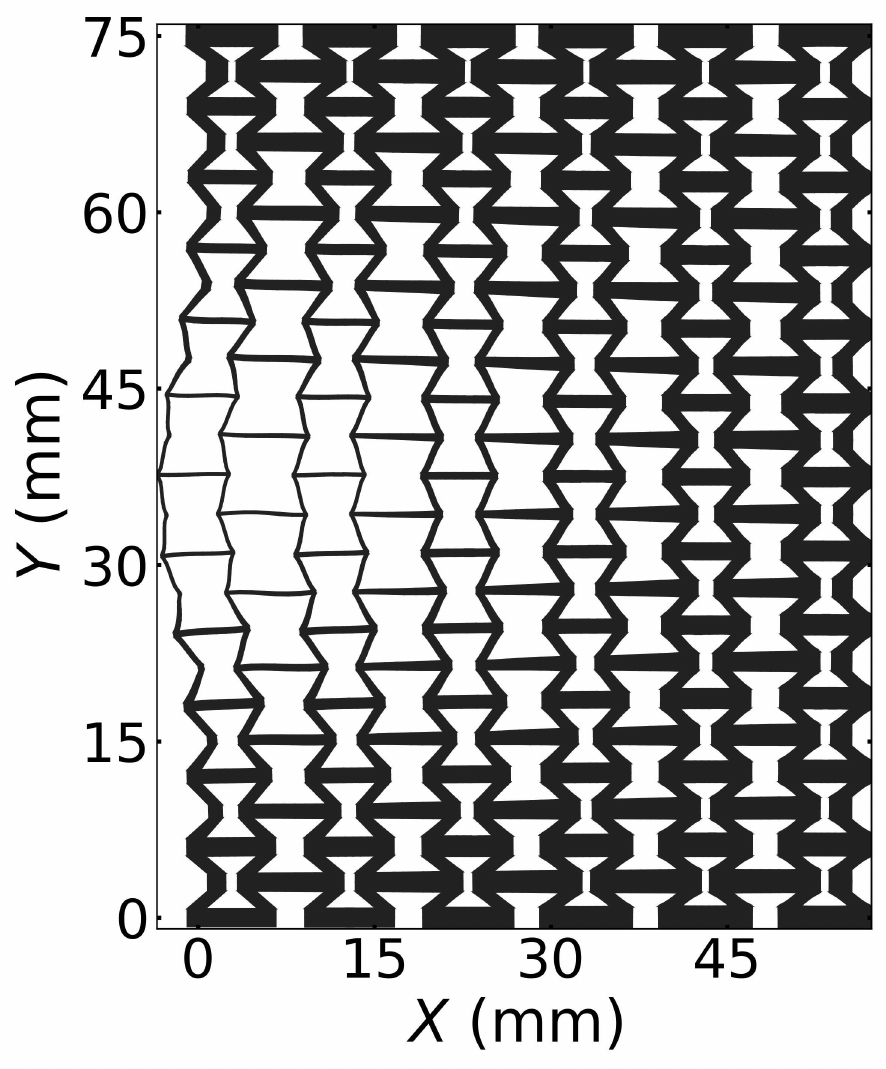}
        \caption{}
        \label{defl30_re_case1}
    \end{subfigure}
    
    \caption{Deformed configuration of the lattice structure composed of the re-entrant unit cells obtained by (a) conventional implementation, (b) GRF with length scale of 30 mm, and (c) GRF with length scale of 40 mm.}
    \label{def_re_case1}
\end{figure}

\begin{figure}[ht!]
    \centering
    \includegraphics[width=0.4\textwidth]{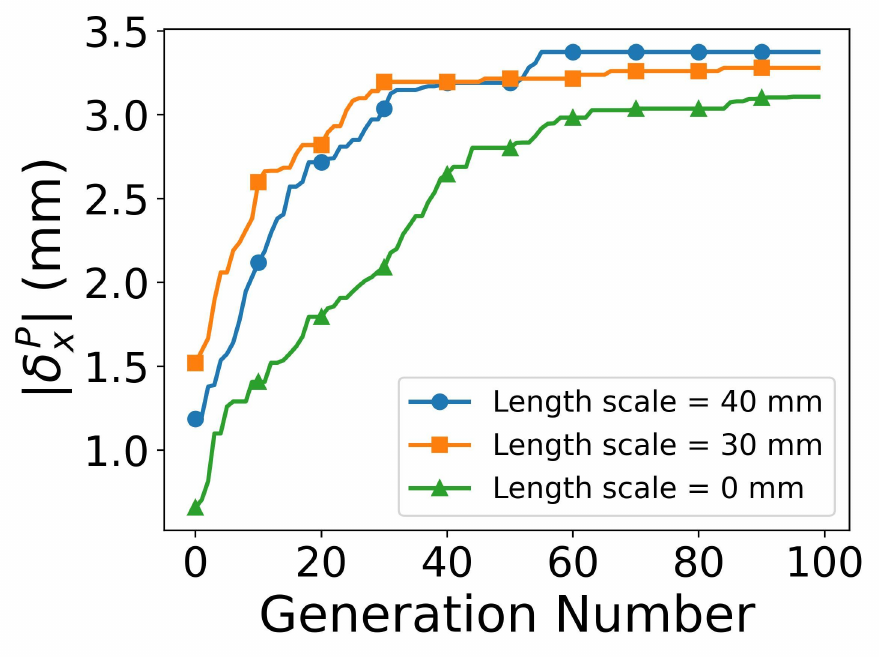}
    \caption{Evolution in the maximum deflection value of the point "P" for the best individual with respect to the GA generation.}
    \label{fitnes_re_case1}
\end{figure}

The optimized profiles with the conventional implementation and the proposed GRF scheme are presented in Fig.~\ref{re_case1}. The first design, Fig.~\ref{random_re_case1}, corresponds to the design obtained using the conventional implementation i.e. by taking design variables to be independent in nature, whereas the remaining profiles are generated using the GRF-based profile design algorithm integrated with a modified GA framework. The corresponding deformed configurations are shown in Fig.~\ref{def_re_case1}. The magnitude of deflection ($\delta_{x}^p$) for these cases is 3.20 mm, 3.30 mm, and 3.37 mm, respectively. As can be observed, these displacement values are comparable across all the optimal profiles and do not differ significantly in magnitude. However, as can be seen from Fig.~\ref{re_case1}, the profiles obtained from the proposed scheme are significantly smoother in terms of thickness transition, compared to the conventional implementation. The evolution of the best profile over GA generations is shown in Fig.~\ref{fitnes_re_case1}.

\begin{figure}[ht!]
    \centering
    \begin{subfigure}[b]{0.32\textwidth}
        \centering
        \includegraphics[width=\textwidth]{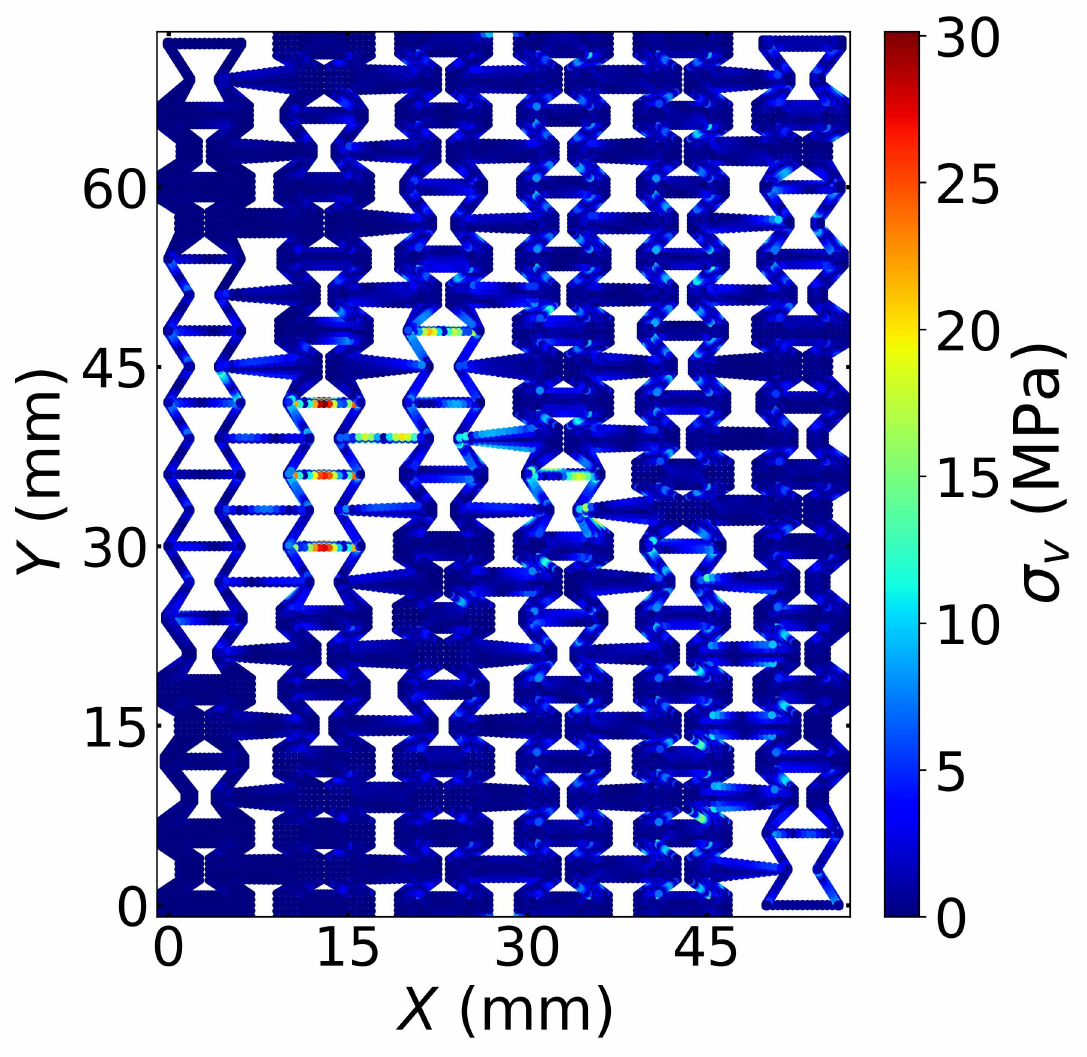}
        \caption{}
        \label{}
    \end{subfigure}
    \hspace{0.05cm}
    \begin{subfigure}[b]{0.32\textwidth}
        \centering
        \includegraphics[width=\textwidth]{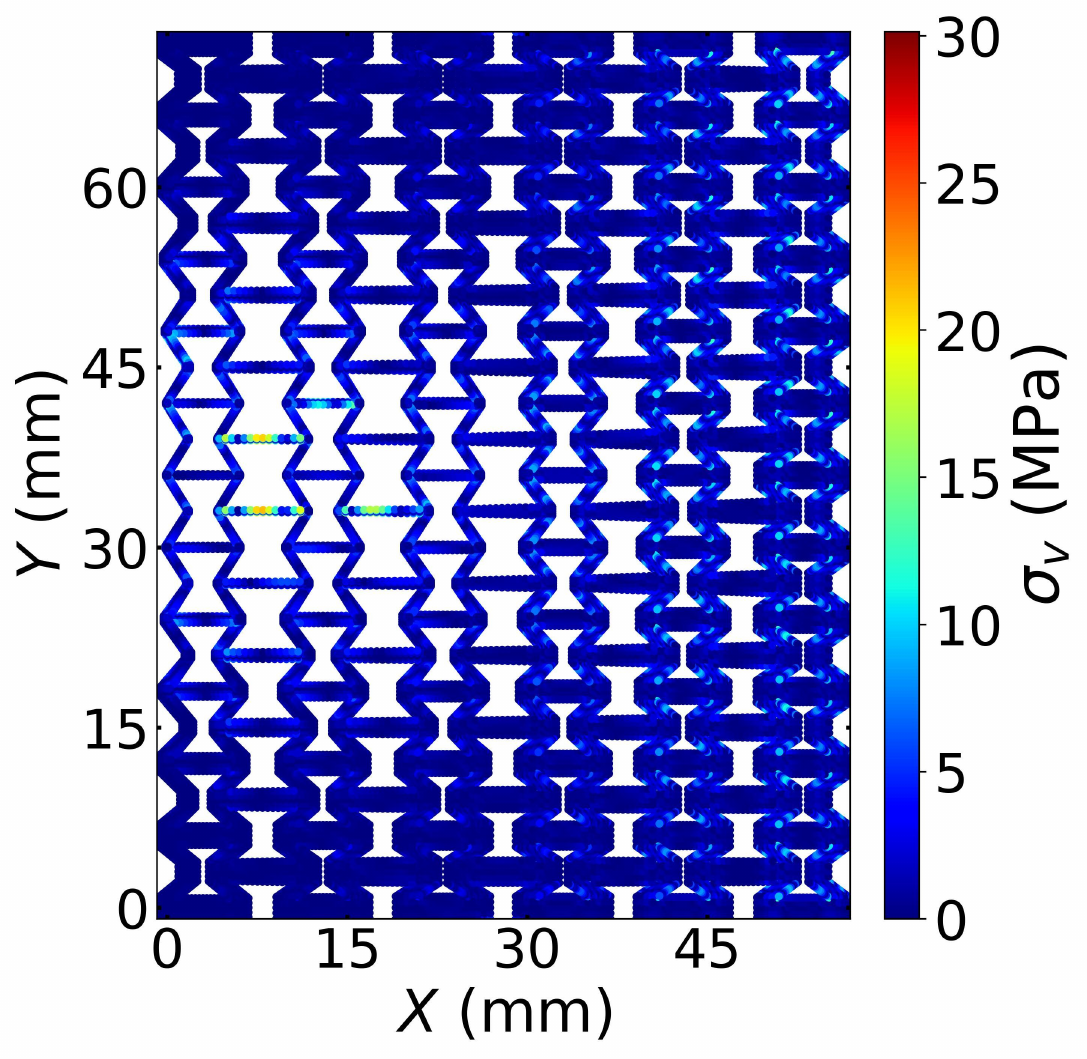}
        \caption{}
        \label{re_case1_l30_stress}
    \end{subfigure}
    \hspace{0.05cm}
    \begin{subfigure}[b]{0.32\textwidth}
        \centering
        \includegraphics[width=\textwidth]{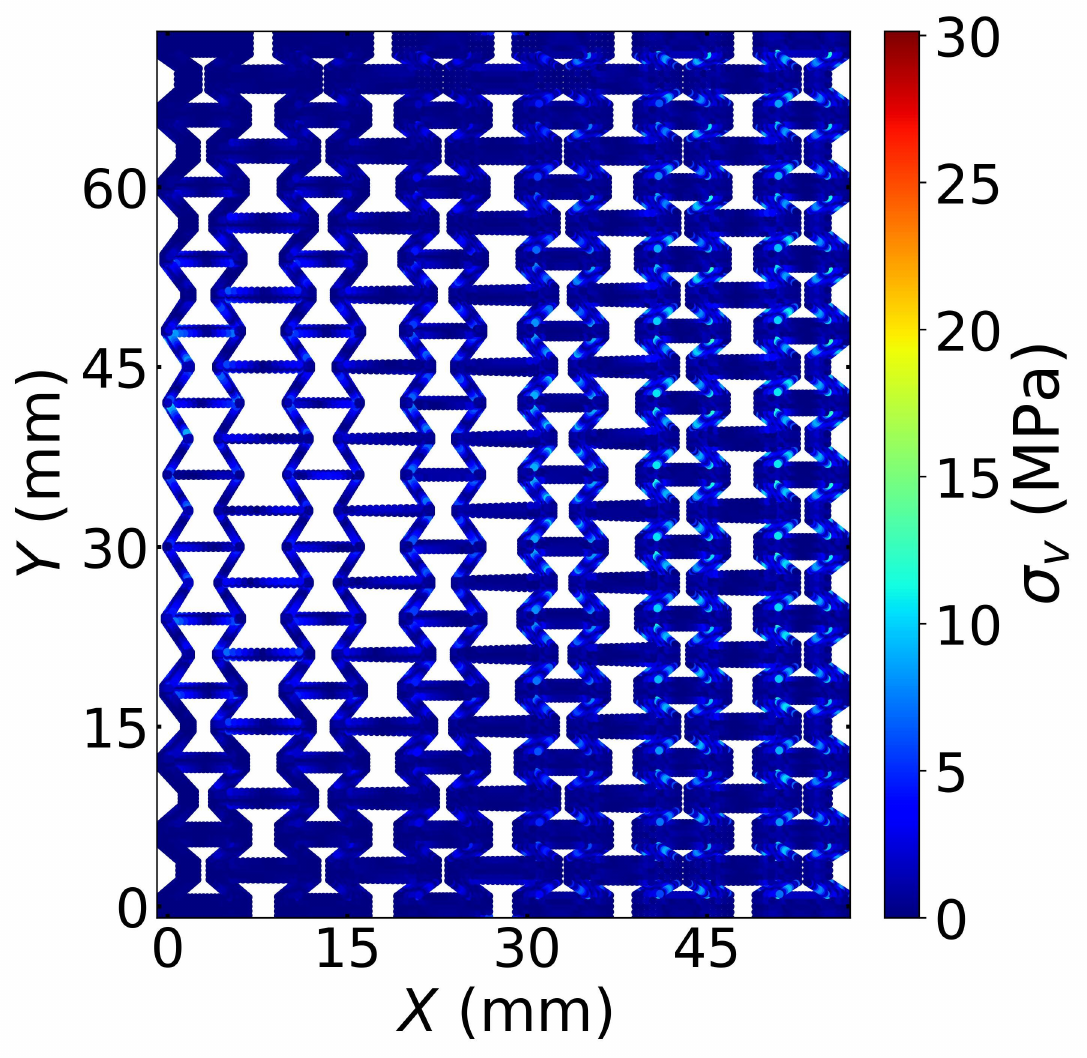}
        \caption{}
        \label{re_case1_random_stress}
    \end{subfigure}
    
    \caption{Von Mises stress distribution within the optimal designs of the lattice structure composed of the re-entrant unit cells obtained by (a) conventional implementation, (b) GRF with length scale of 30 mm, and (c) GRF with length scale of 40 mm.}
    \label{stress_re_case1}
\end{figure}

\begin{figure}[ht!]
    \centering
    \begin{subfigure}[b]{0.32\textwidth}
        \centering
        \includegraphics[width=\textwidth]{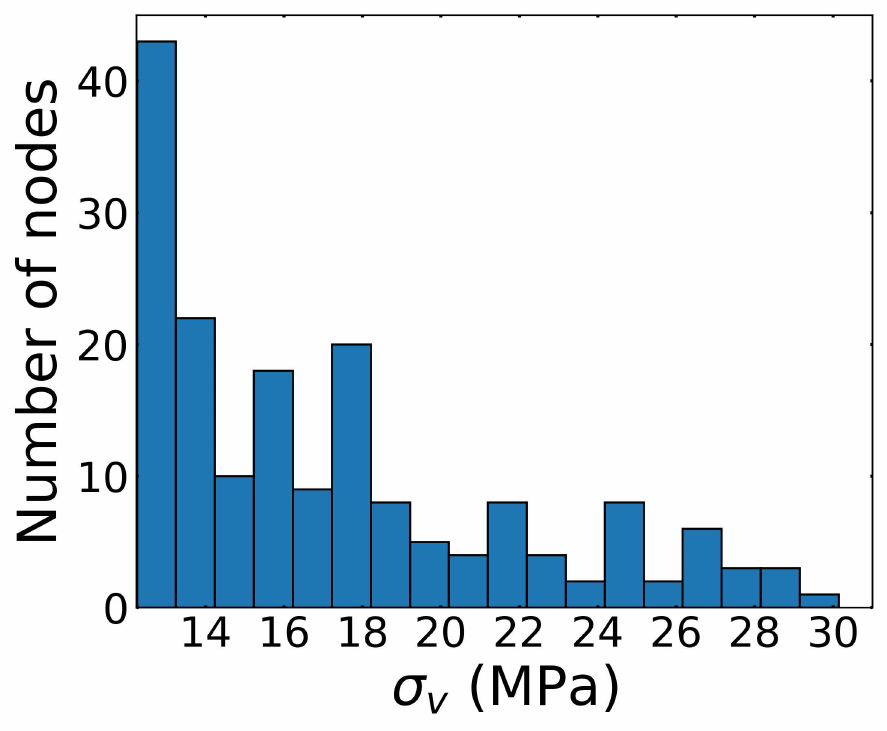}
        \caption{}
        \label{histrandom_re_case1}
    \end{subfigure}
    \hspace{0.05cm}
    \begin{subfigure}[b]{0.32\textwidth}
        \centering
        \includegraphics[width=\textwidth]{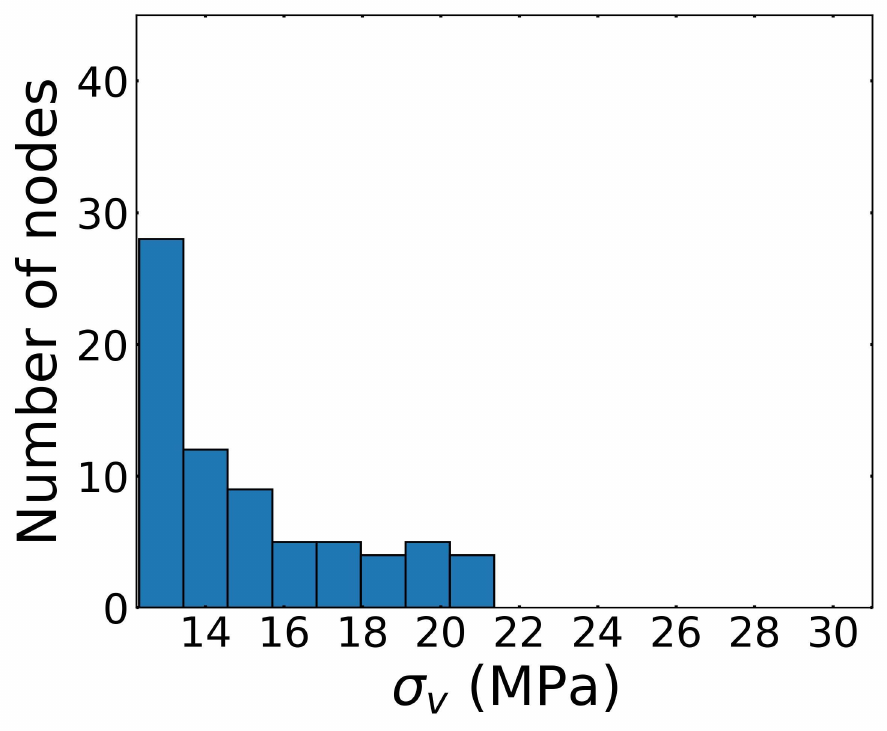}
        \caption{}
        \label{histl20_re_case1}
    \end{subfigure}
    \hspace{0.05cm}
    \begin{subfigure}[b]{0.32\textwidth}
        \centering
        \includegraphics[width=\textwidth]{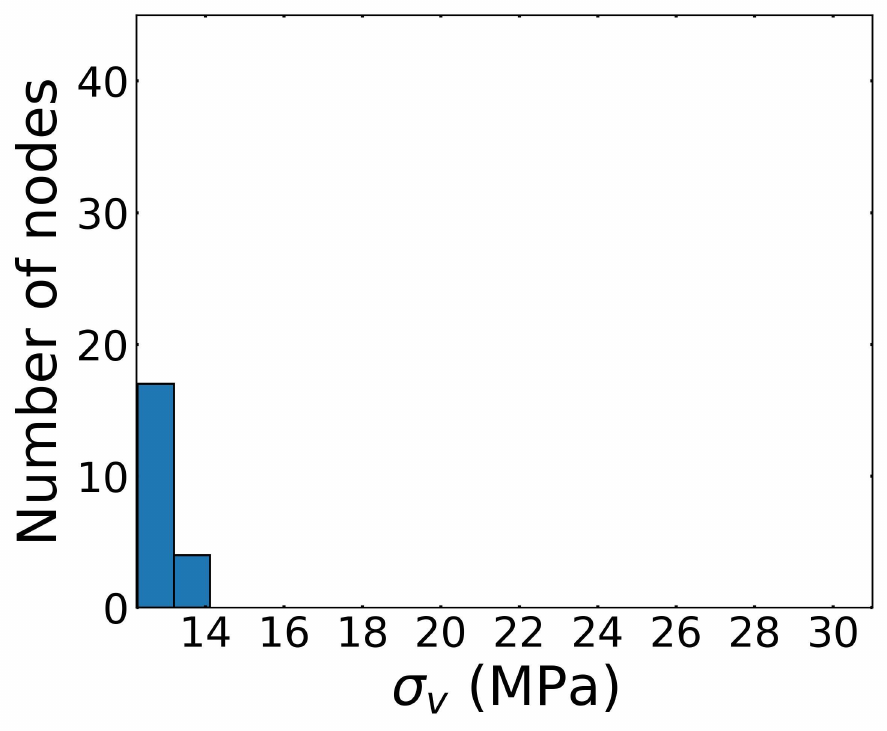}
        \caption{}
        \label{histl30_re_case1}
    \end{subfigure}
    
    \caption{Histogram of the number of  nodes having $\sigma_v$ $\geq$ 12.24 MPa (Note that the reference value of $\sigma_v$ is chosen as the $99.5^{th}$ percentile of the optimal design, considering conventional implementation) for the optimal structure obtained by (a) conventional implementation, (b) GRF with length scale of 30 mm, and (c) GRF with length scale of 40 mm.}
    \label{hist_re_case1}
\end{figure}

We further demonstrate the superiority of the optimum profiles obtained by the proposed scheme by comparing the stress distribution among the profiles presented in Fig.~\ref{stress_re_case1}. The finite element mesh to find the Von Mises stress ($\sigma_v$) consists of 16758 4-node quadrilateral elements. The maximum ($\sigma_v$) developed in these three optimal profiles shows some noticeable variation. For the conventional implementation, the optimized profile's maximum $\sigma_v$ is 30.14~MPa, while this value decreases to 21.35~MPa for the GRF-based optimal profile with a length scale of 30~mm, and further reduces to 14.11~MPa for the profile with a length scale of 40~mm. Furthermore, Fig.~\ref{hist_re_case1} presents a histogram of the number of nodes exceeding \(\sigma^*\), here, \(\sigma^*\) is the $99.5^{th}$ percentile of the $\sigma_v$  in the conventional implementation. As can be observed from the histogram, the smoother designs obtained from the GRF-based proposed scheme are less prone to stress concentration compared to the conventional implementation. In summary, the analysis of all optimal profiles provides strong evidence that the profiles obtained by the GRF-based algorithm are smoother and less prone to stress concentration, without compromising the objective function value.

\subsubsection{Case 2: Maximization of deflection using strut thicknesses and re-entrant angles as design variables.} 

In this case, we consider the more generic version of case 1 above. We now consider two design variables: the strut thickness ($t$) as well as the re-entrant angle ($\theta$) of the unit cells. 

\begin{equation}
\boldsymbol{\alpha}^{(II)} =
\Big\{
t_{p,q},\; \theta_{p,q}
\;\big|\;
(p,q) \in \mathcal{C}
\Big\}.
\end{equation}

For the present problem, material properties, GA parameters and finite element mesh are taken to be the same as those of Case-1. The minimum strut thickness is changed to 0.5 mm, while other thickness parameters such as maximum thickness and minimum average thickness same as previous Case-1. To generate the initial population for the genetic optimization, the GRF parameters are: length scales of 20 mm and 30 mm, while the standard deviation for the strut thickness is taken 0.5 mm, and for the re-entrant angle it is taken $10^\circ$. The re-entrant angle is bounded by $\theta_{\min} = 60^\circ$ and $\theta_{\max} = 100^\circ$. Similar to the thickness, angle values are normalized for each design between the minimum and maximum range. The resulting optimization problem is formulated as follows:

\begin{equation}
\begin{aligned}
&\textbf{maximize:} && \quad -\delta_{x}^{P}(\boldsymbol{\alpha}^{(II)}), 
 \\
&\textbf{subject to:} && \quad  t_{\min} \le t_{p,q} \le t_{\max},
\quad (p,q) \in \mathcal{C}, \\
&&& \quad\theta_{\min} \le \theta_{p,q} \le \theta_{\max},
\quad (p,q) \in \mathcal{C}, \\
&&& \quad\frac{1}{|\mathcal{C}|}
\sum_{(p,q)\in\mathcal{C}} t_{p,q}
\ge \bar{t}_{\max}.
\end{aligned}
\end{equation} 

\begin{figure}[ht!]
    \centering
    \begin{subfigure}[b]{0.27\textwidth}
        \centering
        \includegraphics[width=\textwidth]{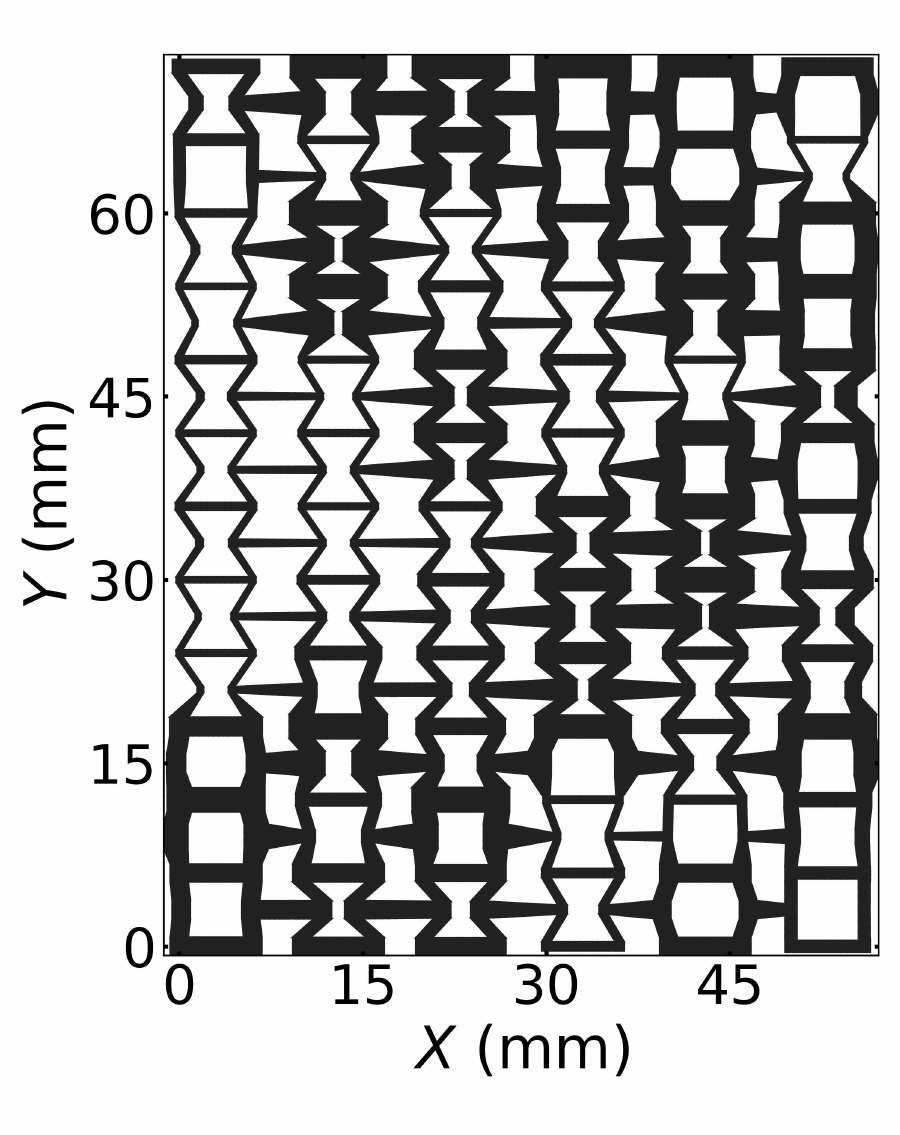}
        \caption{}
        \label{random_re_case2}
    \end{subfigure}
    \hspace{0.7cm}
    \begin{subfigure}[b]{0.27\textwidth}
        \centering
        \includegraphics[width=\textwidth]{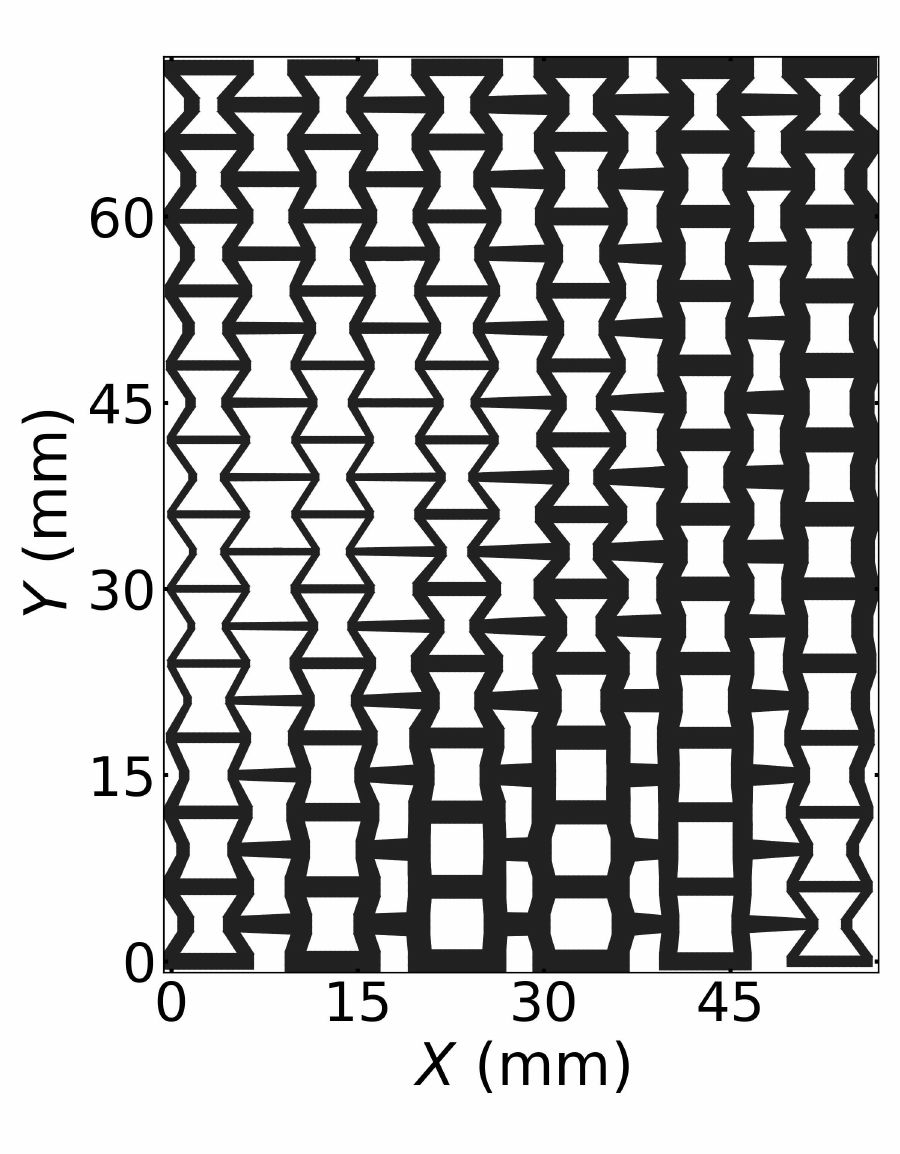}
        \caption{}
        \label{l20_re_case2}
    \end{subfigure}
    \hspace{0.7cm}
    \begin{subfigure}[b]{0.27\textwidth}
        \centering
        \includegraphics[width=\textwidth]{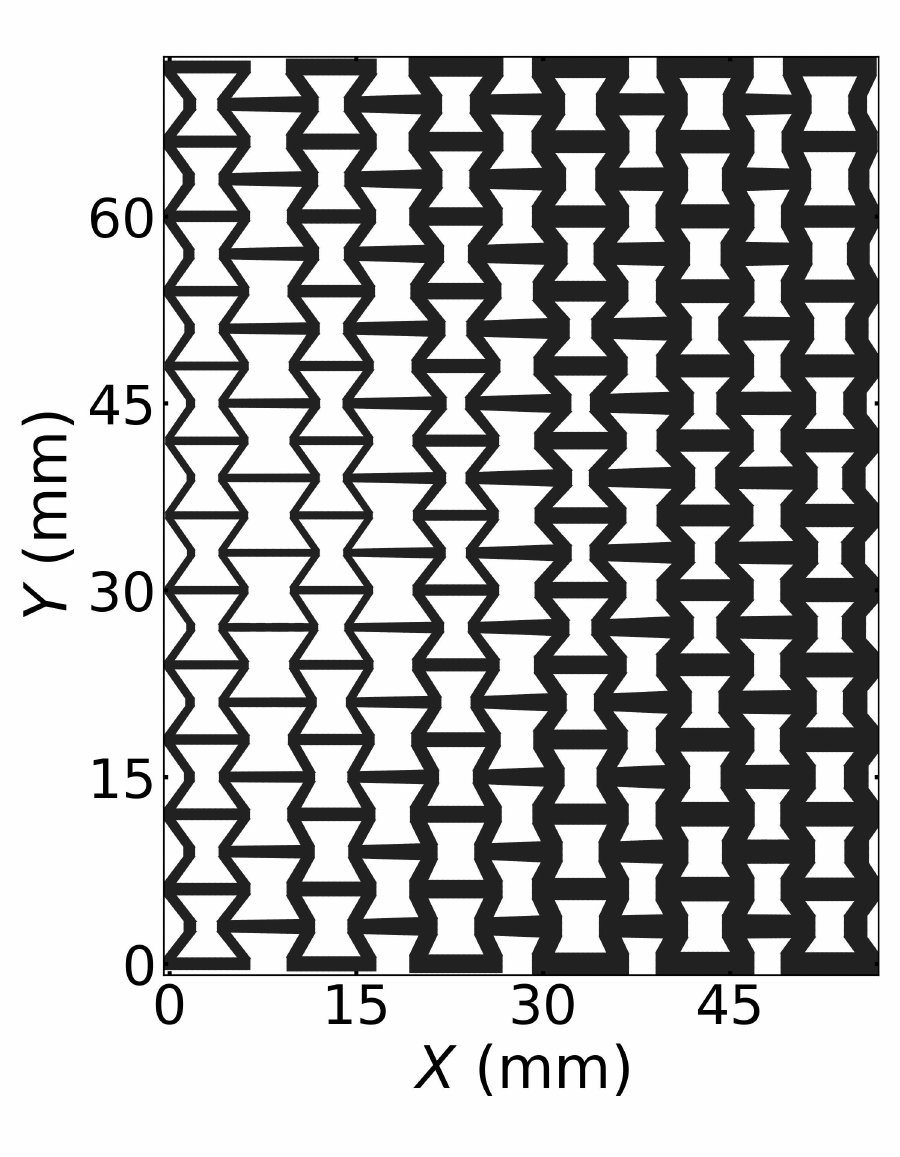}
        \caption{}
        \label{l30_re_case2}
    \end{subfigure}
    
    \caption{Optimized profiles of the lattice structure composed of the re-entrant unit cells generated by (a) conventional implementation, (b) GRF with length scale of 20 mm, and (c) GRF with length scale of 30 mm.}
    \label{re_case2}
\end{figure}

\begin{figure}[ht!]
    \centering
    \begin{subfigure}[b]{0.27\textwidth}
        \centering
        \includegraphics[width=\textwidth]{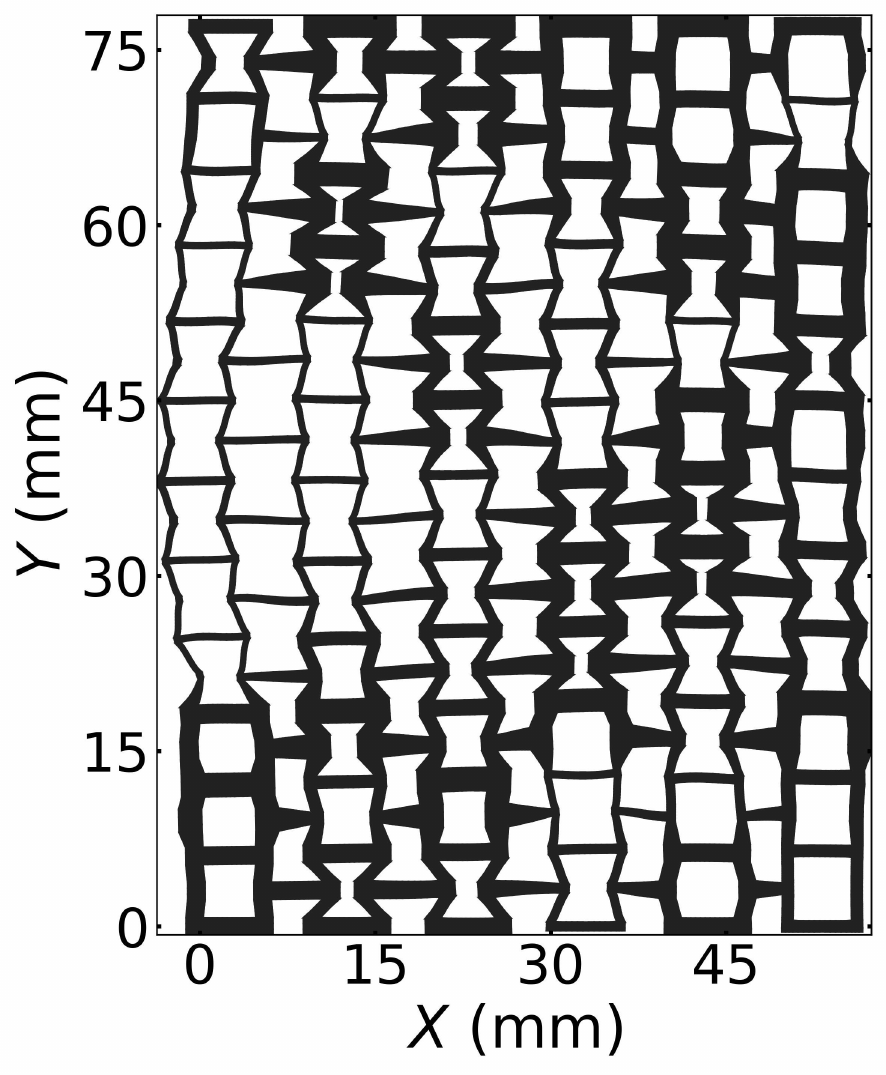}
        \caption{}
        \label{defrandom_re_case2}
    \end{subfigure}
    \hspace{0.7cm}
    \begin{subfigure}[b]{0.27\textwidth}
        \centering
        \includegraphics[width=\textwidth]{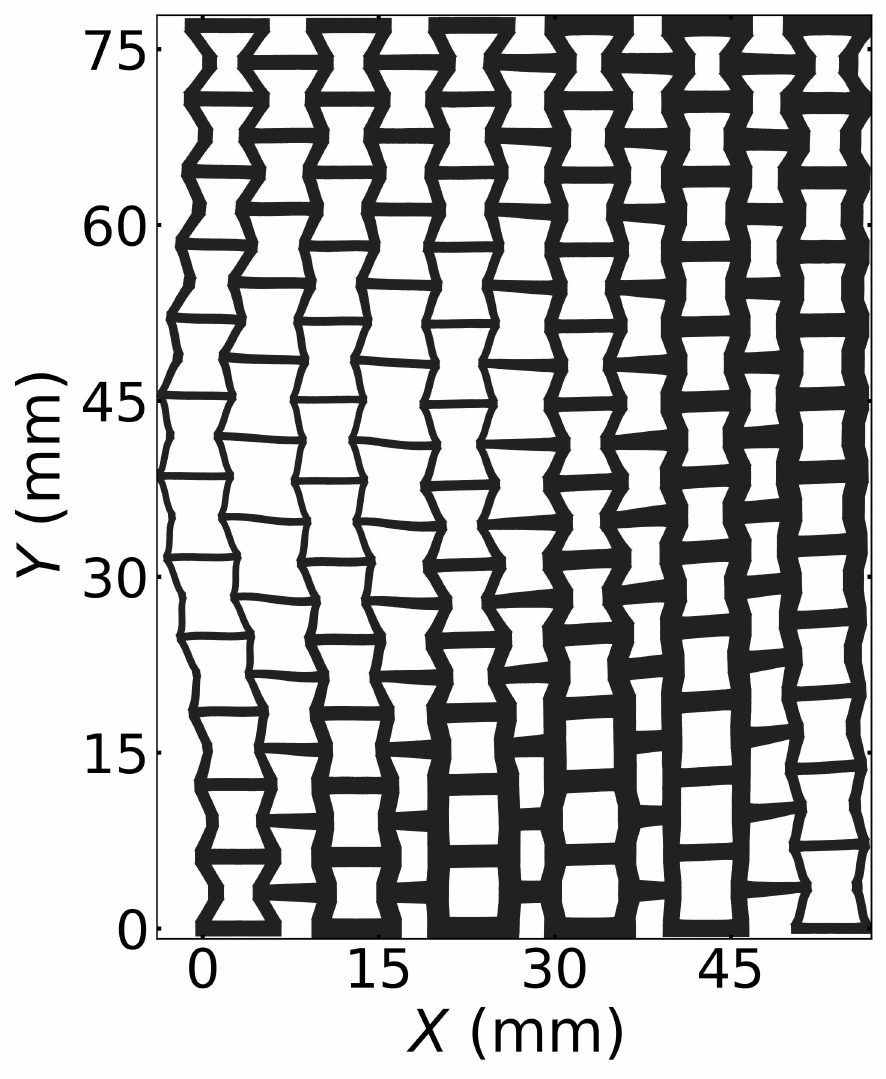}
        \caption{}
        \label{defl20_re_case2}
    \end{subfigure}
    \hspace{0.7cm}
    \begin{subfigure}[b]{0.27\textwidth}
        \centering
        \includegraphics[width=\textwidth]{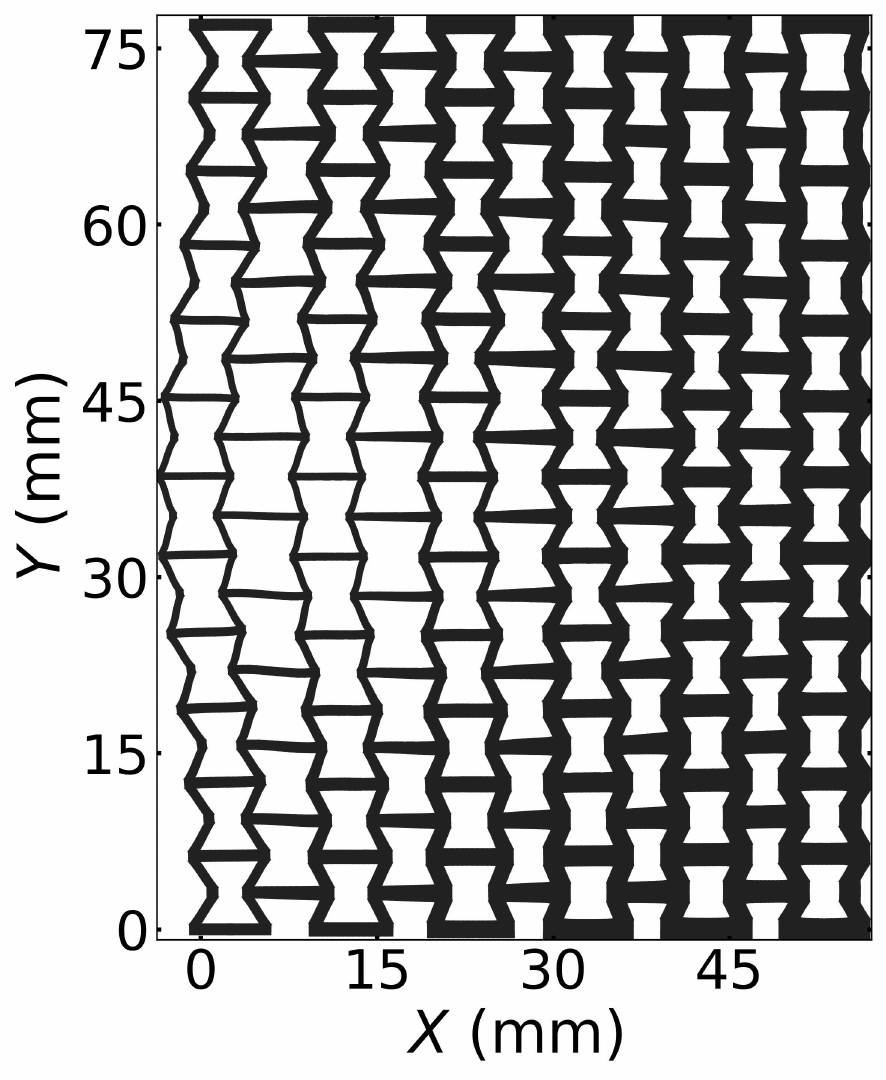}
        \caption{}
        \label{defl30_re_case2}
    \end{subfigure}
    
    \caption{Deformed configuration of the lattice structure composed of the re-entrant unit cells obtained by (a) conventional implementation, (b) GRF with length scale of 20 mm, and (c) GRF with length scale of 30 mm.}
    \label{def_re_case2}
\end{figure}

The optimized profiles obtained from the standard implementation and proposed optimization approach are shown in Fig.~\ref{re_case2}. The deformed geometries corresponding to these optimal geometries are shown in Fig.~\ref{def_re_case2}. Note that these deformed geometries correspond to the displacement applied along the Y-axis and boundary conditions shown in the Fig~\ref{reentrant_case1_diagram}. The displacement at point "P" is 3.30 mm for the optimal profile with conventional implementation, 3.64 mm for the optimum profile with a length scale of 20 mm, and 3.46 mm for the profile with a length scale of 30 mm. As can observed, similar to the case-1, in all the cases the value of the optimum objective function is quite close in all the cases. However, the optimal profiles obtained by the GRF-based algorithm integrated with the modified GA framework exhibit a smoother transition in re-entrant angle and unit cell thickness than those obtained by the conventional implementation.

\begin{figure}[ht!]
    \centering
    \includegraphics[width=0.4\textwidth]{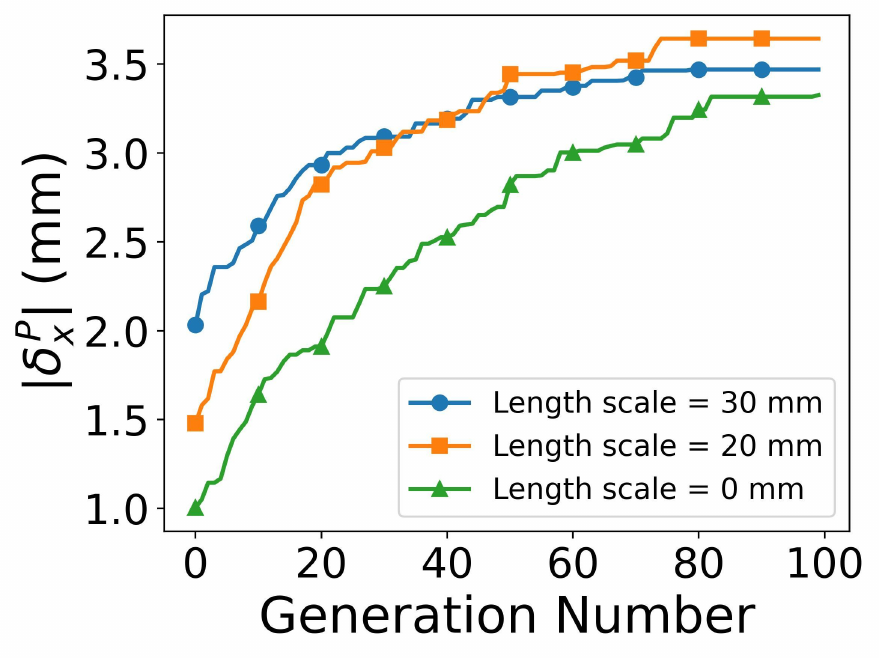}
    \caption{Evolution in the maximum deflection value of the point "P" for the best individual with respect to the GA generation.}
    \label{fitnes_re_case2}
\end{figure}

\begin{figure}[ht!]
    \centering
    \begin{subfigure}[b]{0.32\textwidth}
        \centering
        \includegraphics[width=\textwidth]{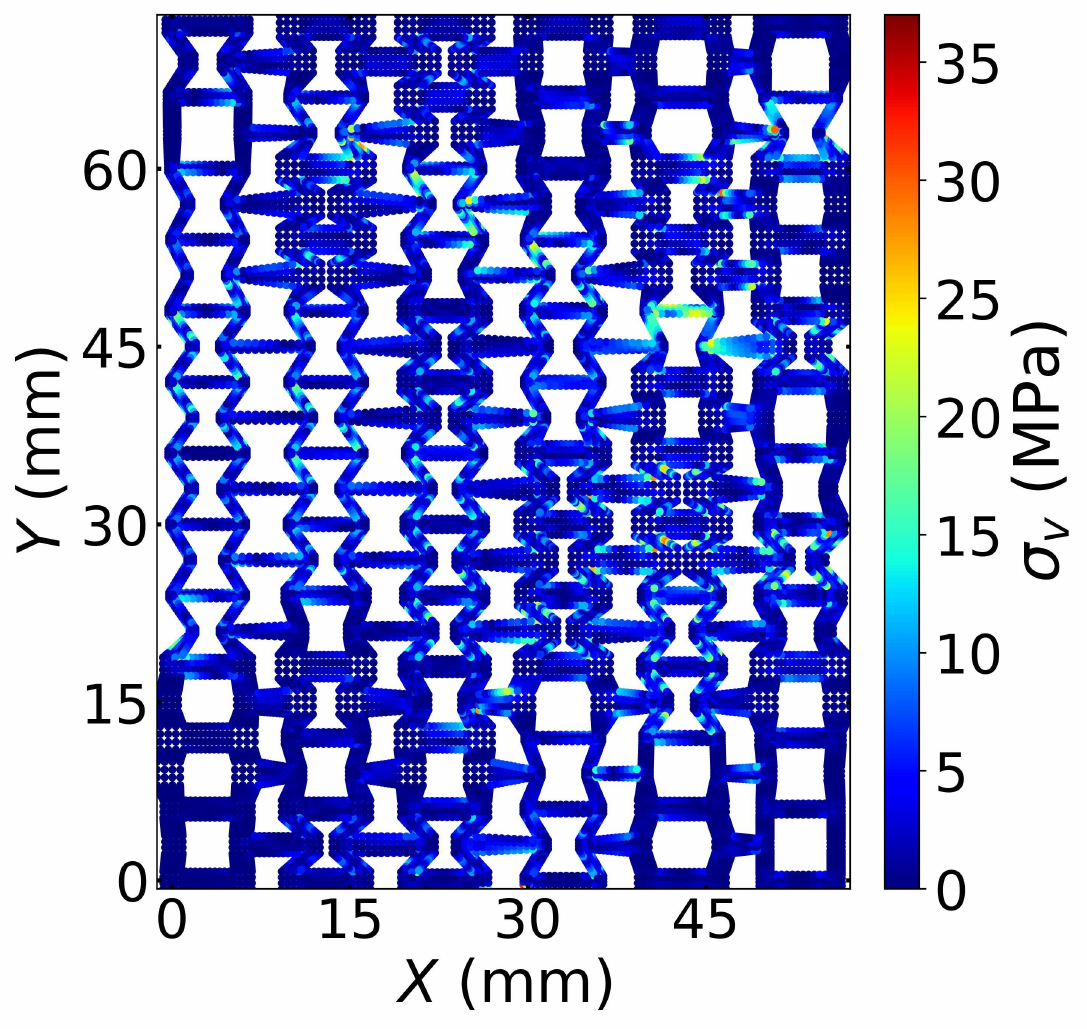}
        \caption{}
        \label{stressrandom_re_case2}
    \end{subfigure}
    \hspace{0.05cm}
    \begin{subfigure}[b]{0.32\textwidth}
        \centering
        \includegraphics[width=\textwidth]{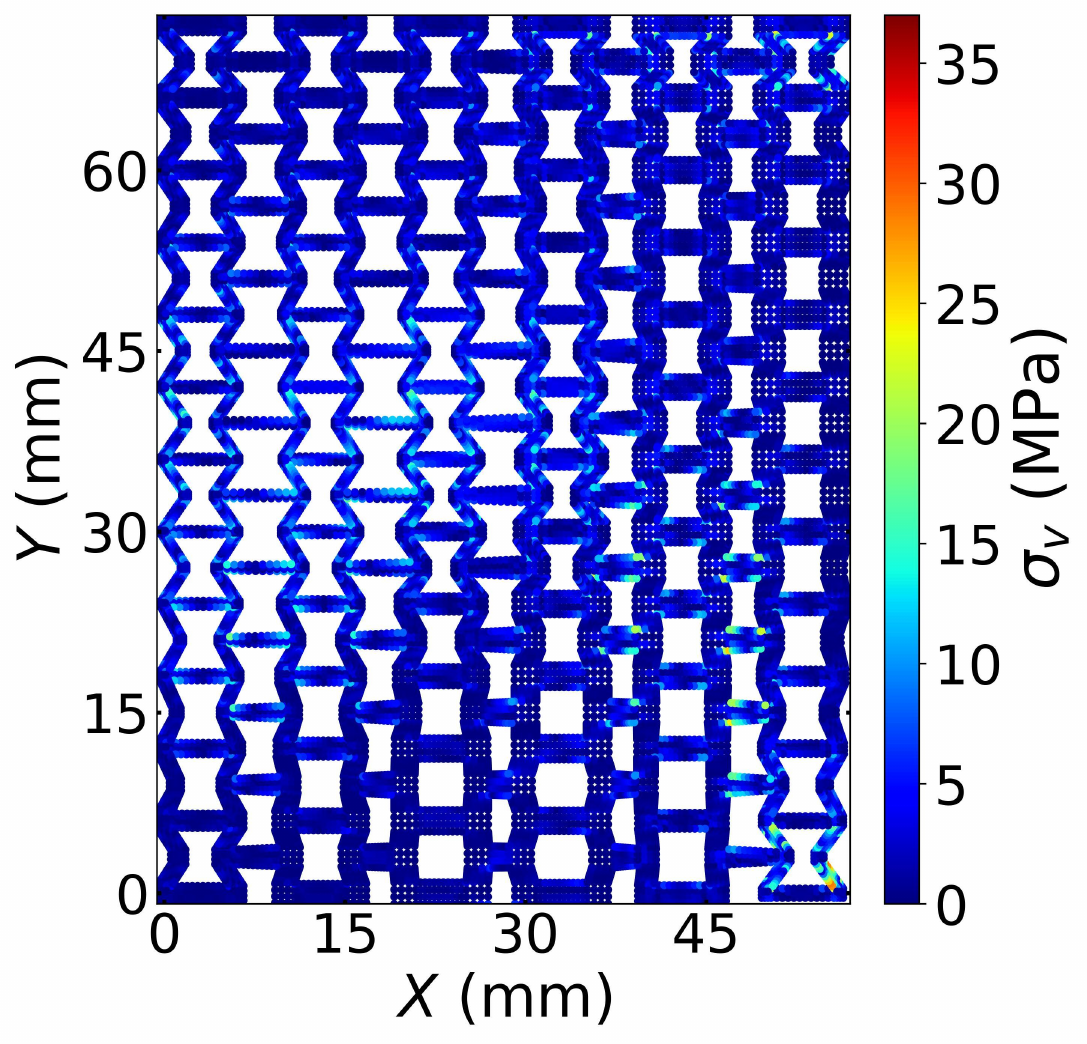}
        \caption{}
        \label{stressl201_re_case2}
    \end{subfigure}
    \hspace{0.05cm}
    \begin{subfigure}[b]{0.32\textwidth}
        \centering
        \includegraphics[width=\textwidth]{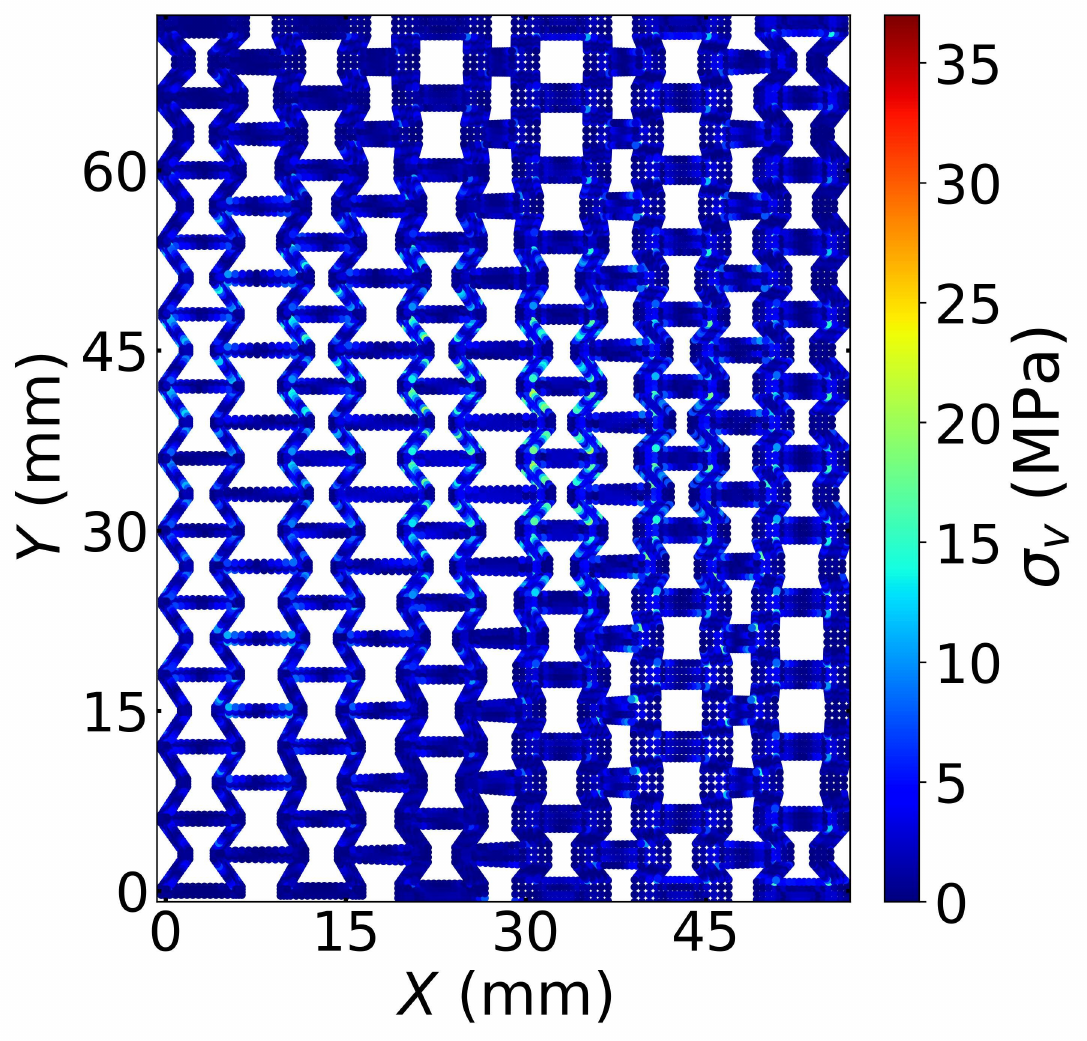}
        \caption{}
        \label{stressl30_re_case2}
    \end{subfigure}
    
    \caption{Von Mises stress distribution within the optimal designs of the lattice structure composed of the re-entrant unit cells obtained by (a) conventional implementation, (b) GRF with length scale of 20 mm, and (c) GRF with length scale of 30 mm.}
    \label{stress_re_case2}
\end{figure}

\begin{figure}[ht!]
    \centering
    \begin{subfigure}[b]{0.32\textwidth}
        \centering
        \includegraphics[width=\textwidth]{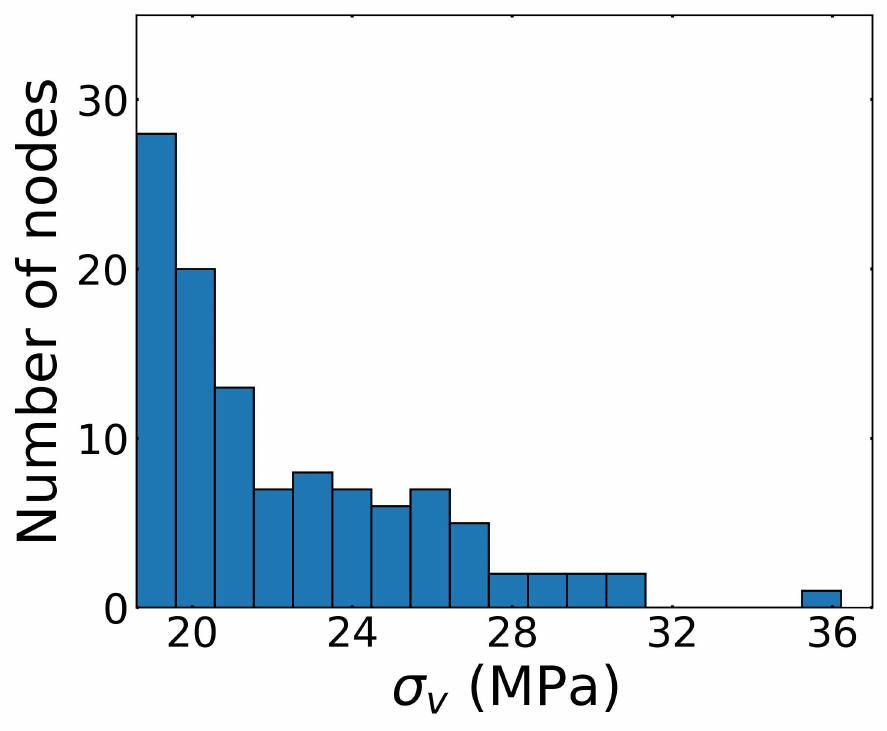}
        \caption{}
        \label{histrandom_re_case2}
    \end{subfigure}
    \hspace{0.05cm}
    \begin{subfigure}[b]{0.32\textwidth}
        \centering
        \includegraphics[width=\textwidth]{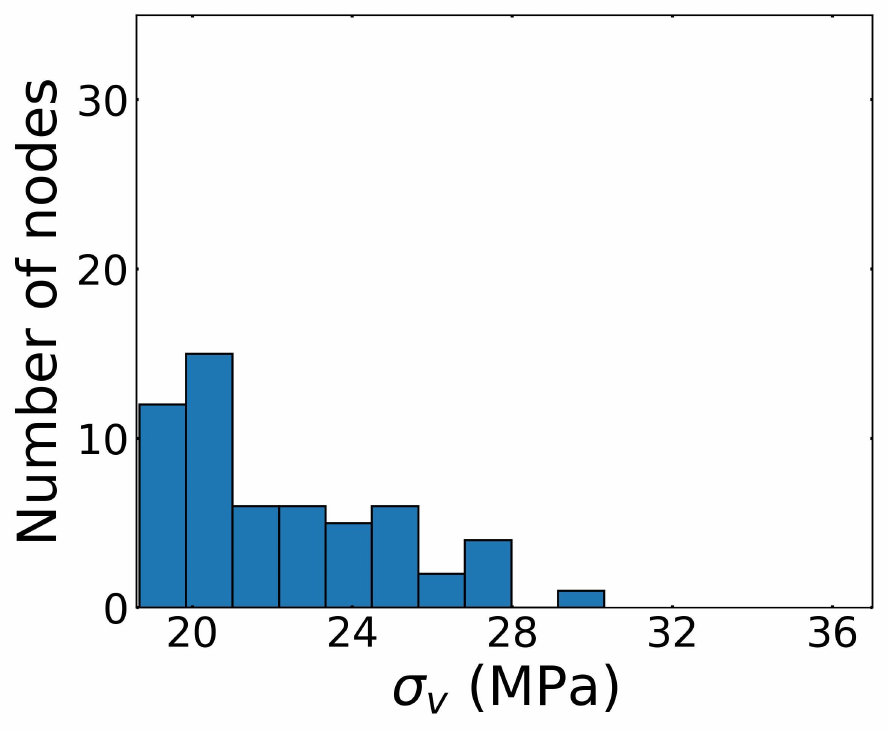}
        \caption{}
        \label{histl201_re_case2}
    \end{subfigure}
    \hspace{0.05cm}
    \begin{subfigure}[b]{0.32\textwidth}
        \centering
        \includegraphics[width=\textwidth]{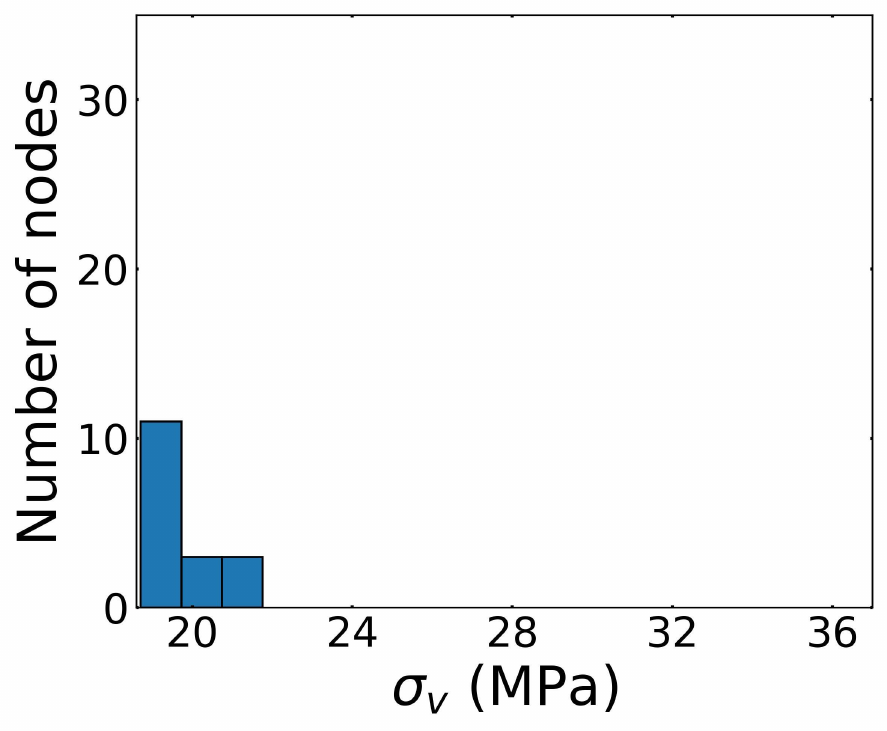}
        \caption{}
        \label{histl30_re_case2}
    \end{subfigure}
    
    \caption{Histogram of the number of  nodes having $\sigma_v$ $\geq$ 18.83 MPa (Note that the reference value of $\sigma_v$ is chosen as the $99.5^{th}$ percentile of the optimal design, considering conventional implementation) for the optimal structure obtained by (a) conventional implementation, (b) GRF with length scale of 20 mm, and (c) GRF with length scale of 30 mm.}
    \label{hist_re_case2}
\end{figure}

  The sudden transition in the design variable leads to higher stress concentrations within the structure, as can be seen from Fig.~\ref {stress_re_case2}, where the stress distribution for the optimum profiles is shown. Furthermore, the histogram of the $\sigma_v$ values (Fig.~\ref{hist_re_case2}) greater than \(\sigma^*\), where \(\sigma^*\) is 99.5 percentile of conventional implementation. This histogram demonstrates that GRF-based optimum profiles are less susceptible to stress concentration compared to the conventional implementation.  Also, the maximum value of $\sigma_v$ in the profile obtained by the conventional implementation is 36.2 MPa, while it reduces to 30.2 MPa and 20.7 MPa for the GRF-based optimal profiles. Also, Fig~\ref{fitnes_re_case2} shows the evolution of the best profile values over GA generations.

\subsection{Centered-rectangular unit cell-based lattice structures}

\begin{figure}[ht!]
    \centering
    \includegraphics[width=0.33\textwidth]{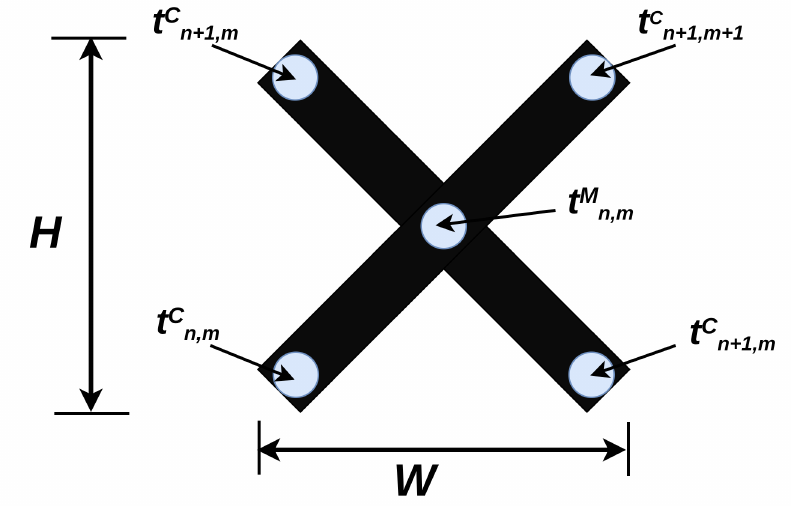}
    \caption{Schematic of a single centered rectangular unit cell located at position ($n,m$) in the lattice structure. where the nodal values are used as parameters governing the thickness of the struts.}
    \label{Bcc_unit_cell}
\end{figure}

This section considers the structures composed of centered rectangular unit cells, as illustrated in Fig. \ref{Bcc_unit_cell}. This section considers four 2-dimensional problems: 1) Cantilever beam subjected to a point load,
2) Cantilever beam under design constraint, 
3) MBB beam structure, and 4) cantilever beam subjected to thermal loading. In all these cases, the strut thickness ($t_k$) is considered as a design variable. 

Note that each two-dimensional structure composed of the $N$ $\times$ $M$ number of unit cells ($N$ along the X-axis and $M$ along the Y-axis), and the thickness of struts is defined by the corner nodes and the mid-cell node of the unit cell, as shown in Fig~\ref{Bcc_unit_cell}. The material has an elastic modulus of 400 MPa and a Poisson's ratio of 0.4. The corresponding index is defined as:
\begin{equation}
T^C
=
\left\{
(i,j)\;\middle|\;
i = 1,\ldots,N+1,\;
j = 1,\ldots,M+1
\right\},
\end{equation}

\begin{equation}
T^M
=
\left\{
(p,q)\;\middle|\;
p = 1,\ldots,N,\;
q = 1,\ldots,M
\right\}.
\end{equation}

Let
\begin{equation}
T = T^C \cup T^M, 
\end{equation}
denote the set of all nodes in the lattice.

The design variables are the strut thicknesses assigned at the lattice nodes, given by
\begin{equation}
t =
\Big\{
t^C_{i,j} \mid (i,j) \in T^C
\Big\}
\;\cup\;
\Big\{
t^M_{p,q} \mid (p,q) \in T^M
\Big\}.
\end{equation}

\subsubsection{Case 1: Design of stiff cantilever beam subjected to a point load}
\label{cantilver_beam_section}

\begin{figure}[ht!]
    \centering
    \includegraphics[width=0.6\textwidth]{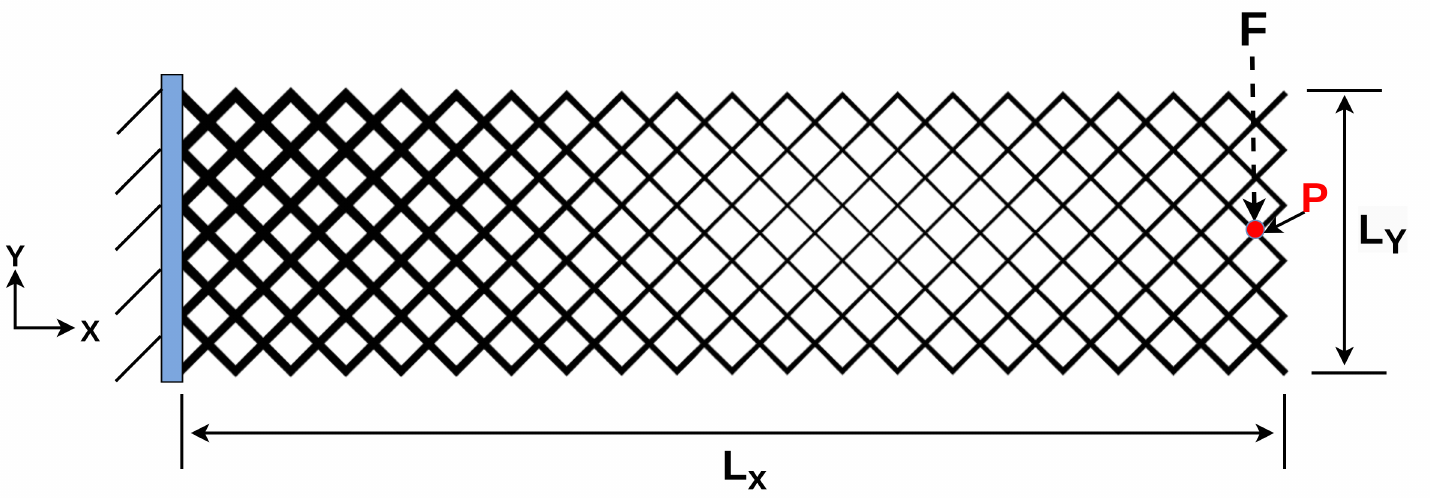}
    \caption{Schematic of the cantilever beam subjected to a point load.}
    \label{cb_diagram}
\end{figure}

In this problem, we are solving a cantilever beam composed of centered rectangular unit cells. The beam is fixed at the left edge and subjected to the force (F) at the mid of the right edge along the negative Y-axis with the intensity of 1,000 N as shown in Fig. \ref{cb_diagram}.  Each unit cell has dimensions of 10 mm along both the width and the height, for a total of 20 × 5 unit cells. The FEM analysis has been carried out using 4-node quadrilateral elements, with a total of 3,026 elements.

Our objective is to minimize the maximum deflection ($\delta_y$) of the cantilever beam along the Y-axis, within the constraint that the mean thickness of the strut is limited to $\bar{t}_{max}$ = 2.25 mm, while the thickness of each individual strut ($t_k$) is bounded by $t_{min}$ = 1.0 mm and  $t_{max}$ = 4.0 mm. The optimization problem is stated as follows:

\begin{equation}
\begin{aligned}
&\textbf{minimize:} && \quad \delta_{y}(t),  \\
&\textbf{subject to:} && \quad t_{\mathrm{min}} \le  t_k \le t_{\mathrm{max}},\; t_k \in T, \\
&&&\quad \frac{1}{|T|}
\sum_{k \in T} t_k \le \bar{t}_{\max}.
\end{aligned}
\end{equation}

For optimization, the initial design space is generated using GRF with three different length scales: 10 mm, 20 mm, and 30 mm. The standard deviation is 1.0 mm, remain same for all three length scales. Further, the design variables are normalized between the minimum and maximum thickness values. The parameters of GA remain same to the previous problems.

\begin{figure}[ht!]
    \centering
    \begin{subfigure}[b]{0.45\textwidth}
        \includegraphics[width=\linewidth]{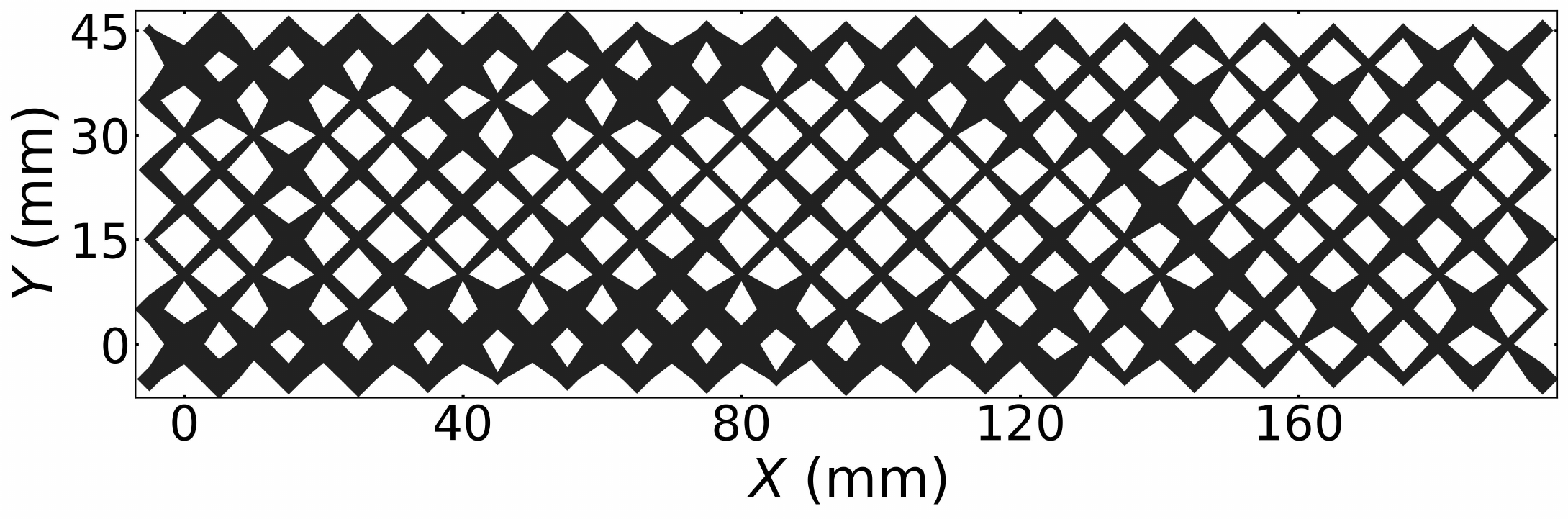}
        \caption{}
        \label{}
    \end{subfigure}
    \hfill
    \begin{subfigure}[b]{0.45\textwidth}
        \includegraphics[width=\linewidth]{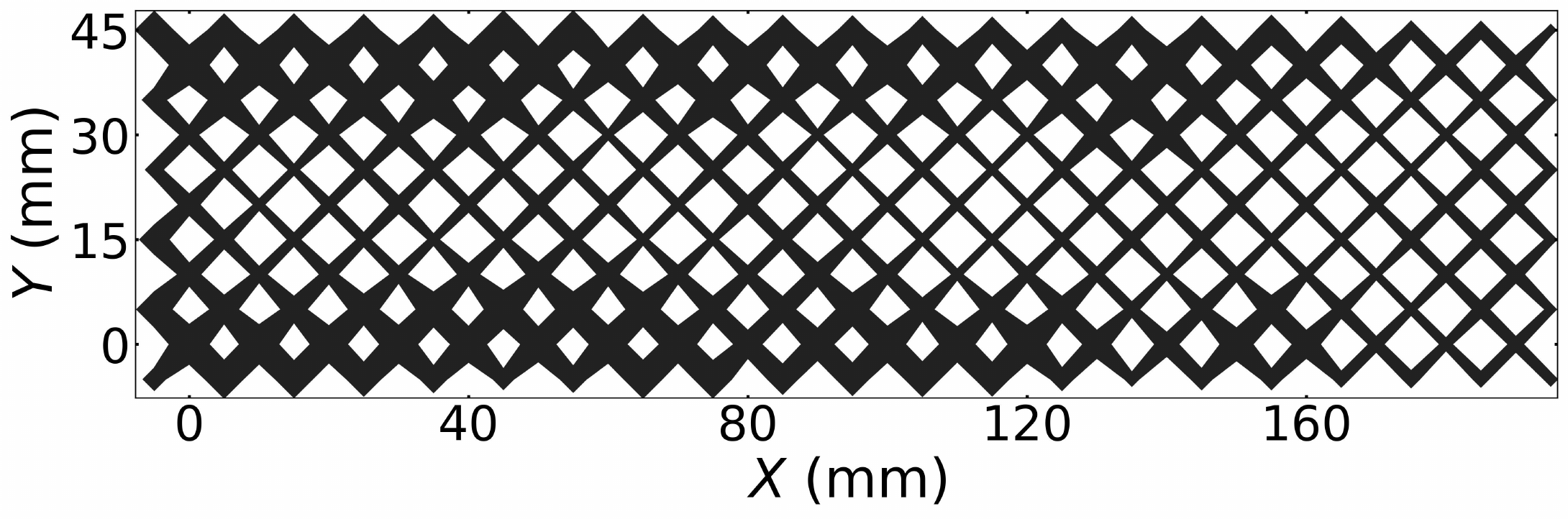}
        \caption{}
        \label{}
    \end{subfigure}

    \vspace{0.5cm}

    \begin{subfigure}[b]{0.45\textwidth}
        \includegraphics[width=\linewidth]{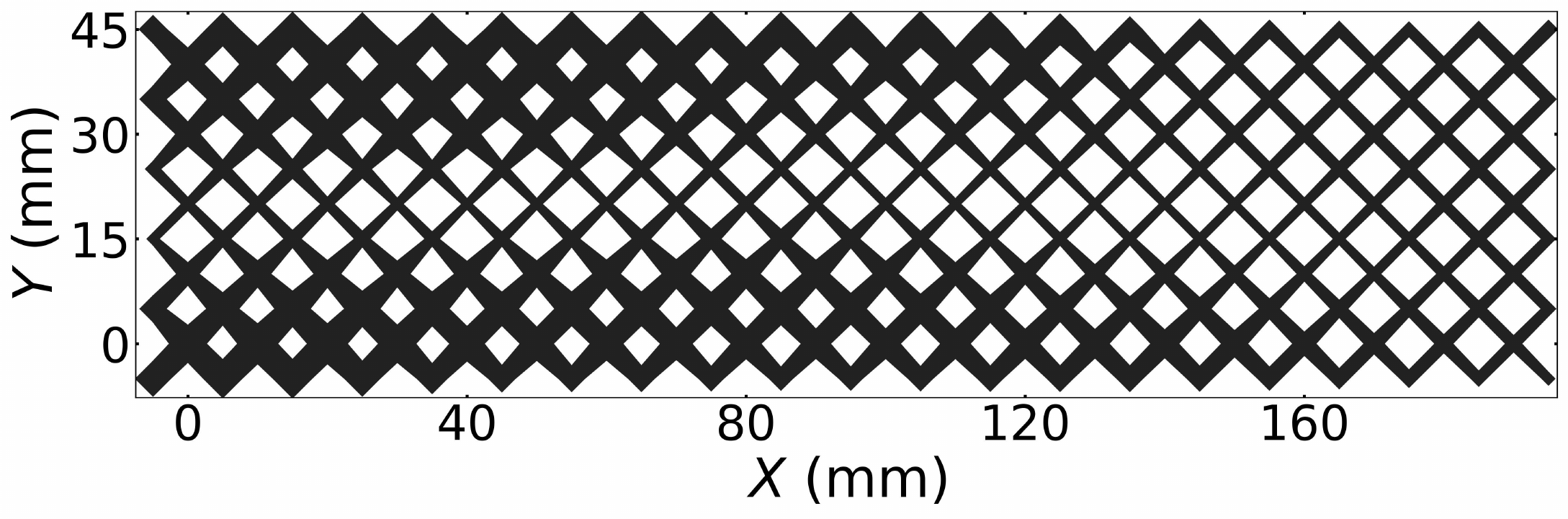}
        \caption{}
        \label{}
    \end{subfigure}
    \hfill
    \begin{subfigure}[b]{0.45\textwidth}
        \includegraphics[width=\linewidth]{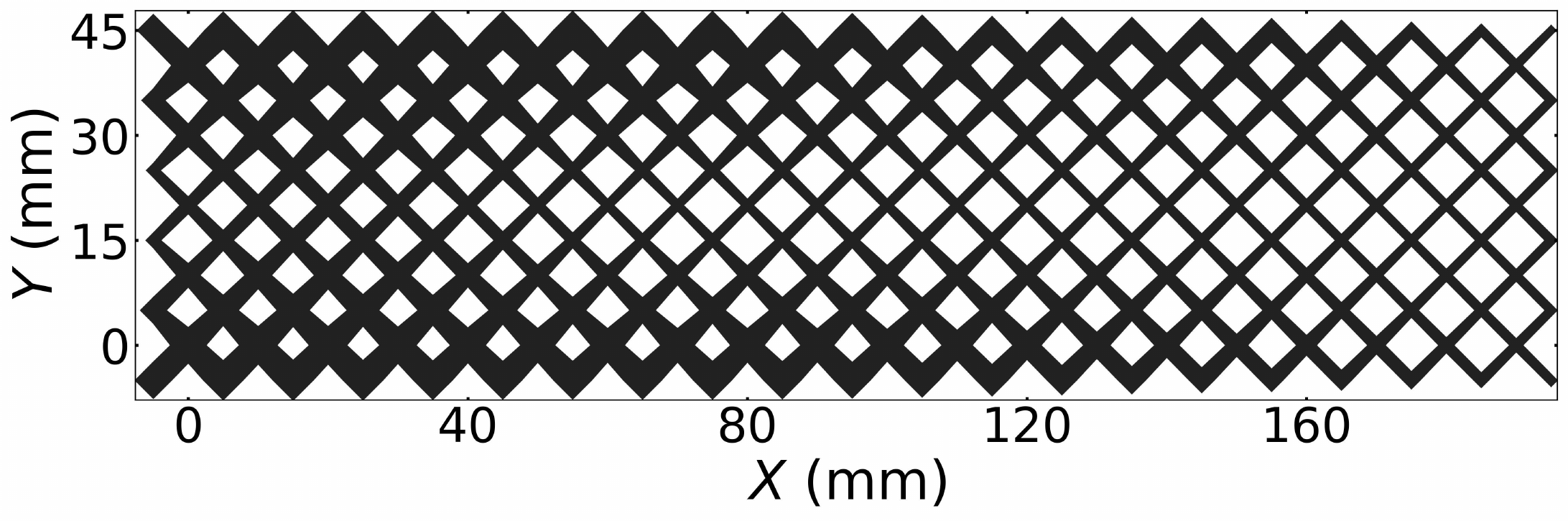}
        \caption{}
        \label{}
    \end{subfigure}

    \caption{Optimized profiles of the cantilever beam composed of the centered rectangular unit cells generated by (a) conventional implementation, (b) GRF with length scale of 10 mm, (c) GRF with length scale of 20 mm, and (d) GRF with length scale of 30 mm.}
    \label{cb}
\end{figure}

\begin{figure}[ht!]
    \centering
    \begin{subfigure}[b]{0.45\textwidth}
        \includegraphics[width=\linewidth]{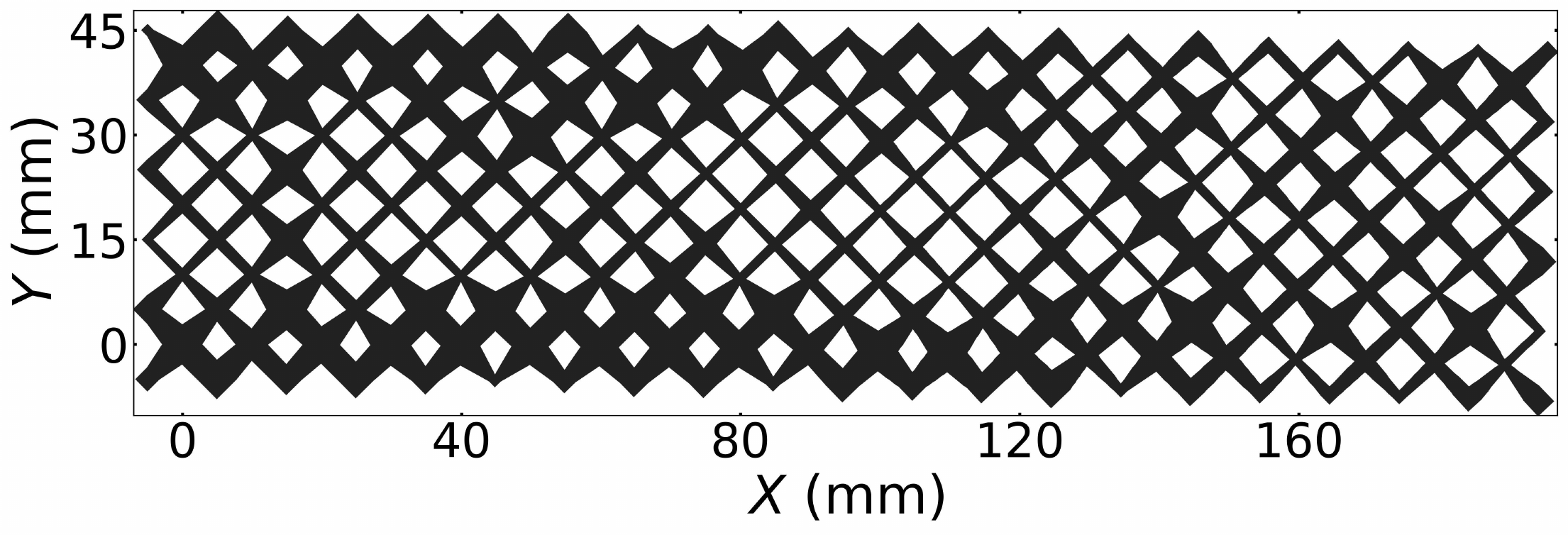}
        \caption{}
        \label{defrandom_cb}
    \end{subfigure}
    \hfill
    \begin{subfigure}[b]{0.45\textwidth}
        \includegraphics[width=\linewidth]{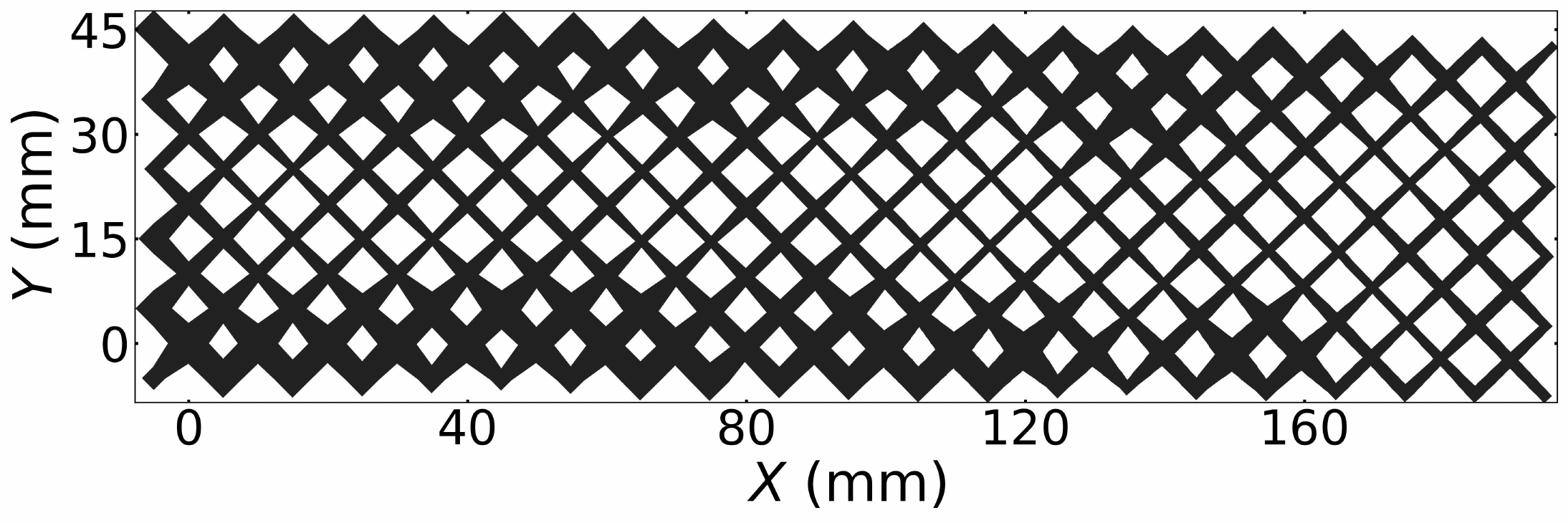}
        \caption{}
        \label{defl10_cb}
    \end{subfigure}

    \vspace{0.5cm}

    \begin{subfigure}[b]{0.45\textwidth}
        \includegraphics[width=\linewidth]{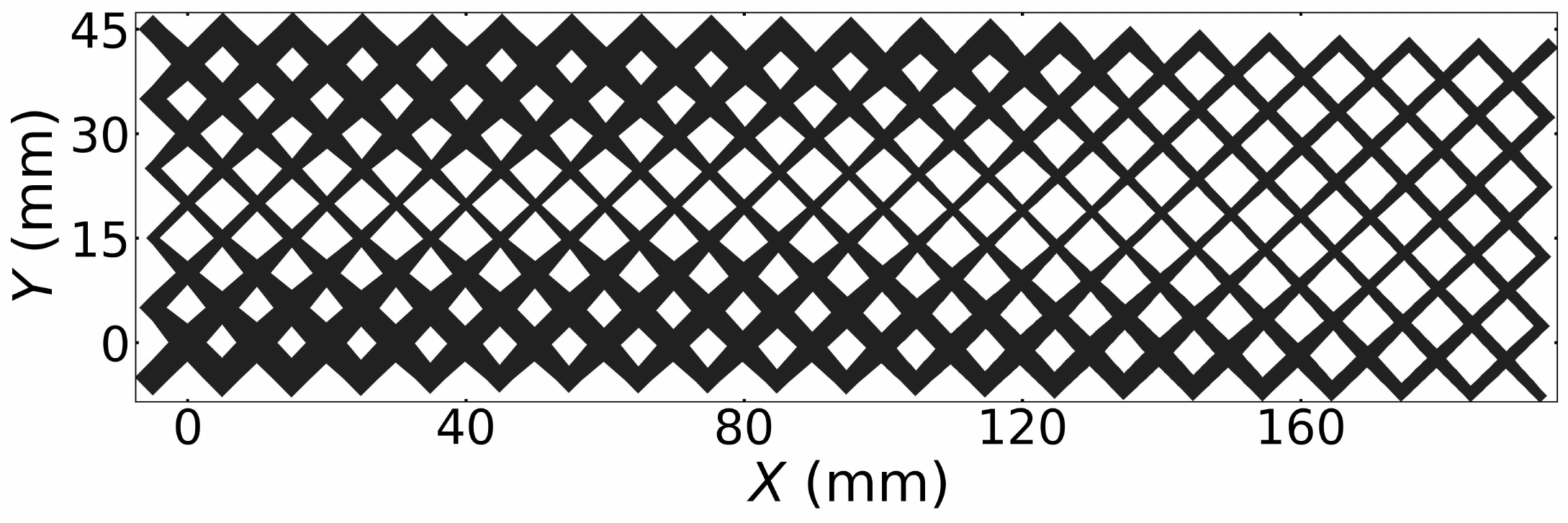}
        \caption{}
        \label{defl20_cb}
    \end{subfigure}
    \hfill
    \begin{subfigure}[b]{0.45\textwidth}
        \includegraphics[width=\linewidth]{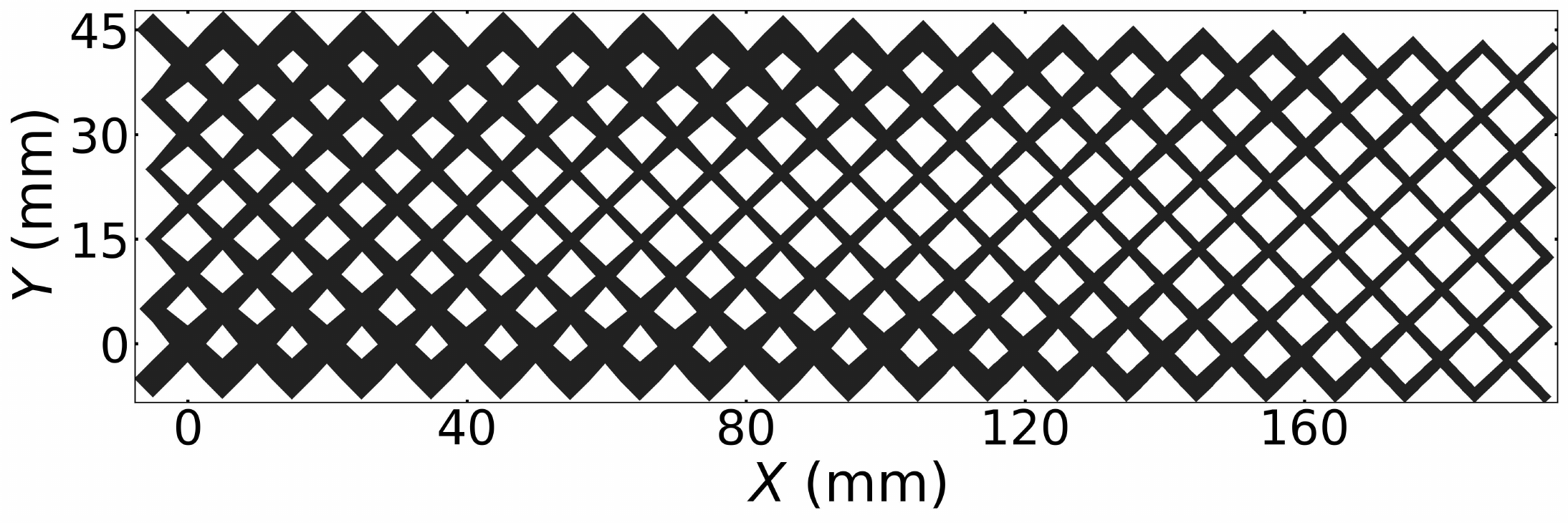}
        \caption{}
        \label{defl30_cb}
    \end{subfigure}

    \caption{Deformed configuration of the cantilever beam composed of the centered rectangular unit cells obtained by (a) conventional implementation, (b) GRF with length scale of 10 mm, (c) GRF with length scale of 20 mm, and (d) GRF with length scale of 30 mm.}
    \label{deformed_cb}
\end{figure}

The optimal profiles obtained by the conventional implementation and the GRF-based algorithm with different length-scales are shown in Fig.~\ref{cb}. The deformed configurations are shown in the Fig~\ref{deformed_cb}, and the evolution of the best profile with GA generation is given in Fig.~\ref{fitness_cb} (Appendix). The value of the maximum deflection for the conventional implementation is 3.17 mm, while for the profiles with the length scale parameters 10 mm, 20 mm, and 30 mm, the values are 2.58 mm, 2.70 mm, and 2.63 mm, respectively. The deflection values across all optimal profiles are similar for different length scales of GRF; however, the optimum profiles obtained from the conventional implementation are slightly worse off.  However, there is a prominent difference across all the profiles in the distribution of strut thickness within the domain. Profiles obtained from conventional implementation exhibit abrupt changes in strut thickness, whereas GRF-based profiles with length scales of 10 mm to 30 mm are smoother.

\begin{figure}[ht!]
    \centering
    \begin{subfigure}{0.32\textwidth}
    \centering
        \includegraphics[width=\textwidth]{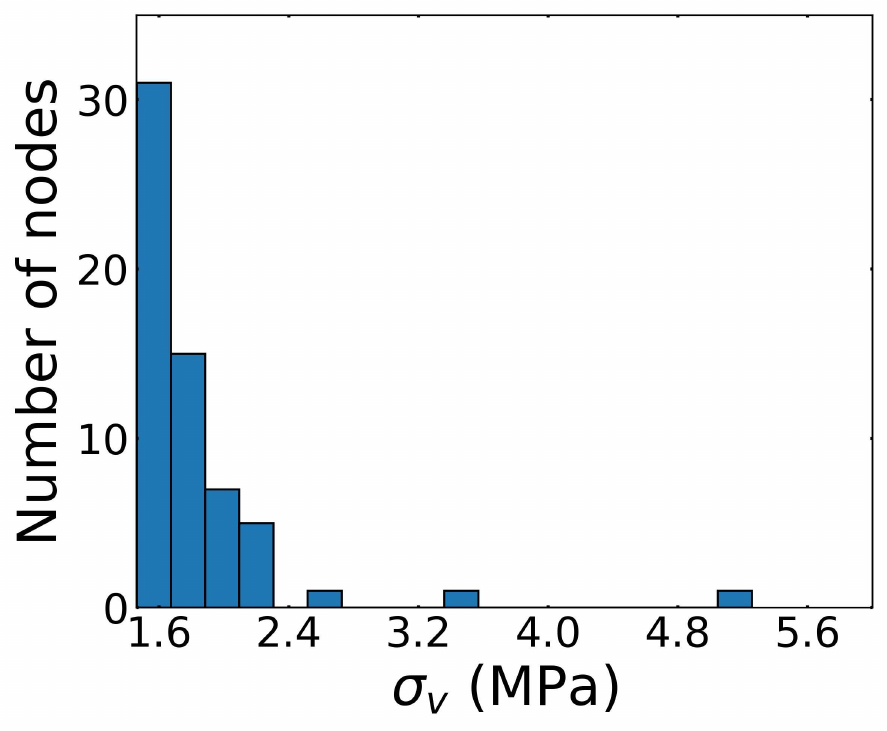}
        \caption{}
        \label{histrandom_cb}
    \end{subfigure}
    \hfill
    \begin{subfigure}{0.32\textwidth}
    \centering
        \includegraphics[width=\textwidth]{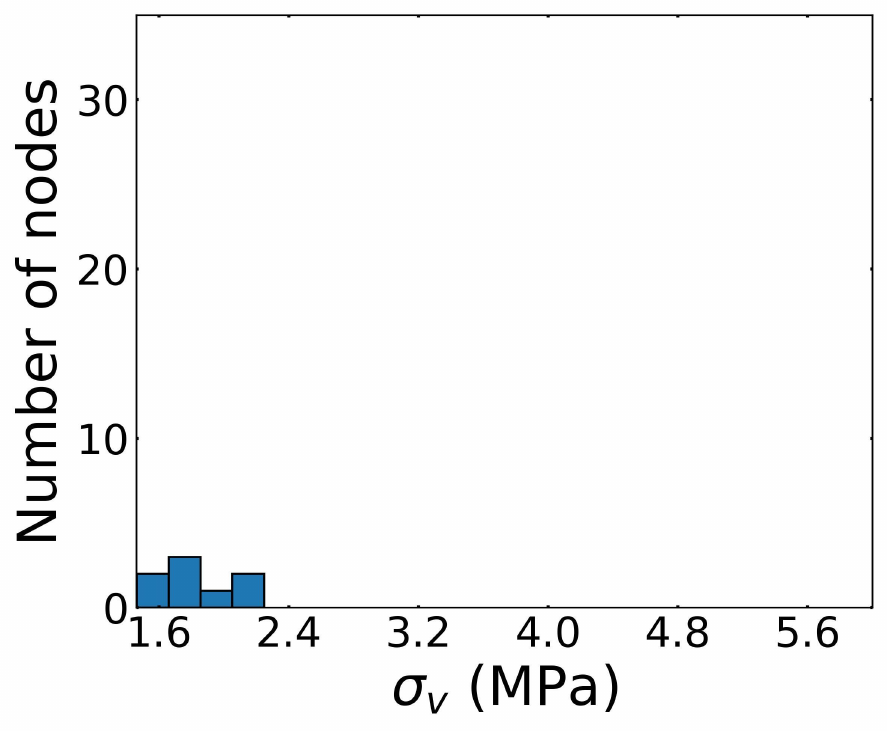}
        \caption{}
        \label{histl20_cb}
    \end{subfigure}
   \hfill
    \begin{subfigure}{0.32\textwidth}
    \centering
        \includegraphics[width=\textwidth]{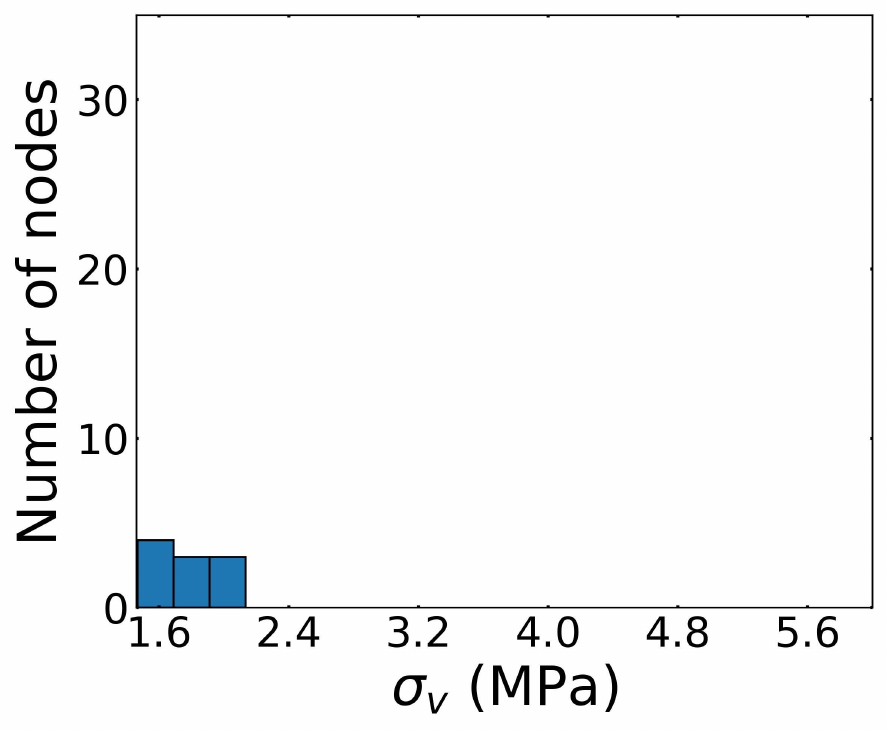}
        \caption{}
        \label{histl30_cb}
    \end{subfigure}

    \caption{Histogram of the number of  nodes having $\sigma_v$ $\geq$ 1.46 MPa (Note that the reference value of $\sigma_v$ is chosen as the $99.5^{th}$ percentile of the optimal design, considering conventional implementation) for the optimal structure obtained by (a) conventional implementation, (b) GRF with length scale of 20 mm, and (c) GRF with length scale of 30 mm.}
    \label{hist_cb}
\end{figure}

The intensity of stress concentration among optimum profiles can be observed from the $\sigma_v$  histogram shown in Fig.~\ref{hist_cb}. The histogram shows the number of nodes where the $\sigma_v\geq \sigma^*$, where \(\sigma^*\) is the $99.5^{th}$ percentile of the conventional implementation.  Additionally, the maximum value of $\sigma_v$ is 5.25 MPa for the conventional implementation's and reduces to 2.06 MPa, 2.24 MPa, and 2.13 MPa for the GRF-based profiles with length scales of 10 mm, 20 mm, and 30 mm, respectively. Note, towards the prediction of accurate stress values, we have taken a finer finite element mesh compared to the one used for the prediction of the displacement field. In particular,  we are using 6,226 4-node quadrilateral elements for stress field prediction.

\subsubsection{Case 2: Design of a stiff Cantilever beam with leftmost unit cells thickness constrained}
\label{cantilver_beam_gpr}

In this example, we consider the setting identical to the previous case \ref{cantilver_beam_section}, with the difference that the thickness of the left-most unit cells is constrained to 4.0 mm ($t_{max}$). Unlike previous cases, where we have deployed GRF for profile generation, here, since boundary values for the design variables are constrained, GPR has been used for profile generation. The optimization problem is stated as follows:

\begin{equation}
\begin{aligned}
&\textbf{minimize:} && \quad \delta_{y}(t),  \\
&\textbf{subject to:} && \quad t_{\mathrm{min}} \le  t_k \le t_{\mathrm{max}},\; t_k \in T, \\
&&&\quad \frac{1}{|T|}
\sum_{k \in T} t_k \le \bar{t}_{\max}.
\end{aligned}
\end{equation}

\begin{figure}[ht!]
    \centering
    \begin{subfigure}[b]{0.45\textwidth}
        \includegraphics[width=\linewidth]{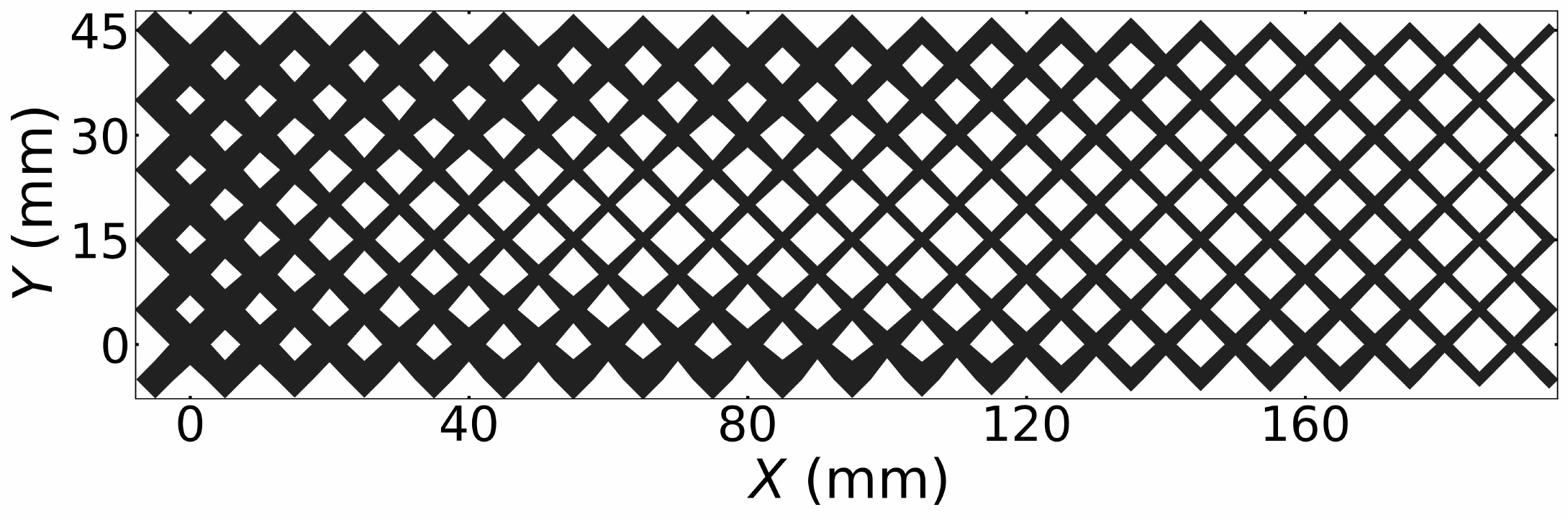}
        \caption{}
        \label{}
    \end{subfigure}
    \hfill
    \begin{subfigure}[b]{0.45\textwidth}
        \includegraphics[width=\linewidth]{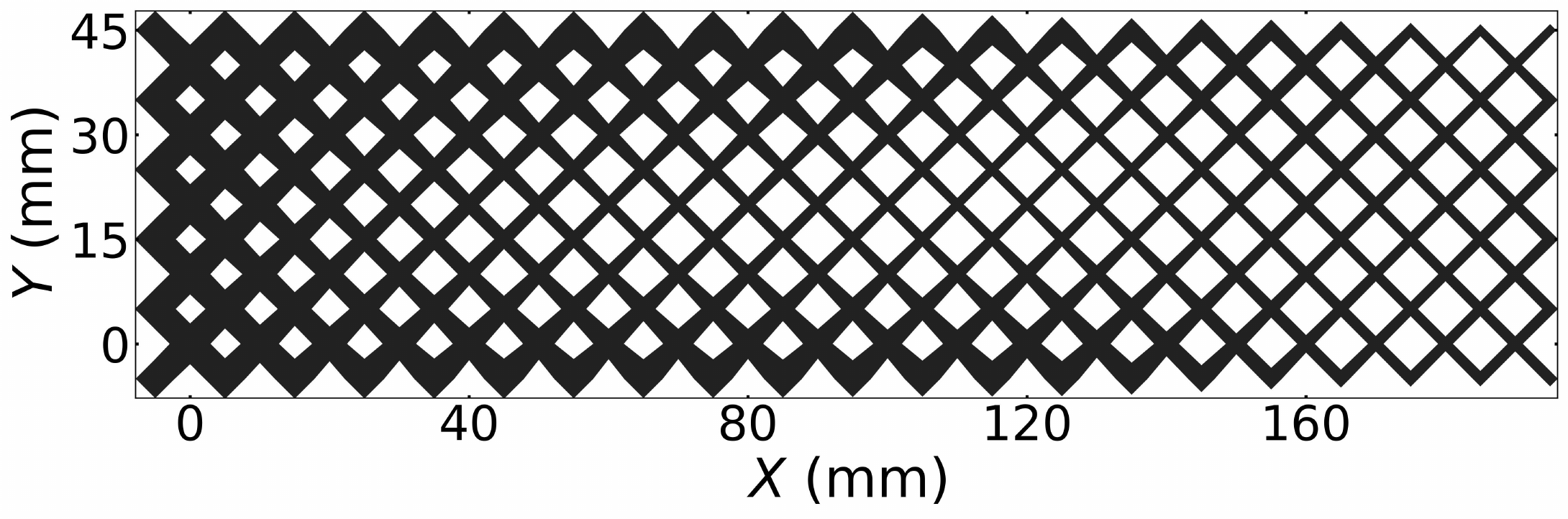}
        \caption{}
        \label{}
    \end{subfigure}
    \caption{Optimized profiles of the cantilever beam under the design constraint of maximum thickness at the leftmost unit cells, generated by (a) GPR with length scale of 20 mm and (b) GPR with length scale of 30 mm.}
    \label{cb_gpr}
\end{figure}

\begin{figure}[ht!]
    \centering
    \begin{subfigure}[b]{0.45\textwidth}
        \includegraphics[width=\linewidth]{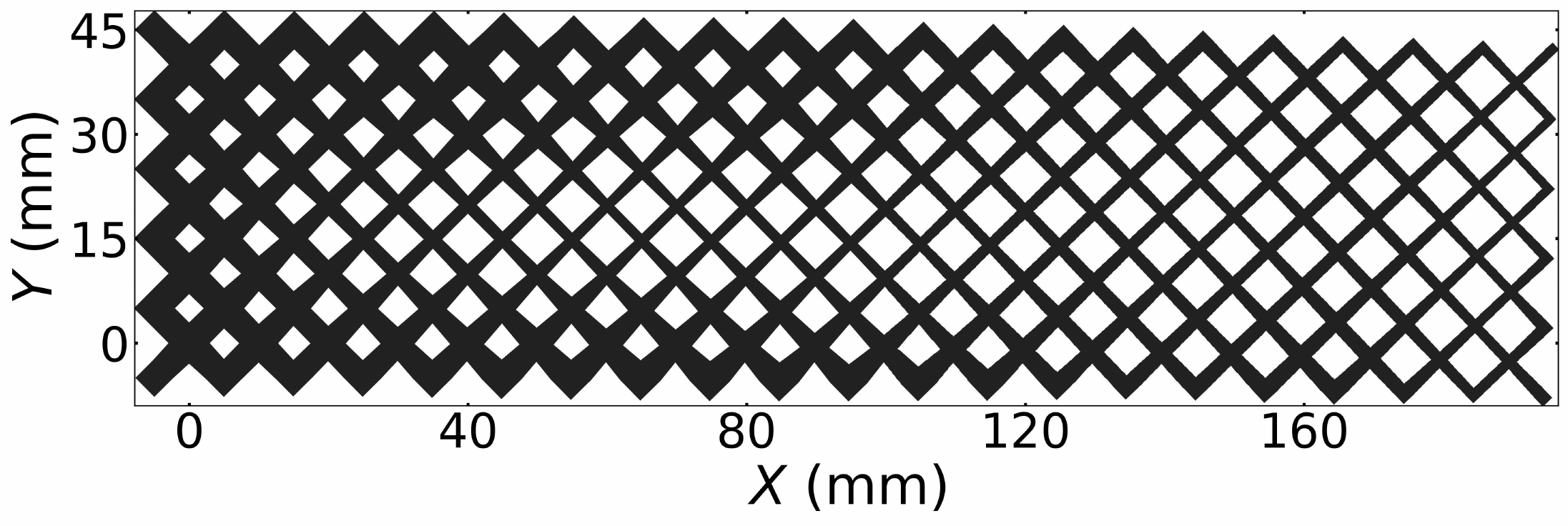}
        \caption{}
        \label{}
    \end{subfigure}
    \hfill
    \begin{subfigure}[b]{0.45\textwidth}
        \includegraphics[width=\linewidth]{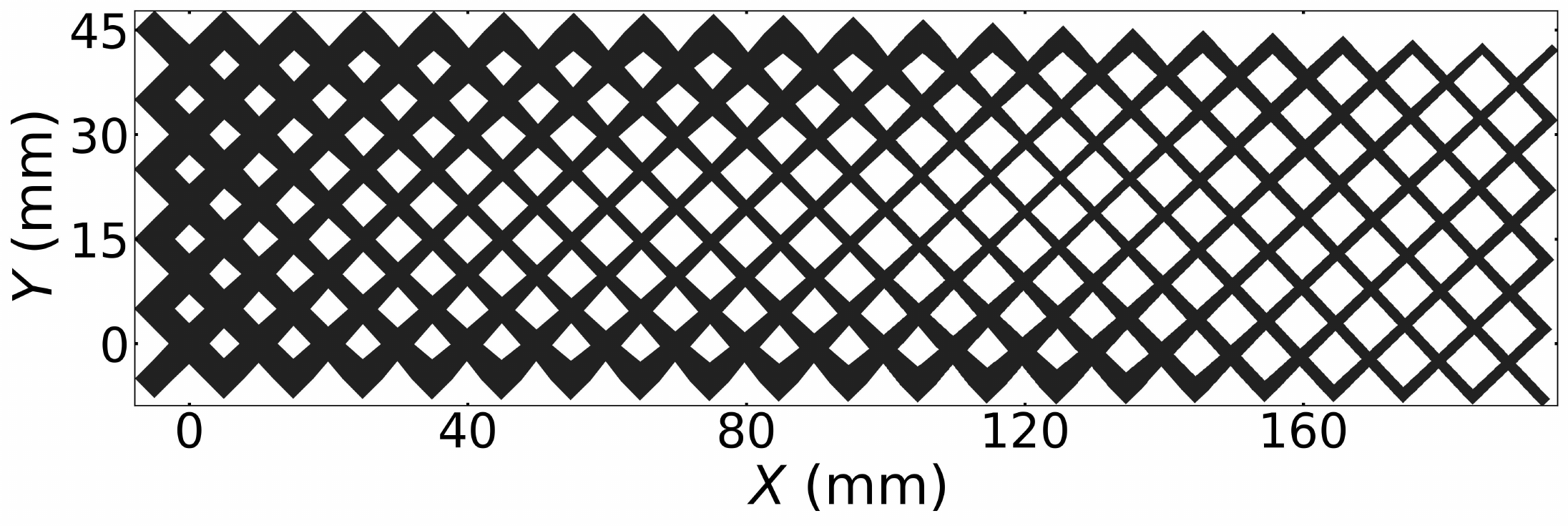}
        \caption{}
        \label{}
    \end{subfigure}

    \caption{Deformed configuration of the cantilever beam under the constraint of maximum thickness at the leftmost unit cells, obtained by (a) GPR with length scale of 20 mm and (d) GPR with length scale of 30 mm.}
    \label{deformed_cb_gpr}
\end{figure}

For optimization, the initial design space is generated using GPR with two different length scales: 20 mm and 30 mm, while the standard deviation is taken as 1.0 mm. The parameters of GA remain same, as given in the Table \ref{GA_Para1}. The optimal profiles obtained by the GPR-based algorithm, subjected to the design constraint, are demonstrated in Fig.~\ref{cb_gpr}. While the deformed configurations are shown in Fig~\ref{deformed_cb_gpr}. The evolution of the best profile with GA generation is given in Fig.~\ref{fitness_cb_gpr} (Appendix). The maximum displacement values are 2.93 mm and 2.99 mm for the optimal profile with length scales of 20 mm and 30 mm. 

For the comparison between the GPR-based optimal profile and the GRF-based optimal profile. The displacement values are slightly inferior in the case of the GPR-based optimal profile compared to the optimum profiles obtained in the previous case. This is expected, since in the present case, we are imposing the additional constraint on the design space. However, these additional constraints on the beam's design help reduce the stress in the optimal profiles. The maximum $\sigma_v$ in the GPR-based profile is 1.78 MPa and 1.75 MPa for the length scales 20 mm and 30 mm, respectively. While in the previous case with the GRF-based profile, the corresponding numbers are 2.24 MPa and 2.13 MPa.

\subsubsection{Case 3: Minimize the deflection of the half MBB beam}
\label{half_mbb_beam}

\begin{figure}[ht!]
    \centering
    \includegraphics[width=0.6\textwidth]{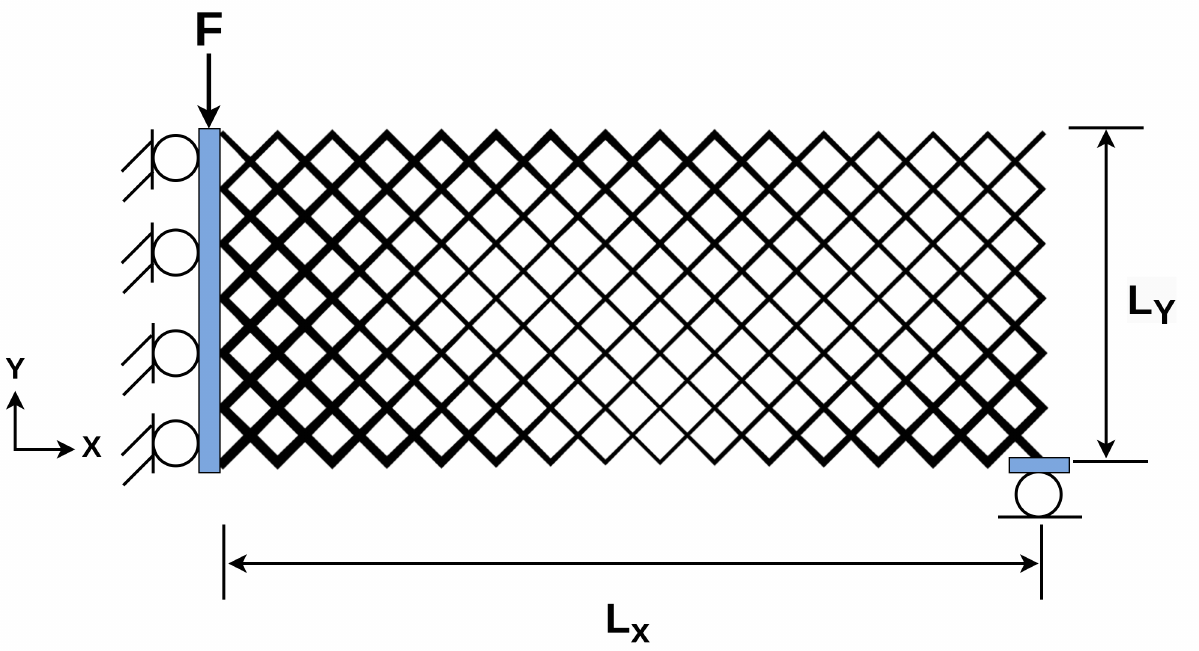}
    \caption{Schematic of the half MBB beam subjected to the point load.}
    \label{MBB_diagram}
\end{figure}

In this problem, we consider the symmetric half of an MBB beam configuration. The left edge of the beam is constrained by a roller-supported boundary condition, and a vertical load (F) of 5.0 KN is applied at the top-left corner. Additionally, the bottom-right corner is subjected to a roller-supported boundary condition, as depicted in Fig. \ref{MBB_diagram}.  The design domain is discretized into 15 unit cells along the X-axis and 6 unit cells along the Y-axis. The dimension of each unit cell is 10 mm along the width and the height.

Our objective is to minimize the maximum deflection ($\delta_y$) of the beam along the Y-axis, while the design variables and constraints are similar to the previous problem (case~\ref{cantilver_beam_section}). The optimization problem is stated as follows:

\begin{equation}
\begin{aligned}
&\textbf{minimize:} && \quad \delta_{y}(t),  \\
&\textbf{subject to:} && \quad t_{\mathrm{min}} \le  t_k \le t_{\mathrm{max}},\; t_k \in T, \\
&&&\quad \frac{1}{|T|}
\sum_{k \in T} t_k \le \bar{t}_{\max},
\end{aligned}
\end{equation}

\begin{figure}[ht!]
    \centering
    \begin{subfigure}[b]{0.45\textwidth}
        \includegraphics[width=\linewidth]{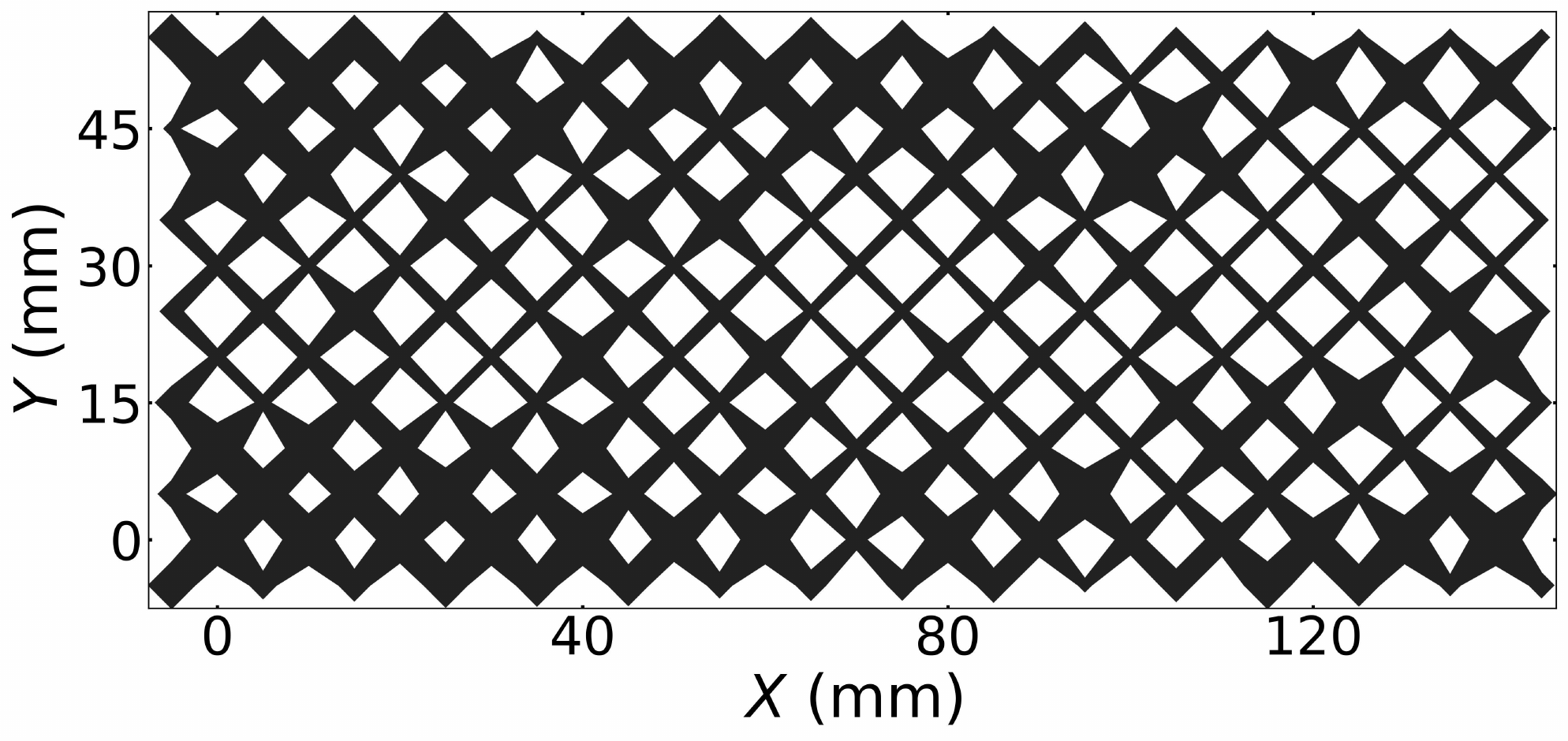}
        \caption{}
        \label{random_mbb}
    \end{subfigure}
    \hfill
    \begin{subfigure}[b]{0.45\textwidth}
        \includegraphics[width=\linewidth]{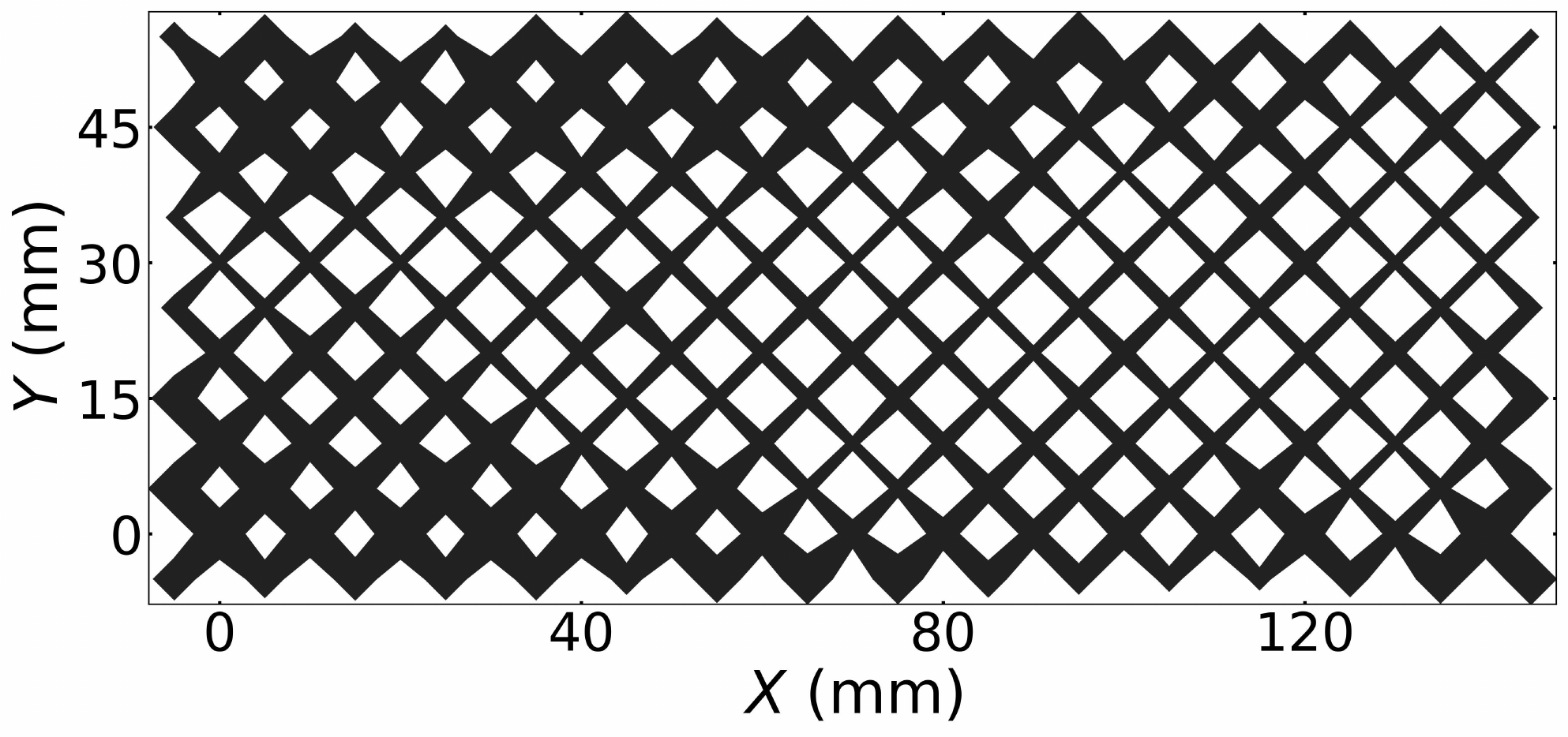}
        \caption{}
        \label{l10_mbb}
    \end{subfigure}

    \vspace{0.5cm}

    \begin{subfigure}[b]{0.45\textwidth}
        \includegraphics[width=\linewidth]{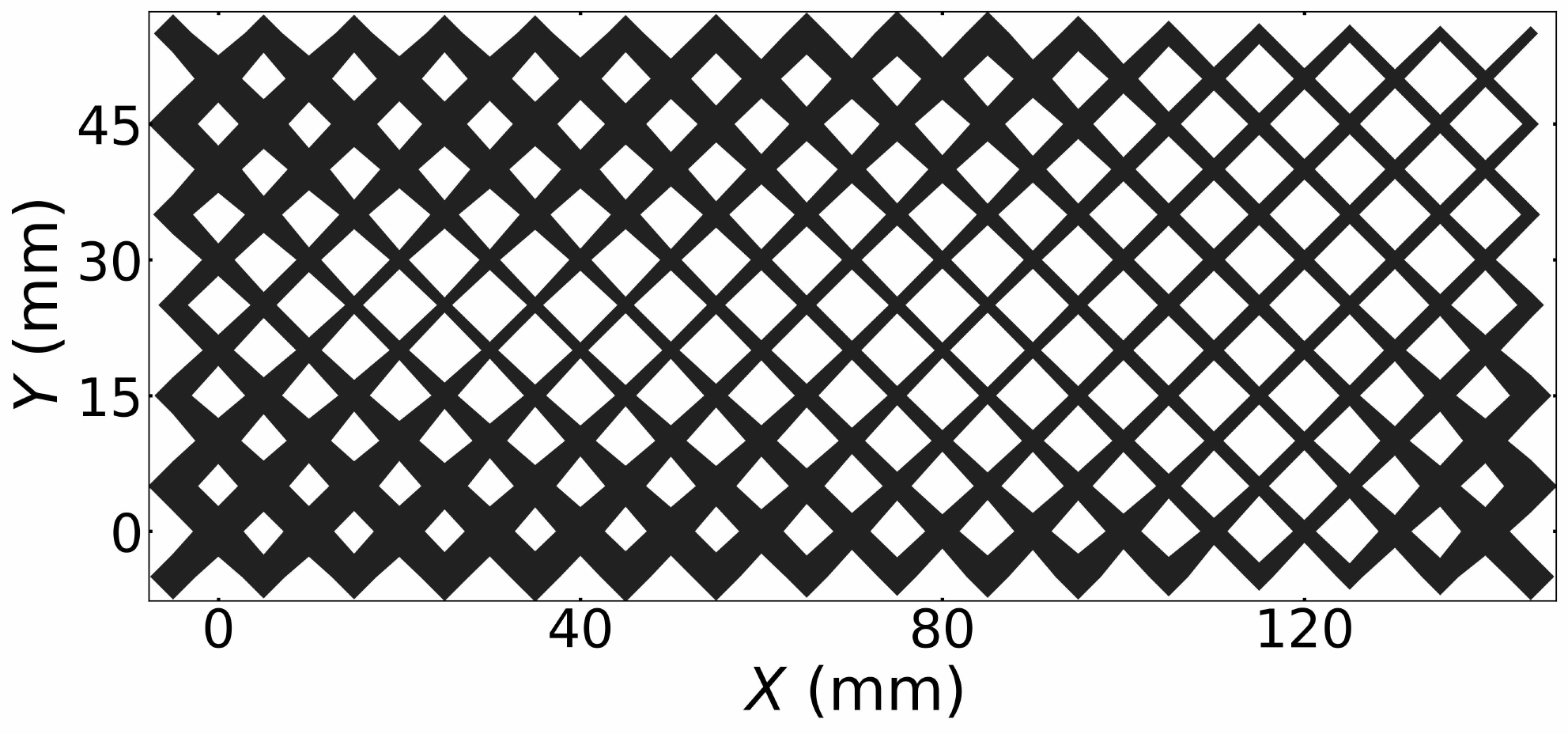}
        \caption{}
        \label{l20_mbb}
    \end{subfigure}
    \hfill
    \begin{subfigure}[b]{0.45\textwidth}
        \includegraphics[width=\linewidth]{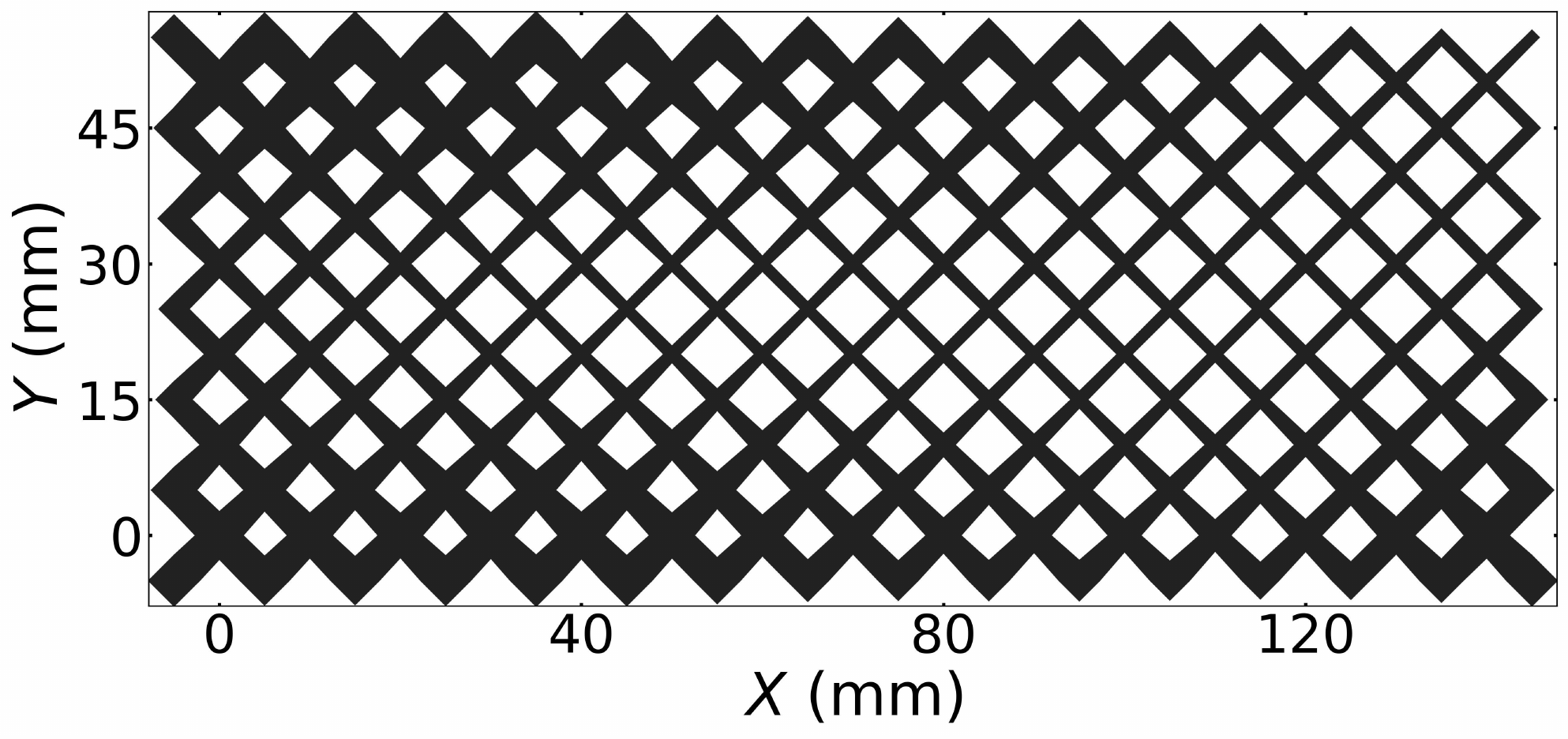}
        \caption{}
        \label{l30_mbb}
    \end{subfigure}

    \caption{Optimized profiles of the half-MBB beam composed of the centered rectangular unit cells generated by (a) conventional implementation, (b) GRF with length scale of 10 mm, (c) GRF with length scale of 20 mm, and (d) GRF with length scale of 30 mm.}
    \label{mbb_case2}
\end{figure}

\begin{figure}[ht!]
    \centering
    \begin{subfigure}[b]{0.45\textwidth}
        \includegraphics[width=\linewidth]{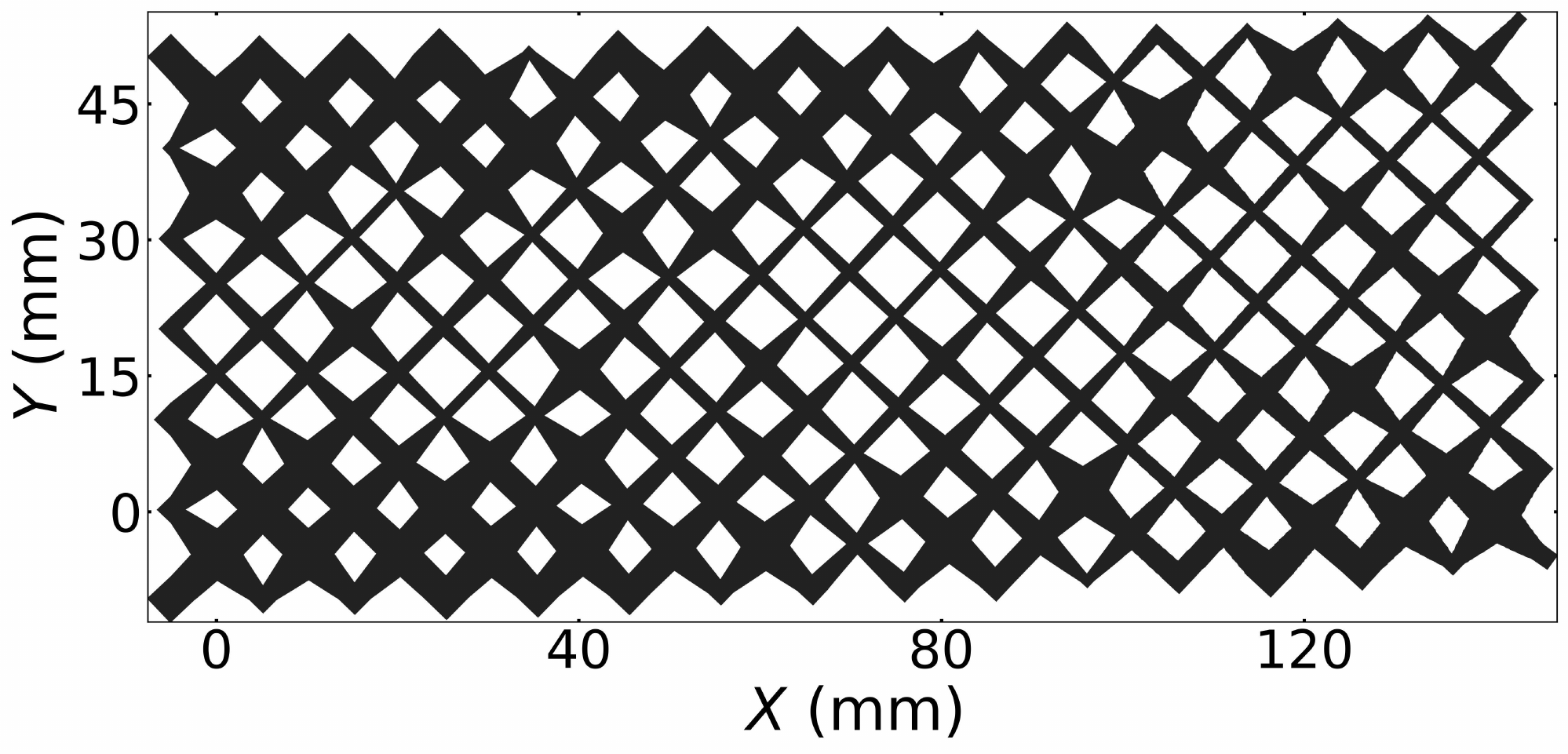}
        \caption{}
        \label{defrandom_mbb}
    \end{subfigure}
    \hfill
    \begin{subfigure}[b]{0.45\textwidth}
        \includegraphics[width=\linewidth]{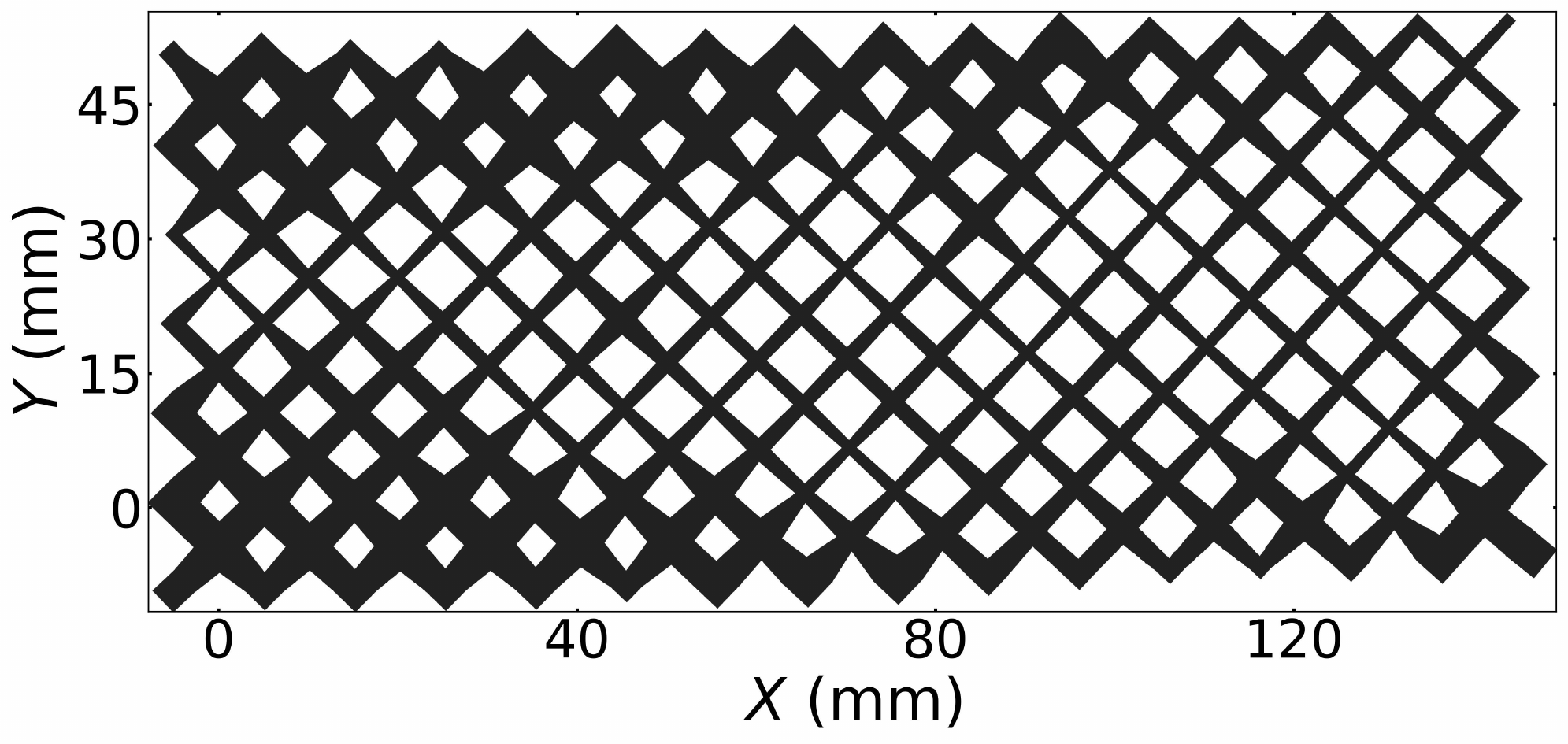}
        \caption{}
        \label{defl20_mbb}
    \end{subfigure}

    \vspace{0.5cm}

    \begin{subfigure}[b]{0.45\textwidth}
        \includegraphics[width=\linewidth]{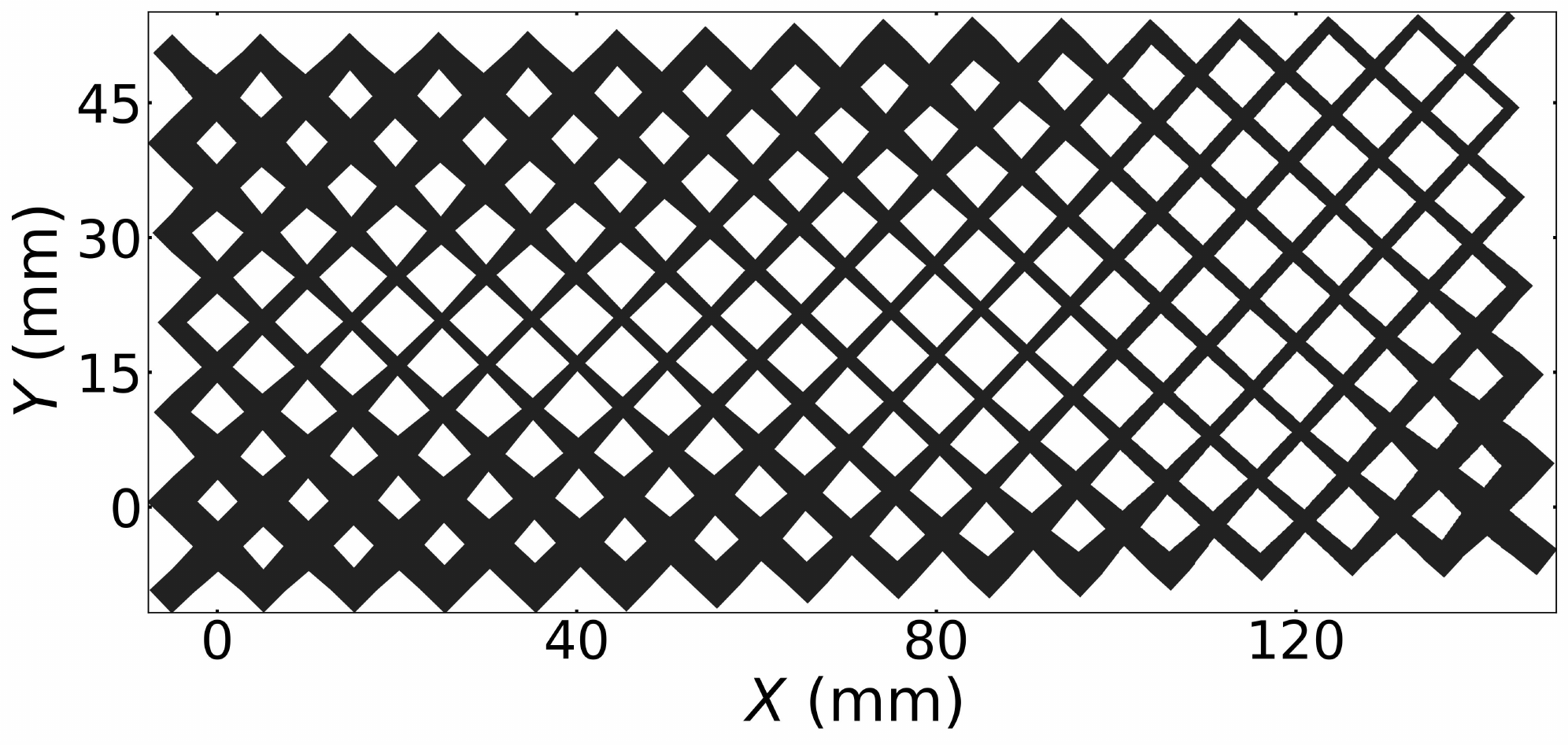}
        \caption{}
        \label{l20_mbb}
    \end{subfigure}
    \hfill
    \begin{subfigure}[b]{0.45\textwidth}
        \includegraphics[width=\linewidth]{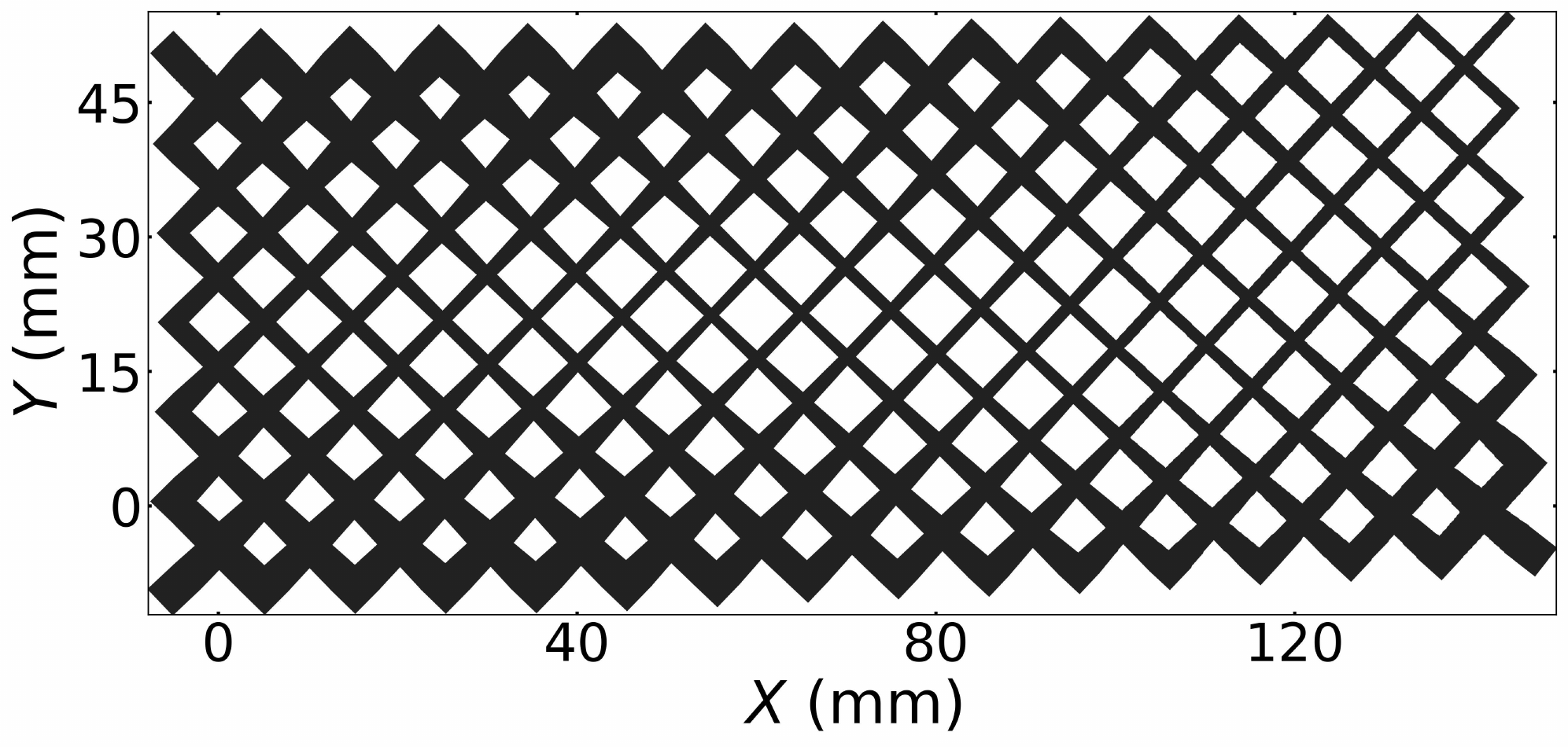}
        \caption{}
        \label{l30_mbb}
    \end{subfigure}

    \caption{Deformed configuration of the half-MBB beam composed of the centered rectangular unit cells obtained by (a) conventional implementation, (b) GRF with length scale of 10 mm, (c) GRF with length scale of 20 mm, and (d) GRF with length scale of 30 mm.}
    \label{deformed_mbb}
\end{figure}

\begin{figure}[ht!]
    \centering
    \begin{subfigure}{0.32\textwidth}
        \centering
        \includegraphics[width=\textwidth]{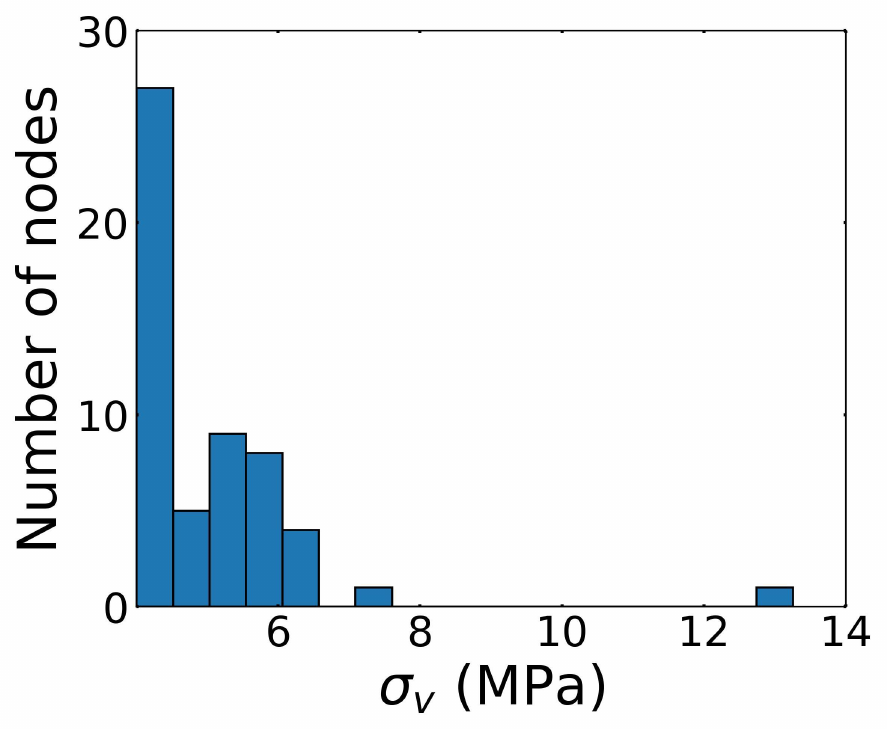}
        \caption{}
        \label{histrandom_mbb}
    \end{subfigure}
    \hfill
    \begin{subfigure}{0.32\textwidth}
        \centering
        \includegraphics[width=\textwidth]{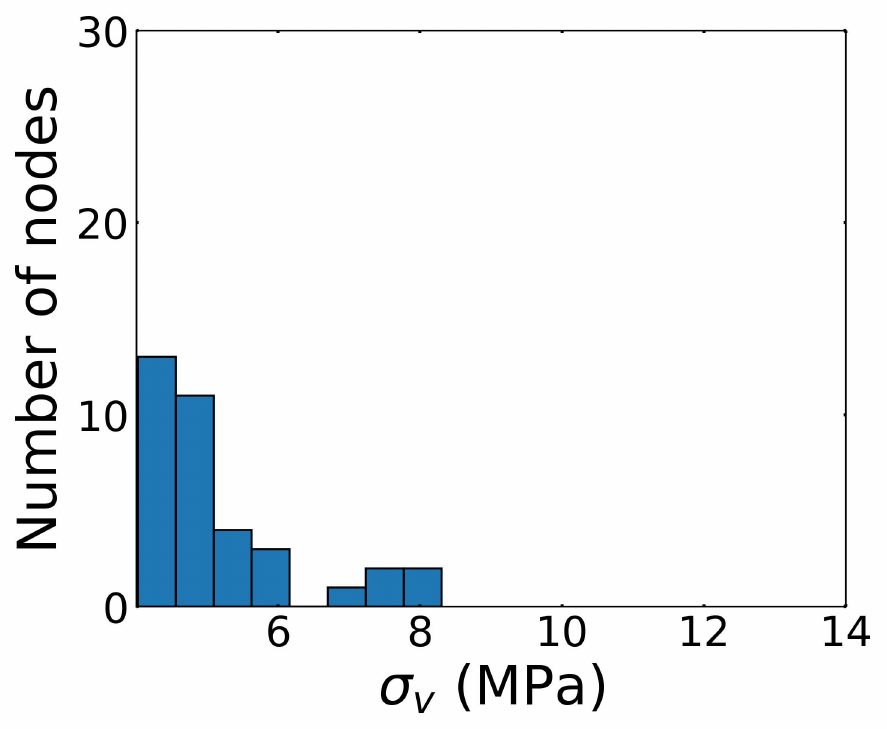}
        \caption{}
        \label{histl20_mbb}
    \end{subfigure}
    \hfill
    \begin{subfigure}{0.32\textwidth}
        \centering
        \includegraphics[width=\textwidth]{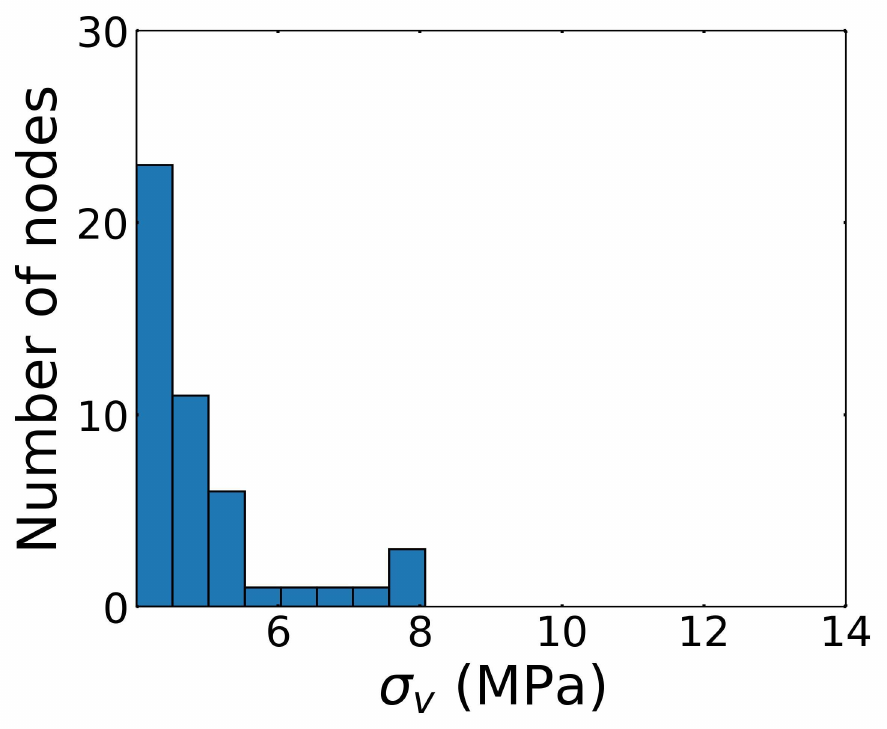}
        \caption{}
        \label{histl30_mbb}
    \end{subfigure}

    \caption{Histogram of the number of  nodes having $\sigma_v$ $\geq$ 4.1 MPa (Note that the reference value of $\sigma_v$ is chosen as the $99.5^{th}$ percentile of the conventional implementation) for the optimal structure obtained by (a) conventional implementation, (b) GRF with length scale of 20 mm, and (c) GRF with length scale of 30 mm.}
    \label{hist_mbb}
\end{figure}

In optimization iterations to calculate the fitness value, we have  used 2,722 4-noded quadrilateral elements for each design. The GA and GRF parameters remain the same as in case 1. The optimal profiles obtained for the half-MBB beam problem are shown in Fig.~\ref{mbb_case2}, and the corresponding deformed configurations are shown in Fig.~\ref{deformed_mbb}. The evolution of the best profile over GA generations is shown in Fig.~\ref{fitness_mbb} (Appendix). The maximum deflection for the conventional implementation profile is 4.9 mm, while for the GRF-based optimum profiles it is  4.58 mm, 4.53 mm, and 4.64 mm with length scales of 10 mm, 20 mm, and 30 mm, respectively. As observed in previous cases, the value of the objective function is in a similar range in all the cases, while GRF provides smoother designs compared to the conventional implementation. The histogram for the $\sigma_v$ $\geq$ $\sigma^*$ is demonstrated in the Fig~\ref{hist_mbb}. Here $\sigma^*$ is the $99.5^{th}$ percentile value of the $\sigma_v$ in the conventional implementation optimal profile. The maximum value of $\sigma_v$ in the conventional implementation optimal profile is 13.25 MPa, and for the optimal profile obtained by the GRF-based scheme is 7.9 MPa, 8.3 MPa, and 8.03 MPa for the length scales 10 mm, 20 mm, and 30 mm, respectively. Note: to obtain the stress field, the finite element analyses with  5,602 elements have been carried out.

\subsubsection{Case 4: Maximize the deflection of the cantilever beam subjected to the thermal loading}

In this problem, we take a 2D cantilever beam subjected to thermal loading. The temperature of the beam is uniformly changed from  0.0\,$^{\circ}\mathrm{C}$ to 500\,$^{\circ}\mathrm{C}$. Material properties are: elastic modulus 400 MPa, Poisson's ratio equals 0.3, and thermal expansion coefficient is $100 \times 10^{-6}\,^\circ\mathrm{C}^{-1}$. The geometry, boundary conditions, and finite element mesh of the beam are similar to the case \ref{cantilver_beam_section}.

Our objective is to maximize the maximum deflection ($\delta_y$) of the cantilever beam, under the constraint of mean and individual strut thickness. The mean thickness of the structure is bounded by $\bar{t}_{\min}$ =  2.25 mm and $\bar{t}_{\max}$ =  2.75 mm, while the thickness of each strut is restricted to the $t_{\min}$ = 1.0 mm and $t_{\max}$ = 4.0 mm. Our optimization objective is stated as follows:

\begin{equation}
\begin{aligned}
&\textbf{maximize:} && \quad \delta_{y}(t),  \\
&\textbf{subject to:} && \quad t_{\mathrm{min}} \le  t_k \le t_{\mathrm{max}},\; t_k \in T, \\
&&&\quad  \bar{t}_{\min} \le \frac{1}{|T|}
\sum_{k \in T} t_k \le \bar{t}_{\max}.
\end{aligned}
\end{equation}

\begin{figure}[ht!]
    \centering
    \begin{subfigure}[b]{0.45\textwidth}
        \includegraphics[width=\linewidth]{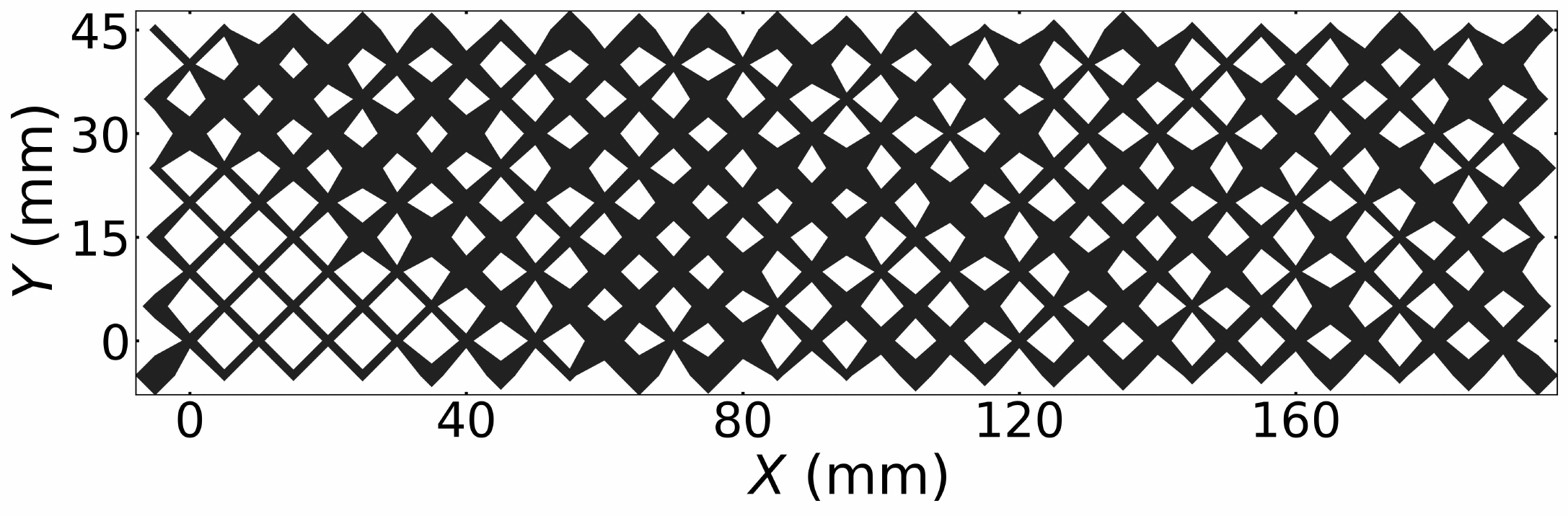}
        \caption{}
        \label{}
    \end{subfigure}
    \hfill
    \begin{subfigure}[b]{0.45\textwidth}
        \includegraphics[width=\linewidth]{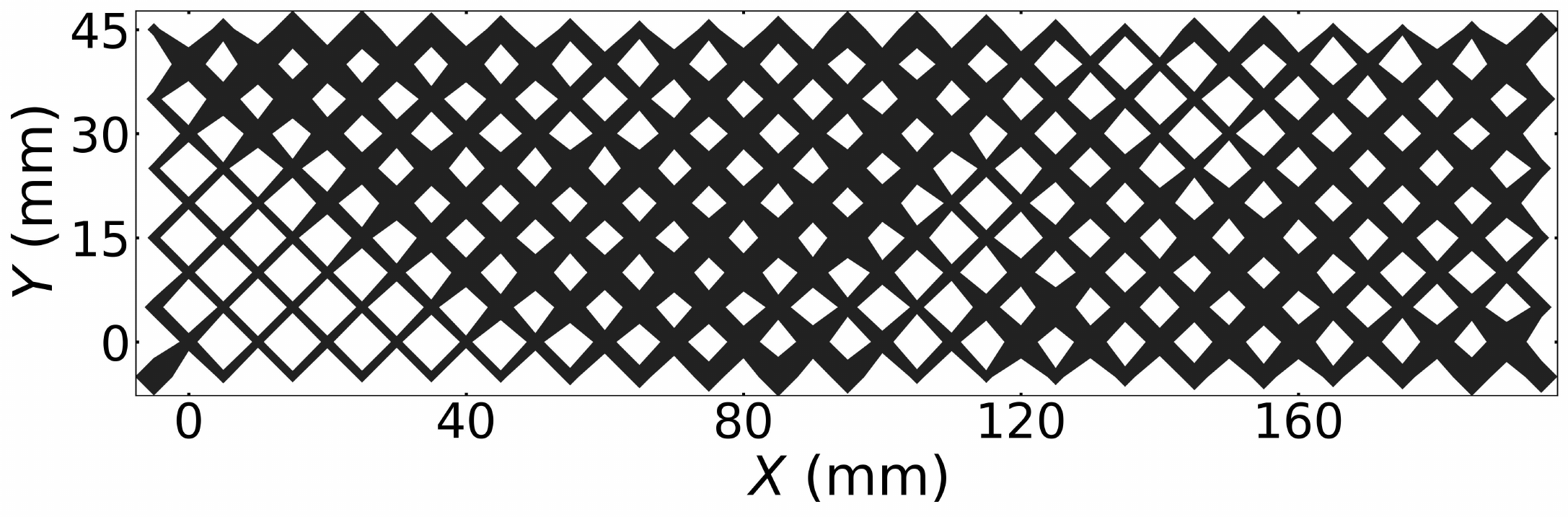}
        \caption{}
        \label{}
    \end{subfigure}

    \vspace{0.5cm}

    \begin{subfigure}[b]{0.45\textwidth}
        \includegraphics[width=\linewidth]{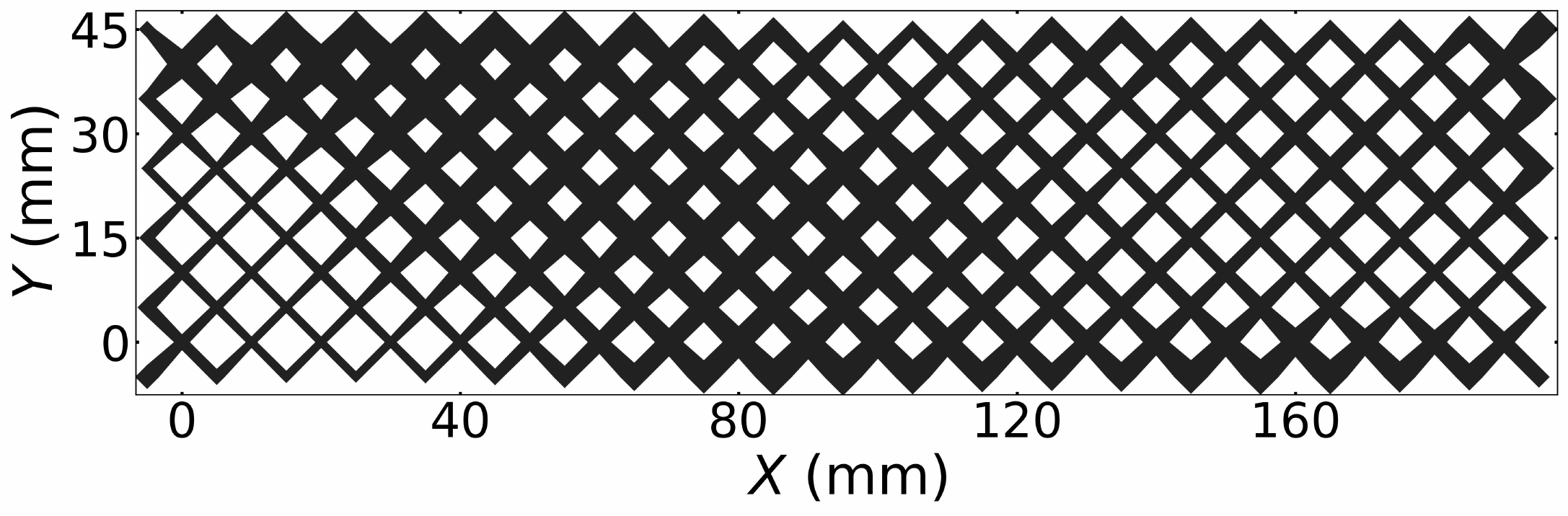}
        \caption{}
        \label{}
    \end{subfigure}
    \hfill
    \begin{subfigure}[b]{0.45\textwidth}
        \includegraphics[width=\linewidth]{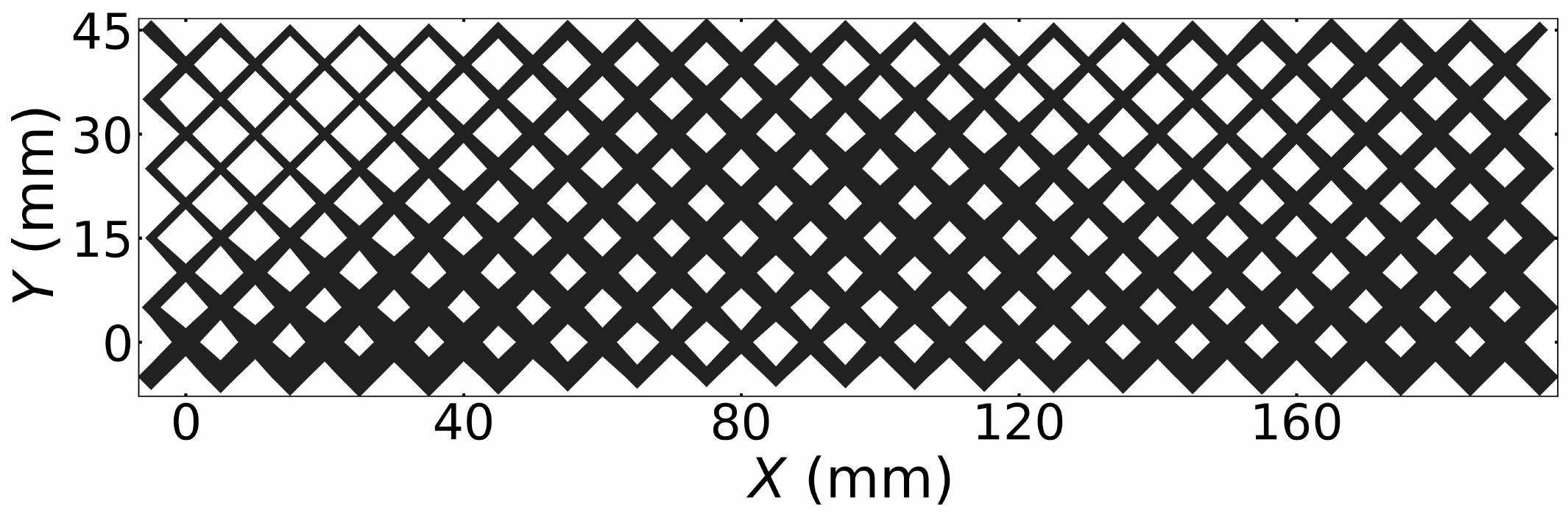}
        \caption{}
        \label{}
    \end{subfigure}

    \caption{Optimized profiles of the cantilever beam under thermal loading, composed of the centered rectangular unit cells generated by (a) conventional implementation, (b) GRF with length scale of 10 mm, (c) GRF with length scale of 20 mm, and (d) GRF with length scale of 30 mm.}
    \label{cb_thermal}
\end{figure}

\begin{figure}[ht!]
    \centering
    \begin{subfigure}[b]{0.45\textwidth}
        \includegraphics[width=\linewidth]{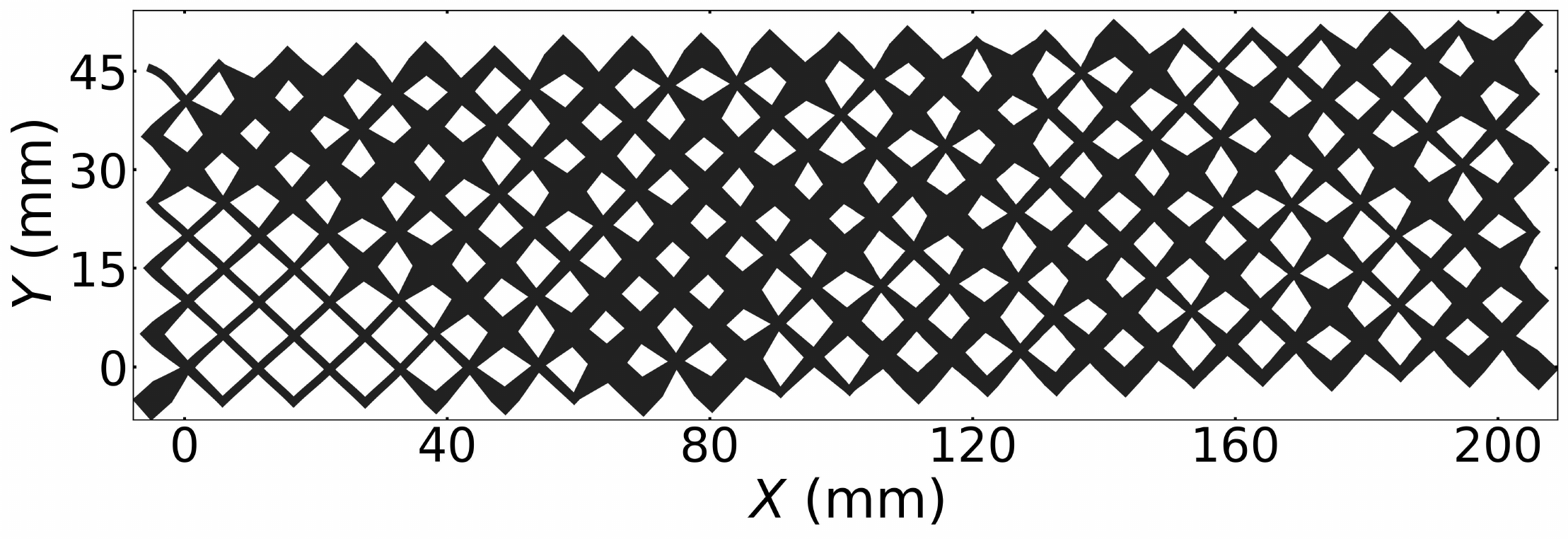}
        \caption{}
        \label{}
    \end{subfigure}
    \hfill
    \begin{subfigure}[b]{0.45\textwidth}
        \includegraphics[width=\linewidth]{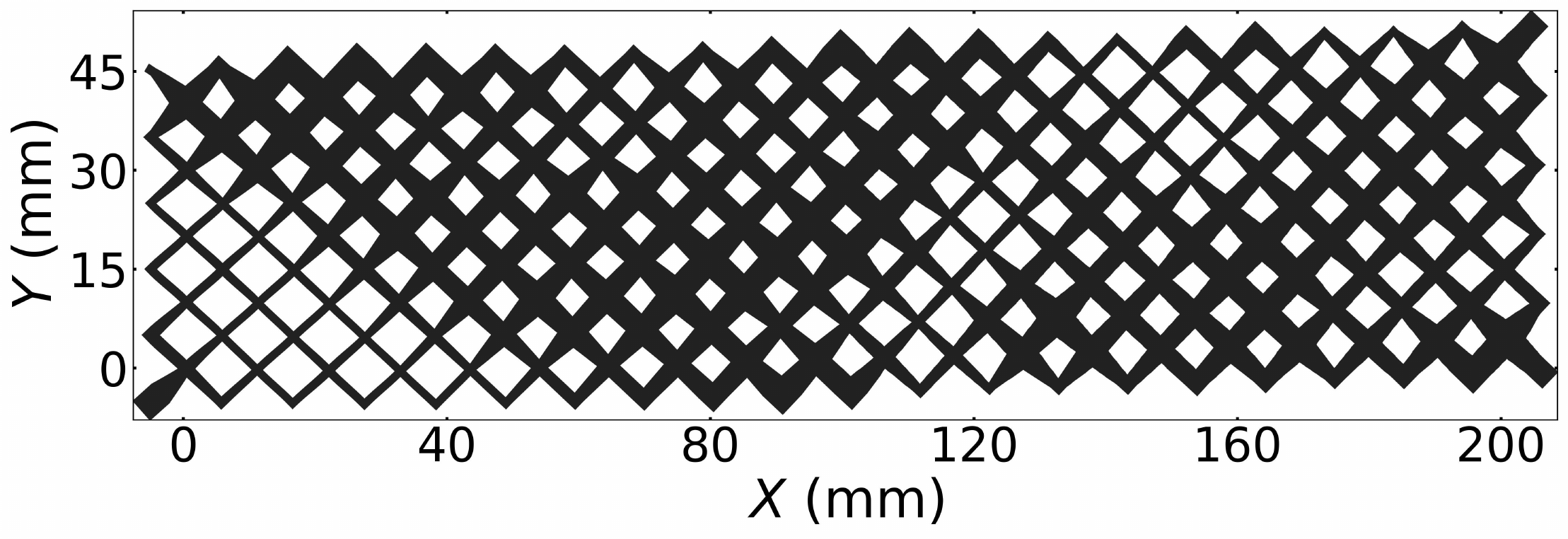}
        \caption{}
        \label{}
    \end{subfigure}

    \vspace{0.5cm}

    \begin{subfigure}[b]{0.45\textwidth}
        \includegraphics[width=\linewidth]{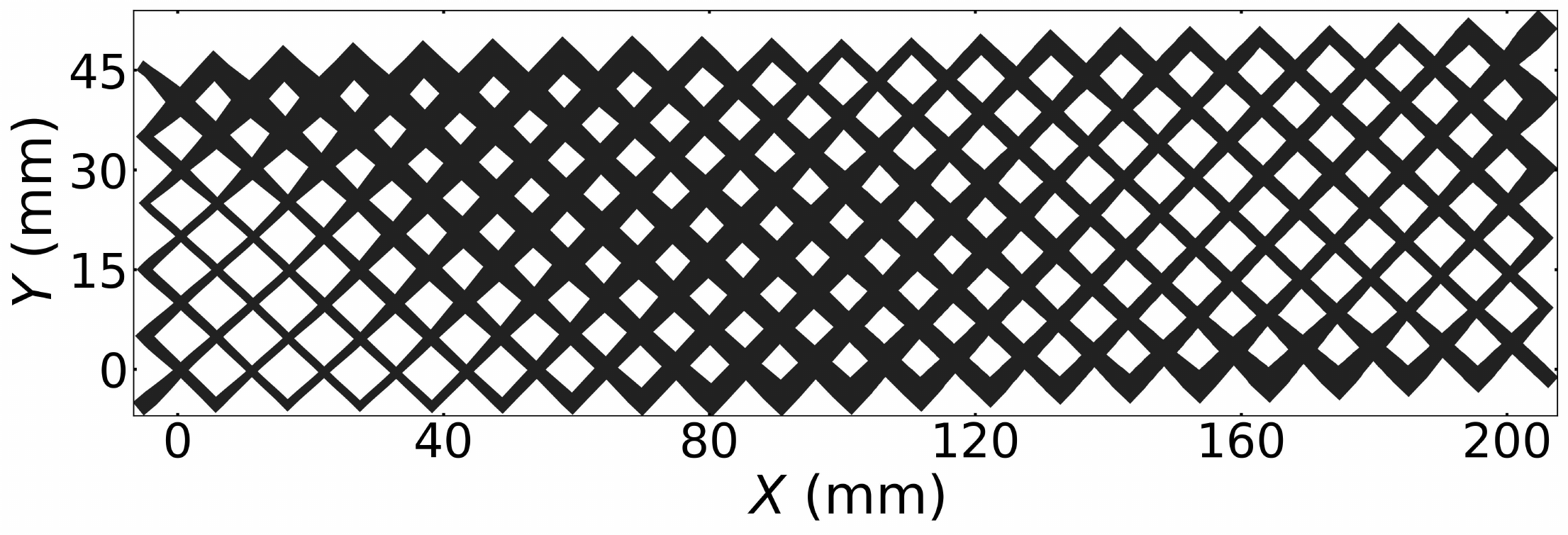}
        \caption{}
        \label{}
    \end{subfigure}
    \hfill
    \begin{subfigure}[b]{0.45\textwidth}
        \includegraphics[width=\linewidth]{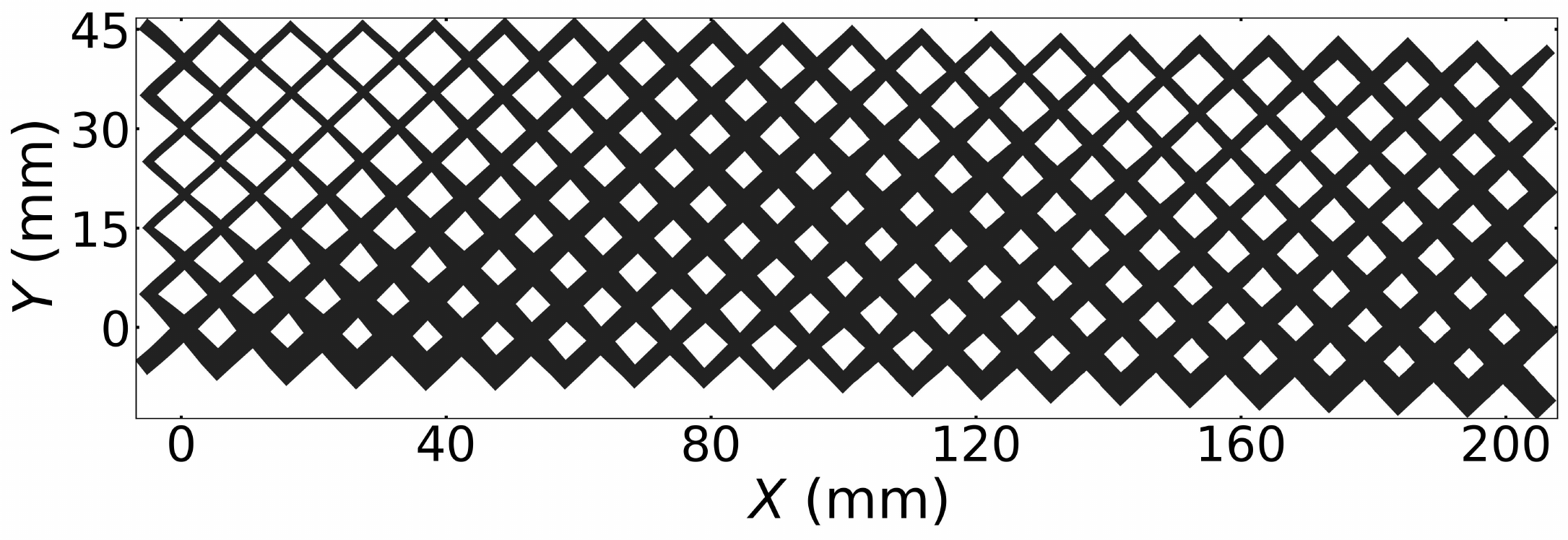}
        \caption{}
        \label{}
    \end{subfigure}

    \caption{Deformed configuration of the cantilever beam under thermal loading, composed of the centered rectangular unit cells obtained by (a) conventional implementation, (b) GRF with length scale of 10 mm, (c) GRF with length scale of 20 mm, and (d) GRF with length scale of 30 mm.}
    \label{deformed_cb_thermal}
\end{figure}

\begin{figure}[ht!]
    \centering
    \begin{subfigure}{0.32\textwidth}
     \centering
        \includegraphics[width=\linewidth]{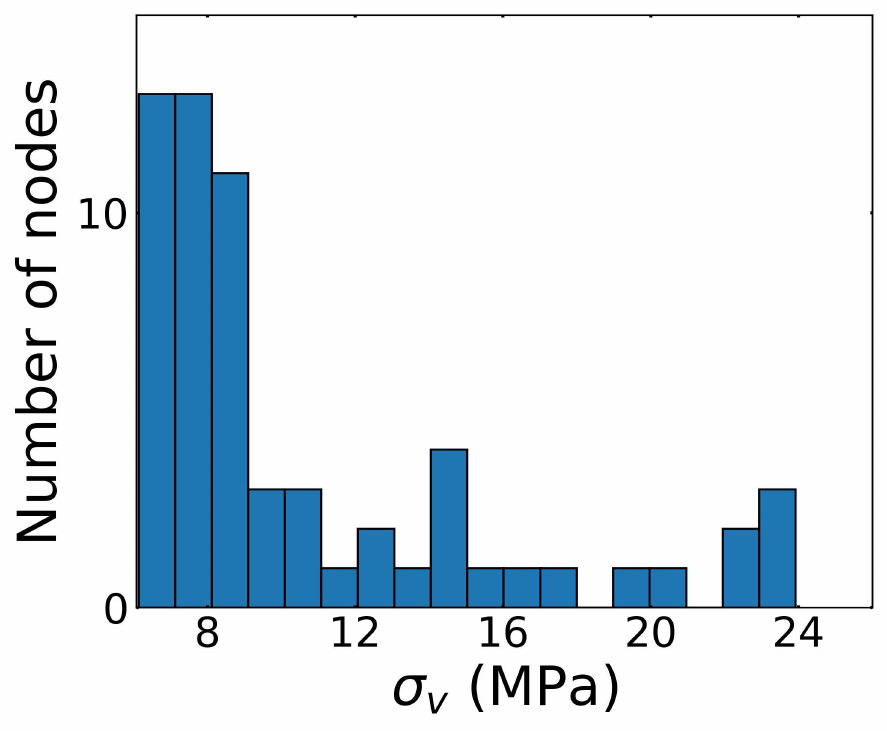}
        \caption{}
        \label{}
    \end{subfigure}
    \hfill
    \begin{subfigure}{0.32\textwidth}
     \centering
        \includegraphics[width=\linewidth]{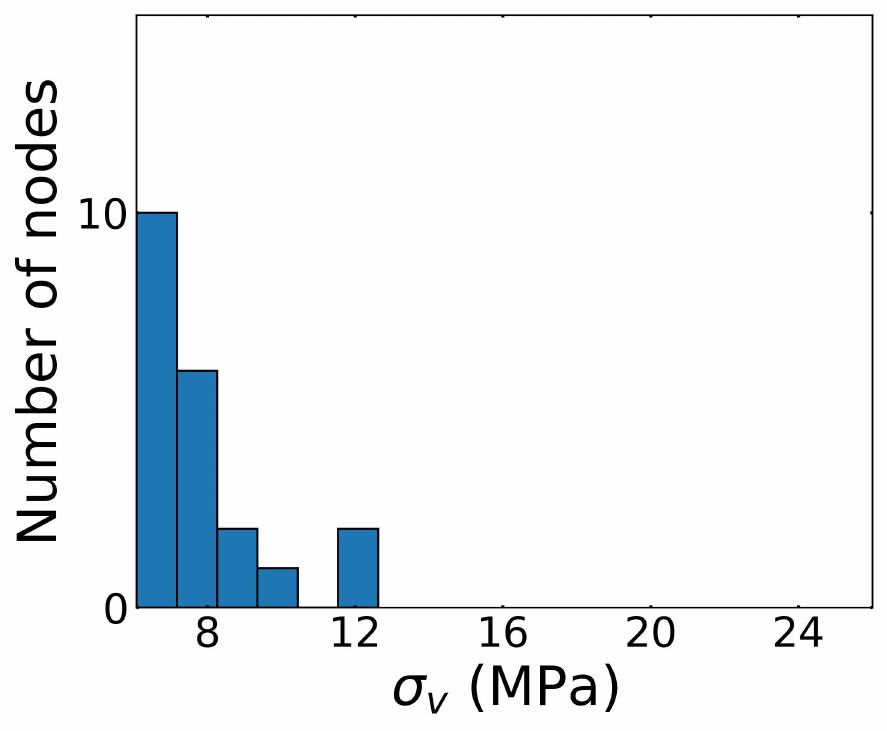}
        \caption{}
        \label{}
    \end{subfigure}
    \hfill
    \begin{subfigure}{0.32\textwidth}
     \centering
        \includegraphics[width=\linewidth]{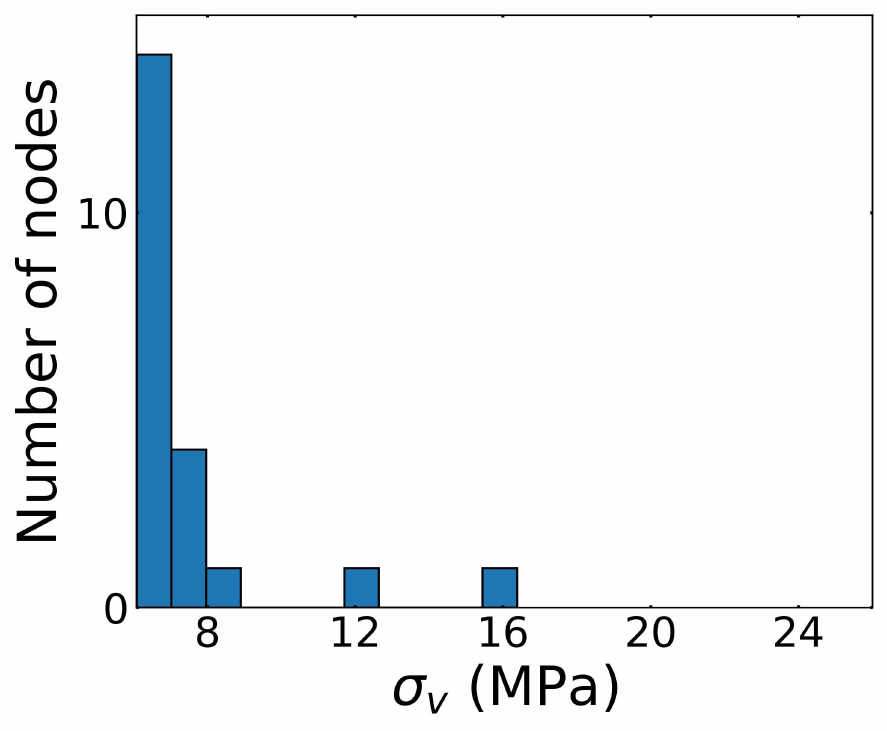}
        \caption{}
        \label{}
    \end{subfigure}

    \caption{Histogram of the number of  nodes having $\sigma_v$ $\geq$ 6.1 MPa (Note that the reference value of $\sigma_v$ is chosen as the $99.5^{th}$ percentile of the optimal design, considering conventional implementation) for the optimal structure obtained by (a) conventional implementation, (b) GRF with length scale of 20 mm, and (c) GRF with length scale of 30 mm.}
    \label{hist_cb_thermal}
\end{figure}

The optimized profiles are demonstrated in Fig.~\ref{cb_thermal}, and the corresponding deformed configuration is shown in Fig.~\ref{deformed_cb_thermal}. The evolution of the best profile with GA generation is shown in Fig.~\ref{fitness_cb_thermal} (Appendix). The conventional implementation optimal profile exhibits a maximum deflection of 6.96 mm, followed by 6.84 mm, 6.30 mm, and 6.10 mm for the 10 mm, 20 mm, and 30 mm length-scale profiles, respectively. The histogram of the $\sigma_v$ $\geq$ $\sigma^*$ is shown in Fig.~\ref{hist_cb_thermal}. Where $\sigma^*$ is the $99.5^{th}$ percentile value of the $\sigma_v$ in the conventional implementation.
The maximum $\sigma_v$ for the conventional implementation profile is 23.92 MPa, while for the optimal profile based on the GRF-scheme is 20.11 MPa, 12.64 MPa, and 16.35 MPa for the length scales 10 mm, 20 mm, and 30 mm, respectively. 

In general, numerical examples show that the objective function values from conventional implementation and from GRF-based optimization are close to each other. However, the GRF-based optimization scheme consistently provides smoother designs, which, in turn, reduce the design's susceptibility to stress concentration compared to the conventional implementation.

\section{Conclusions}
\label{Conclusion}

In this work, we proposed an optimization framework for the functionally graded lattice structures. The approach includes a GRF/GPR-based FGL structure design-generation scheme and a projection operator that maintains the smoothness of the design parameters during each optimization iteration. The efficacy of the proposed framework is demonstrated through various numerical problems consisting of centered rectangular and re-entrant unit cells. The results clearly demonstrate the superiority of the proposed scheme for designing FGL structures, as it enables a smooth transition of the design parameters throughout the domain. This continuous variation enhances structural strength by minimizing abrupt changes and reducing stress concentration within the structure.

\FloatBarrier
\appendix

\section{}  
\setcounter{figure}{0}

\begin{figure}[h!]
    \centering

    \begin{subfigure}[b]{0.4\textwidth}
        \centering
        \includegraphics[width=\textwidth]{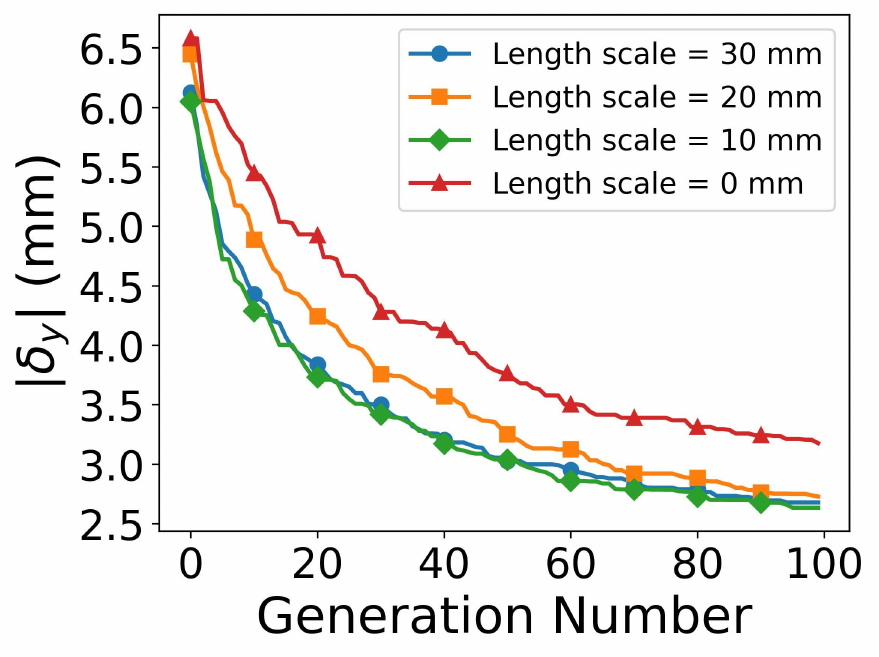}
        \caption{}
        \label{fitness_cb}
    \end{subfigure}
    \hspace{1.5cm}
    \begin{subfigure}[b]{0.4\textwidth}
        \centering
        \includegraphics[width=\textwidth]{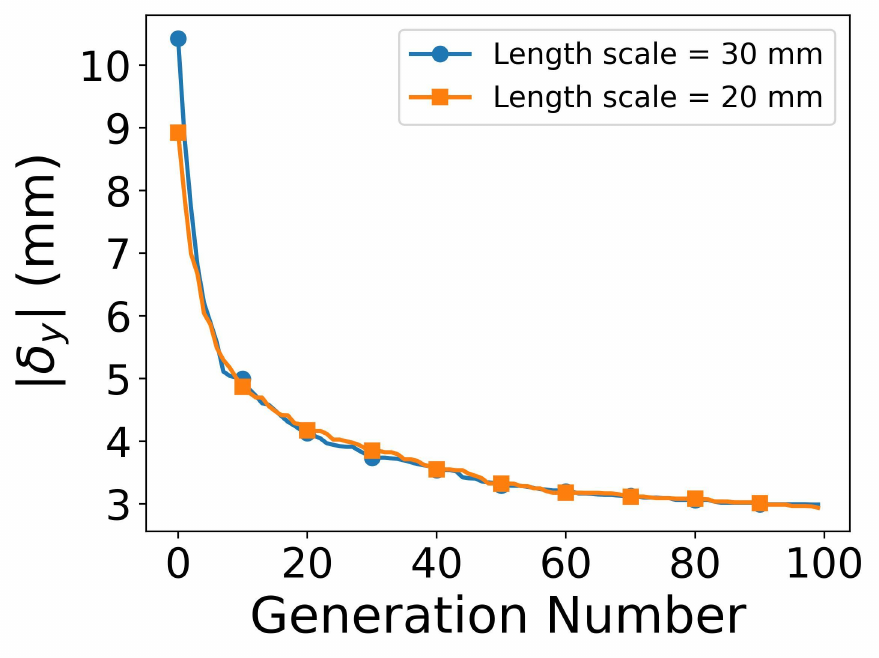}
        \caption{}
        \label{fitness_cb_gpr}
    \end{subfigure}

    \vspace{0.5cm}

    \begin{subfigure}[b]{0.4\textwidth}
        \centering
        \includegraphics[width=\textwidth]{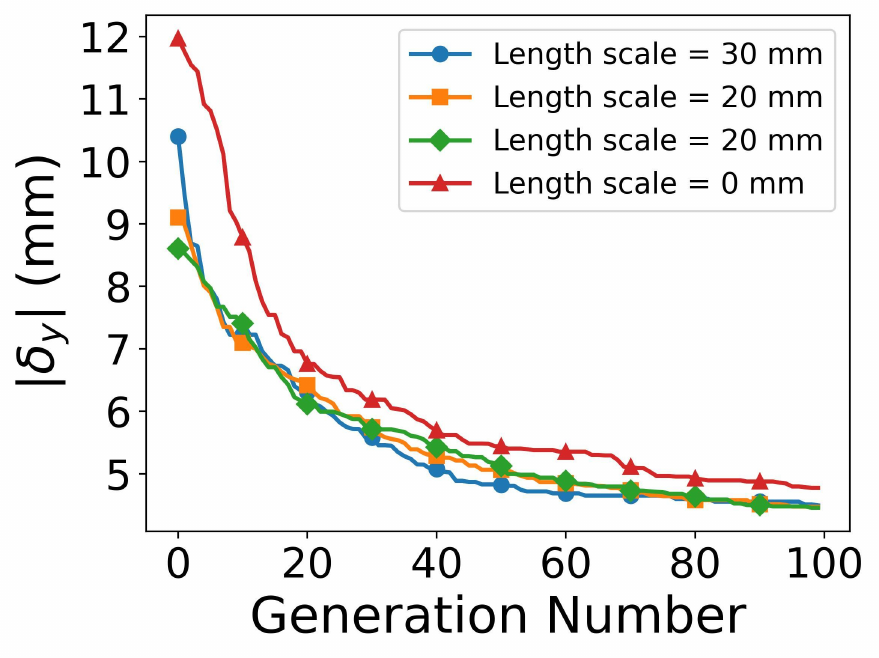}
        \caption{}
        \label{fitness_mbb}
    \end{subfigure}
    \hspace{1.5cm}
    \begin{subfigure}[b]{0.4\textwidth}
        \centering
        \includegraphics[width=\textwidth]{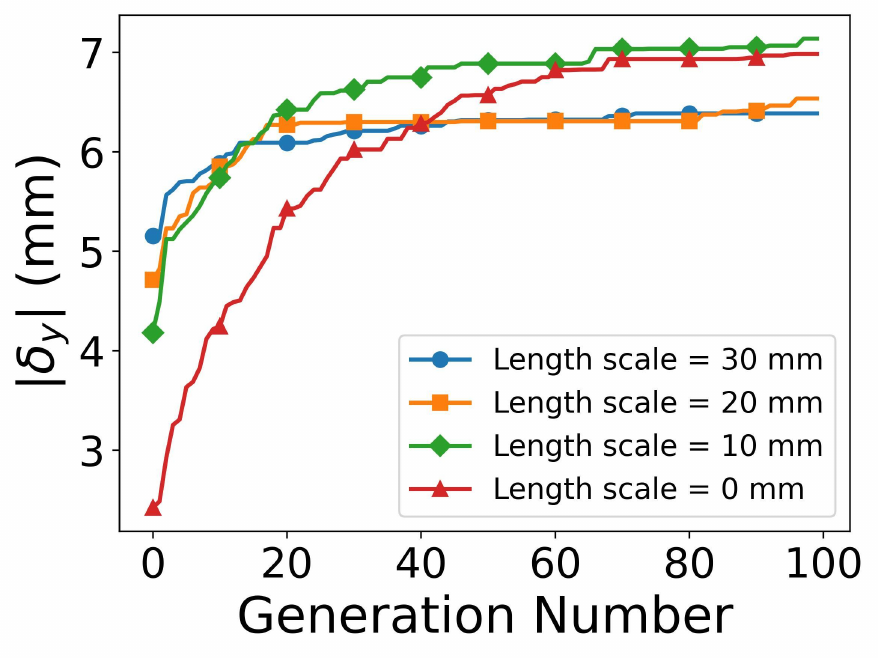}
        \caption{}
        \label{fitness_cb_thermal}
    \end{subfigure}

    \caption{Evolution in the deflection value for the best individual with respect to the GA generation, for the problem (a) cantilever beam subjected to the point load, (b) cantilever beam under the constraint of the maximum thickness at the leftmost unit cells, (c) half-MBB beam under point load, and (d) cantilever beam subjected to the thermal loading.}
    \label{}
\end{figure}

\bibliographystyle{elsarticle-num}
\bibliography{ref}
\end{document}